\documentclass[a4paper,12pt,english]{book}

\usepackage{amsmath,amssymb}
\usepackage[english]{babel}
\usepackage{tikz}
\usepackage[normalem]{ulem}
\usepackage{indentfirst}
\usepackage[T1]{fontenc}
\usepackage[utf8]{inputenc}
\usepackage{graphicx}
\usepackage{graphics}
\usepackage{xcolor}
\usepackage{epigraph, varwidth}
\usepackage{pdfpages}
\usepackage{setspace}
\usepackage[font=small,labelfont=bf]{caption}
\usepackage{pgfplots}
\usepackage{epstopdf}
\usepackage{subfig}
\usepackage{afterpage}
\usepackage{array}
\usepackage{emptypage}
\usepackage[retainorgcmds]{IEEEtrantools}
\usepackage[acronym,nonumberlist,sort=use]{glossaries}
\usepackage{enumitem}
\usepackage[colorlinks=true, allcolors=black]{hyperref}
\usepackage{caption}
\usepackage[super]{nth}
\usepackage{cite}
\usepackage{import}
\usepackage{varwidth}

\newcommand{\be}{\begin{equation}}
\newcommand{\ee}{\end{equation}}
\newcommand{\bal}{\begin{aligned}}
\newcommand{\eal}{\end{aligned}}
\newcommand{\beq}{\begin{eqnarray}}
\newcommand{\eeq}{\end{eqnarray}}
\newcommand{\ba}{\begin{array}}
\newcommand{\ea}{\end{array}}

\newcommand{\bi}{\begin{itemize}}
\newcommand{\ei}{\end{itemize}}
\newtheorem{theorem}{Theorem}
\newcommand{\bt}{\begin{theorem}}
\newcommand{\et}{\end{theorem}}

\usepackage[bottom=2cm,top=3cm,left=3cm,right=2cm]{geometry}

\newglossary{abbrev}{abs}{abo}{List of Abbreviations}
\newglossaryentry{IC}{
    name        = IC,
    description = Initial condition,
    type        = abbrev
}
\newglossaryentry{SM}{
    name        = SM,
    description = Standard map,
    type        = abbrev
}
\newglossaryentry{CTRW}{
    name        = CTRW,
    description = Continuous time random walk,
    type        = abbrev
}
\newglossaryentry{FDE}{
    name        = FDE ,
    description = Fractional diffusion equation,
    type        = abbrev
}
\newglossaryentry{UPO}{
    name        = UPO,
    description = Unstable periodic orbit ,
    type        = abbrev
}
\newglossaryentry{BM}{
    name        = BM,
    description = Boozer map or Single-null divertor map,
    type        = abbrev
}
\newglossaryentry{UM}{
    name        = UM,
    description = Ullmann map or Ergodic magnetic limiter map,
    type        = abbrev
}
\newglossaryentry{$ER$}{
    name        = $\boldsymbol{ER}$,
    description = Escape rate,
    type        = abbrev
}
\newglossaryentry{RM}{
    name        = RM,
    description = Recurrence matrix,
    type        = abbrev
}
\newglossaryentry{RP}{
    name        = RP,
    description = Recurrence plot,
    type        = abbrev
}
\newglossaryentry{RQA}{
    name        = RQA,
    description = Recurrence quantification analysis,
    type        = abbrev
}
\newglossaryentry{$RR$}{
    name        = $\boldsymbol{RR}$,
    description = Recurrence rate,
    type        = abbrev
}
\makenoidxglossaries

\begin{document}

\pagenumbering{gobble}

\includepdf[pages=-]{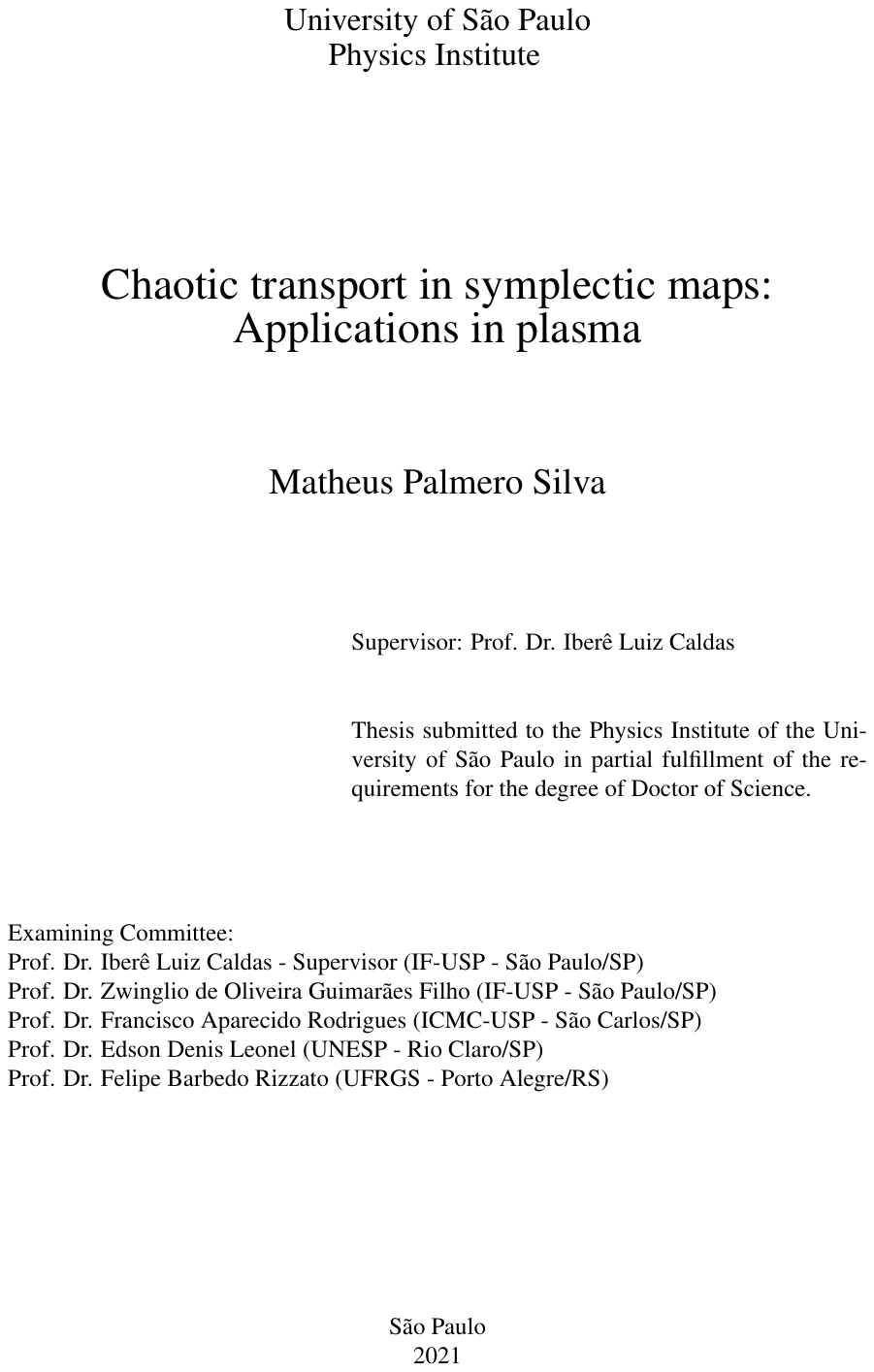}


\begin{center}
\section*{Acknowledgments}
\end{center}

I might be wrong about this etymological analysis (since I am not a linguist and do not speak German), but here it goes:

\vspace{0.5cm}

{\it Dankbarkeit: dankbar + -keit} is the German word for {\it gratefulness}; {\it Dankbar: dank + -bar}, that translates to {\it grateful} is composed of {\it dank} that is essentially thanks, gratitude or appreciation, and (curiously) the suffix {\it -bar}, related to the suffix {\it -able} in English, that means you are capable of performing this specified action. Moreover, the suffix {\it -keit} is related to the suffix {\it -ness}, meaning a state or condition. So, in a free translation, {\it Dankbarkeit} is the (difficult, underestimated and sometimes wrongly associated with a self-improvement trend preached by "hashtag culture") ability to be in the state of gratitude (an antidote to dissatisfaction - \url{https://youtu.be/WPPPFqsECz0}).

\vspace{0.5cm}

If I assume (only for this moment) that I have this ability (many times lost in rush and anxiety), I am extremely grateful for all elements and circumstances that brought me here. I thank God because I believe He aligned these factors in my life;

\vspace{0.5cm}

I thank my mom, Luciane, and my dad, José, for their never-ending trust, love and kindness;

\vspace{0.5cm}

I thank my girlfriend, Raquel, for choosing to be my partner and for her love, patience and faith in me;

\vspace{0.5cm}

I thank CREW, my group of best friends, for the important happy moments of distraction and friendship;

\vspace{0.5cm}

I thank my supervisor, Iberê, for the highest belief in my competence and for teaching me that there always something nice to say. I thank my research friends, in special Vitor, Gabriel, David, Everton and Jean, for all the discussions and contributions to my professional growth;

\vspace{0.5cm}

I thank my German supervisor, Igor, for valuable discussions and for teaching me other approaches to science. I thank colleagues and professors from the Department of Physics at Adlershof for their attention and hospitality;

\vspace{0.5cm}

I thank the Institute of Physics, from USP, for the infrastructure and support. And I thank the São Paulo Research Foundation (FAPESP), processes 2018/03000-5 and 2020/12478-6, for the financial support.

\vspace{0.5cm}

My state of gratitude is due to all of you.

\newpage
\vspace*{\fill}
\epigraph{"Chaos isn't a pit. Chaos is a ladder."}{Petyr Baelish, Game of Thrones}

\newpage
\vspace*{\fill}
\begin{flushright}
\begin{varwidth}{\linewidth}\centering
\includegraphics[scale=0.25]{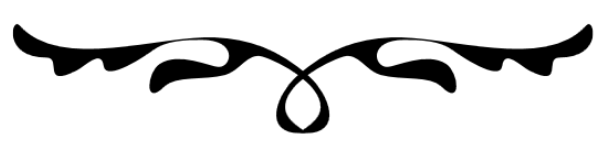}\\
To the memory of my grandfather, Dorival Palmero.
\end{varwidth}
\end{flushright}

\chapter*{Abstract}

\thispagestyle{empty}
Chaotic transport is related to the complex dynamical evolution of chaotic trajectories in Hamiltonian systems, which models various physical processes. In magnetically confined plasma, it is possible to qualitatively describe the configuration of the magnetic field via the phase space of suitable symplectic maps. These phase spaces are of mixed type, where chaos coexists with regular motion, and the complete understanding of the chaotic transport is a challenge that, when overcomed, may provide further knowledge into the behaviour of confined fusion plasma. In this research, we focus our investigation on properties of chaotic transport in mixed phase spaces of two symplectic maps that model the magnetic field lines of tokamaks under distinct configurations. The single-null divertor map, or Boozer map, models the field lines of tokamaks with poloidal divertors. The ergodic magnetic limiter map, or Ullmann map, models the magnetic configuration of tokamaks assembled with ergodic magnetic limiters. We propose two numerical methods developed to study and illustrate the differences between the transient behaviour of open field lines in both models while considering induced magnetic configurations that either enhance or restrain the escaping field lines. The first analysis shows that the spatial organisation of invariant manifolds creates fitting transport channels for the open field lines, influencing the average dynamical evolution in the selected magnetic configurations. The second identifies trajectories that widely differs from the average chaotic behaviour, specifically detecting the stickiness phenomenon, which can be related to additional confinement regions in the nearest surroundings of magnetic islands in the plasma edge. These analyses may, ultimately, assist in selecting optimal experimental parameters to achieve specific goals in tokamak discharges. 

\vspace{.5cm}
\textbf{Keywords}: Hamiltonian systems; Chaos; Tokamak.

{\normalsize\tableofcontents}
\thispagestyle{empty}

{\normalsize\listoffigures}
\thispagestyle{empty}

\glsaddall
{\printnoidxglossary[type=abbrev]}
\thispagestyle{empty}

\mainmatter

\setcounter{page}{13}

\chapter{Introduction}
\label{chap1:intro}

The study of dynamics improves our comprehension of natural phenomena. Ranging from the movement of celestial bodies to the behaviour of particles at the atomic level, dynamics appear everywhere since we are far from the stillness of absolute zero. In this sense, movement is a fundamental aspect of nature that can help us understand the underlying principles that govern everything. By investigating the various forms of dynamics, we can deepen our knowledge of the systems that make up the world around us. 

In physics, we adopt the concept of dynamical systems from mathematics to describe the evolution of any given system over time \cite{Arrowsmith1990,Hirsch2013}. Dynamical equations often provide a suitable mathematical framework for understanding the relations between the spatial displacement of any moving entity and its time evolution. Depending on the system under consideration, these equations can take various forms. When investigating complex systems such as the weather \cite{Lorenz1963, Donges2009}, the human brain \cite{Bullmore2009,Bassett2011}, and ecosystems \cite{Jorgensen1992,Baurmann2007}, dynamical models often require non-linear approaches to deal with unexpected outcomes. 

In contrast to linear systems, which exhibit an expected response to external stimuli, non-linear systems can display unexpected dynamical behaviour. These systems are often characterised by intricate feedback mechanisms, where small changes in one part can cause significant changes in the overall system's evolution. This notion is inherently related to chaos theory, an interdisciplinary area of scientific study focused on underlying patterns and deterministic laws of dynamical systems that are highly sensitive to initial conditions \cite{Strogatz2000,Ott2002,Alligood2012,Wiggins2003}.

Chaotic dynamics, from the classical mechanics' point of view, is closely connected to Hamiltonian systems, an important class of dynamical systems. Near-integrable Hamiltonian systems, in particular, can exhibit rich dynamical scenarios with the coexistence of periodic and chaotic motion \cite{Lichtenberg1992}. The Hamiltonian description appropriately describes different problems in several fields of physics. Examples include celestial mechanics, investigating the motion of natural satellites around planets \cite{Contopoulos2004}; Fluid dynamics, modelling mixing and transport of fluids \cite{Morrison1998}; Acoustic and optics applications, computing trajectories of propagating waves \cite{ElBoudouti2009} and; Plasma physics, modelling the configuration of magnetic field lines for the magnetic confinement of fusion plasma \cite{Evans2015}. The volume-preserving property of Hamiltonian systems is one of their most essential features, making them suitable for this wide range of applications.

In particular cases of interest, many near-integrable Hamiltonian systems can be reduced to non-linear area-preserving maps that, in specific setups, are classified as \emph{symplectic maps} \cite{Meiss1992}. Essentially, symplectic maps can emulate the general dynamical behaviour of higher dimensional dynamical systems via discrete mapping functions that relate variables at future \emph{states} to their past ones. These iterative relations are most suitable for numerical algorithms and simulations. In this study, we numerically explore different features of non-linear and chaotic dynamics provided by symplectic maps that models weakly perturbed Hamiltonian systems, specifically in the application context of magnetic confinement of fusion plasma.  

Fusion is the process of merging two or more atomic nuclei to form a heavier nucleus, releasing a tremendous amount of energy in the process. Plasma fusion occurs when atomic nuclei are heated to extremely high temperatures, causing them to ionise forming plasma and, in this state, the positively charged nuclei can overcome their natural repulsion and come close enough to undergo fusion reactions. These reactions power the sun and other stars in the universe. Scientists and engineers have been working to harness this energy on Earth since the 30s \cite{Harms2002,Clery2014}, as it has the potential to provide a virtually limitless source of clean and sustainable power. Fusion can change the world's energy paradigm, offering a safe, abundant, and environmentally friendly alternative to traditional sources. Despite the challenges associated with achieving sustainable fusion energy, research and development efforts are ongoing, and significant progress has been made towards realising the goal of fusion power \cite{Krakowski1991,Nakai2004,Loarte2014,Wurzel2022}. One of the most promising approaches to plasma fusion is known as \emph{magnetic confinement} fusion, which involves the use of strong magnetic fields to confine and heats the plasma. A classical example of a machine designed to handle this problem is the tokamak.

The tokamak, transliteration from the Russian acronym for "Toroidal Chamber with Magnetic Coils", is one of the most promising machines of magnetically confined plasma to achieve thermonuclear fusion \cite{Wesson2004,Ikeda2010,Horton2015}. In essence, a tokamak is a toroidal metallic chamber, immersed in a suitable configuration of magnetic fields. Those fields are intended to confine the charged particles of ionised gas at high temperatures inside the chamber. To efficiently produce energy, important problems are still under investigation, expected to have further definitive results over the next few years in experiences on the ITER \cite{Horton2015,Aymar2002,Li2016} tokamak, an international project for the most ambitious and expensive machine in human history.  

In modern tokamaks, like ITER, a well-investigated subject for controlling the magnetic confinement is the topology of the magnetic field lines \cite{Robinson1993,Stacey2010,Ongena2016}, so-called \emph{magnetic configuration} of the system. This configuration is constantly perturbed by natural or induced oscillations to control the plasma \cite{Connor1993}. In the plasma core, located at the centre of the tokamak, magnetic field lines form toroidal magnetic surfaces that are stable for the duration of a typical discharge. However, in the plasma edge located near the inner wall of the tokamak chamber, resonant magnetic perturbations may break those stable surfaces forming unstable regions with open field lines that can escape the confinement, dragging particles from the plasma towards the inner wall of the tokamak chamber. This process, when not controlled, may damage the machine.

Through the lens of nonlinear dynamics, open magnetic field lines are closely related to chaotic orbits \cite{Morrison2000, Abdullaev2014} wandering through the chaotic portion of the system's \emph{phase space}. The former is a mathematical space in which all possible states of a system are represented. In this sense, modelling magnetic field lines via Hamiltonian maps is to investigate magnetic configurations given by the phase spaces of the models. Moreover, open field lines in tokamaks can be controlled by applying external electric currents or magnetic devices that alter the configuration on the plasma edge, altering the phase spaces of the models as well.

Experimental evidence \cite{Finken2006,Evans2006} and theoretical models \cite{Kroetz2012,Ciro2016} suggest that controlling the open magnetic field lines is one of the key factors for managing the transport of particles at the plasma edge. This transport is considered to be \emph{anomalous} \cite{Horton2014,Negrete2006} because it deviates from the expectations of normal diffusion. Investigating the anomalous transport through phase space analysis of the field line maps is a theoretical/numerical approach that may provide insight into the confined plasma behaviour and plasma-wall interactions \cite{Janeschitz2001,Samm2010}. 

In this research, we focus on understanding aspects of \emph{chaotic transport}, i.\ e.\ how is the dynamical behaviour of open field lines, or chaotic trajectories, considering their evolution before escaping the system. For that, we conduct thorough numerical investigations of \emph{mixed phase spaces} in symplectic maps that model tokamaks under two distinct perturbation regimes, or different magnetic configurations caused by the introduction of external devices. Essentially, by enhancing our comprehension of these \emph{transient} dynamical behaviours, we may, ultimately, assist in selecting optimal experimental parameters to achieve specific goals in tokamak discharges. 

The selected symplectic maps are the \emph{single-null divertor map}, also known as Boozer map \cite{Punjabi1992}, which models the configuration of the magnetic field in tokamaks with \emph{poloidal divertors}, and the \emph{ergodic magnetic limiter map}, or Ullmann map \cite{Ullmann2000}, that models the magnetic field lines of tokamaks assembled with \emph{ergodic magnetic limiters}. We propose a methodology, based on the rate of escaping field lines, that defines appropriate values for the \emph{perturbation strengths} to either enhance or restrict the escape in both systems. With that, we employ two different numerical methods to investigate and illustrate the intrinsic differences between the magnetic configurations that enhance or restrict escaping field lines.

Finally, the numerical methods developed to assist our understanding of specific properties of chaotic transport in the field lines maps are based on analyses of \emph{transient motion} and \emph{recurrences}. First, we show that the transient dynamical behaviour of open field lines, experienced by chaotic trajectories before escaping the systems, is constantly influenced by the spatial organisation of \emph{invariant manifolds} \cite{Ozorio1990} in the models' mixed phase spaces. The intertwined manifolds create and destroy transport channels associated with the enhancement or restriction of escaping field lines. Furthermore, we show a recurrence-based detection for the \emph{stickiness phenomenon} \cite{Contopoulos2010} experienced by chaotic trajectories. This detection allows us to identify field lines that widely differ from the general behaviour, suggesting the existence of additional temporal confinement regions in the nearest surroundings of magnetic islands in the plasma edge.    

This thesis is organised as follows. In Chapter \ref{chap2:hamiltonian}, we review and introduce the main concepts of Hamiltonian systems that are relevant to our work, discussing chaotic dynamics and features of chaotic transport using a well-studied pragmatic example of symplectic maps. In Chapter \ref{chap3:models}, we present and discuss in detail the selected models to study tokamaks under two different magnetic perturbations. We also explain a methodology, based on the rate of escaping field lines, that defines appropriate values for the perturbation strengths used throughout our numerical analyses. Chapters \ref{chap4:transient} and \ref{chap5:recurrence} present, therefore, our methods designed to investigate the aforementioned features of chaotic transport, namely the dynamical influence of invariant manifolds and the stickiness phenomenon. Finally, in Chapter \ref{chap6:conclusions}, we draw the conclusions regarding our investigations, presenting also a compilation of our scientific productions, including the manuscripts and the open source code that resulted from this research.  

\chapter{Hamiltonian systems and Chaos}
\label{chap2:hamiltonian}

The second Maxwell's equation $\nabla \cdot \mathbf{B} = 0$, based on Gauss's law for magnetism, essentially allows a Hamiltonian description for the magnetic field. In this chapter, we review some fundamental concepts in Hamiltonian systems, focusing on symplectic maps and their application in modelling magnetic field lines. Then, we briefly discuss integrability and the rise of chaotic dynamics in near-integrable Hamiltonian systems.

After the initial review, we further explain important concepts for our investigations throughout this thesis such as phase space analysis, initial condition sensitivity of chaotic trajectories and the important notion of chaotic transport, all considering a pragmatic example of symplectic non-linear maps: The Chirikov standard map.

\section{Hamiltonian dynamics}
\label{sec:hamilto}
A \emph{dynamical system} that can be characterised by a scalar function $\mathcal{H}({\boldsymbol {q}},{\boldsymbol {p}})$, known as the Hamiltonian, is classified as a \emph{Hamiltonian system} \cite{Arrowsmith1990}. The system's state is specified by its \emph{generalised momentum} ${\boldsymbol {p}}$ and \emph{position} ${\boldsymbol {q}}$, where both are real-valued vectors with the same dimension $N$. Hamilton's equations determine $\boldsymbol {p}(t)$ and $\boldsymbol {q}(t)$ by the following

\begin{equation}
    \dot{\boldsymbol {p}} = -\frac{\partial \mathcal{H}}{\partial \boldsymbol {q}} ~, ~~\dot{\boldsymbol {q}} = \frac{\partial \mathcal{H}}{\partial \boldsymbol {p}}~. 
    \label{eq:hamilton}
\end{equation}The evolving states over time trace paths (or flows) commonly known as \emph{trajectories} (or orbits) in a 2$N$-dimensional \emph{phase space}. In that sense, the phase space is a mathematical space in which all possible states of a system are represented. 

A noteworthy outcome of Hamilton's Eqs.\ (\ref{eq:hamilton}) is that all Hamiltonian flows preserve volume in phase space, as per the well-known Liouville's theorem \cite{Ott2002}. This property ensures that phase space volumes are incompressible and that the system lacks attractors, a fundamental condition for modelling magnetic field lines.

In addition to the volume-preserving property, Hamiltonian systems also preserve a \emph{loop action}, or Poincaré's invariant \cite{MacKay2020}. This fact is used in the construction of Poincaré sections, where the dynamics of a $N$-dimensional continuous-time system can be emulated by a $(N-1)$-dimensional conjugate system. In that sense, it is fair to assume the existence of a special kind of transformation $\mathbf{T}$ that is, by construction, discrete in time intervals and responsible to \emph{map} a point $\boldsymbol{x}$ in the Poincaré section at time $t=n$ to its next iteration at $t=n+1$. Hence, one defines a discrete dynamical system given by the following mapping function 

\begin{equation}
    \boldsymbol{x}_{n+1} = \mathbf{T}(\boldsymbol{x}_n)~.
    \label{eq:map}
\end{equation}It is possible to show \cite{Moser2020} that the preservation of the loop action implies that a map constructed for any Hamiltonian flow is defined as a \emph{symplectic map}. In practical terms for us, symplectic maps possess two important features namely {\it (i)} they can be derived from a suitable generating function and; {\it (ii)} the area/volume-preserving property is ensured with the determinant of the Jacobian matrix equals unity.  

To aid the comprehension of these concepts let us consider a practical example involving magnetic field lines. This example introduces another key dynamical invariant known as the \emph{rotation number} which is fundamentally linked to a crucial quantity for tokamaks known as the \emph{safety factor}.

\subsection*{Magnetic field in Tokamaks}
In general terms, magnetic field lines are parallel curves, at any point in space, to a given arbitrary magnetic field $\mathbf{B}$, condition written in terms of the line element $\mathrm{d}\mathbf{l}$ as 

\begin{equation}
    \mathbf{B} \times \mathrm{d}\mathbf{l} = 0~.
    \label{eq:field_lines}
\end{equation}Considering initially $\mathbf{B}$ and $\mathrm{d}\mathbf{l}$ in a cylindrical coordinate system $(r,\theta,z)$ it follows

\begin{equation}
    \frac{\mathrm{d}r}{B_r}=\frac{r\mathrm{d}\theta}{B_{\theta}}=\frac{\mathrm{d}z}{B_z}~,
    \label{eq:dr/db}
\end{equation}which we rewrite in the form 

\begin{equation}
    \frac{\mathrm{d}r}{\mathrm{d}z}=\frac{B_r(r,\theta,z)}{B_z(r,\theta,z)}~,~~
    \frac{\mathrm{d}\theta}{\mathrm{d}z}=\frac{B_\theta(r,\theta,z)}{B_z(r,\theta,z)}~.
    \label{eq:rewrite}
\end{equation}Assuming now that we are interested in problems with a certain axial symmetry, it is possible to find an exact \emph{integral of movement} $H=H(r,\theta,z)$ that defines a family of surfaces $z=z(r,\theta,H)$ called \emph{magnetic surfaces} on which all magnetic field lines lay in. It can also be demonstrated that by imposing the integral $H$, typically defined as magnetic flux, the system of Eqs.\ (\ref{eq:rewrite}) can be reformulated in the Hamiltonian form, with the variable $z$ taking on the role of time. This is particularly better visualised on toroidal coordinates $(r,\theta,\phi)$ that suitably describe the tokamak system, where the azimuthal coordinate $\phi$, periodic in $2\pi$, can now play the role of time.

\begin{figure}[h]
    \centering
    \includegraphics[scale=0.60]{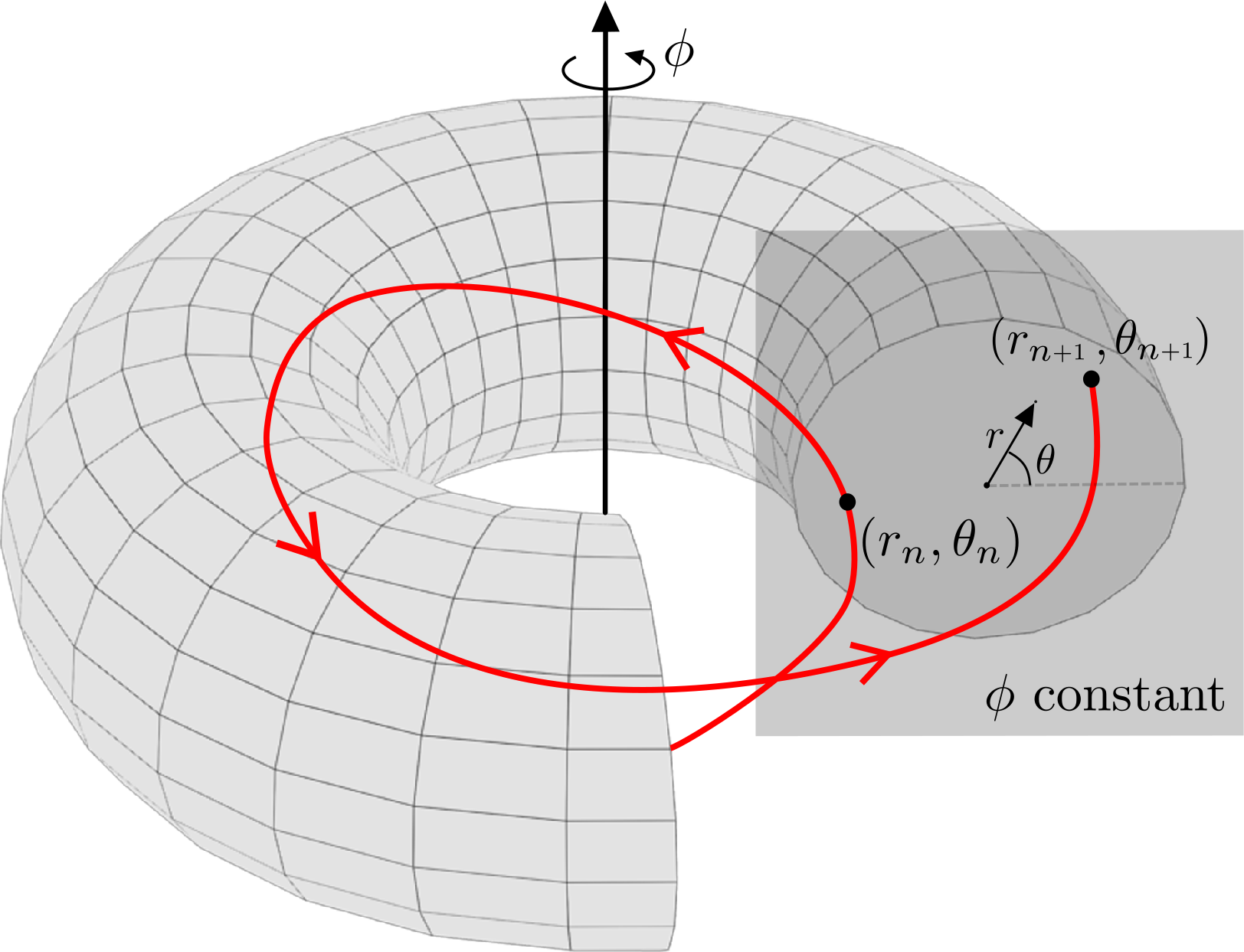}
    \caption[Schematic representation of mapping in tokamaks]{Schematic representation of mapping process for the magnetic field lines in a tokamak. The red curve represents one field line in a toroidal turn crossing the poloidal section at constant $\phi$ (in grey) in two distinct positions and moments. Both radial and poloidal coordinates $(r,\theta)$ are also drawn over the poloidal section.}
    \label{fig:coordinates_and_mapping}
\end{figure}

Indeed, relations described in Eq.\ (\ref{eq:rewrite}) allow the calculation of the path followed by a magnetic field line considering an arbitrary magnetic field $\mathbf{B}$ via integration. Since $z$ (or $\phi$) corresponds to a canonical time variable, the configuration of magnetic field lines in a tokamak can be represented via \emph{return maps}; The trajectories are obtained through continuous analysis, but we restrict to the values of their coordinates only when they return to a defined region, thereby fixing one variable. This procedure is similar to the construction of the aforementioned Poincaré sections where the dynamics are constrained in tori. Figure \ref{fig:coordinates_and_mapping} displays a representation of this procedure.

A return map over the periodic coordinate $\phi$ consists essentially of a Poincaré map \cite{Henon1982} over section $\phi = c$ to $\phi = c+2n\pi$, where $c$ is a constant and $n \in \mathbb{Z^+}$, defining what is also known as a \emph{stroboscopic map}. The pair $(r_n,\theta_n)$ denotes the coordinates over the defined poloidal section at time $t=n$, i.\ e.\ coordinates of the $n$-th intersection between a followed field line and the poloidal surface. Then, there exists an analytical transformation $T$ that relates the position of the trajectory at future time $(n+1)$ to its past position at time $n$ as follows  

\begin{equation}
    (r_{n+1}, \theta_{n+1}) = T(r_n,\theta_n)~,
    \label{eq:general_map}
\end{equation}defining a map for all intersection points of field lines with the poloidal surface, once provided with an initial value $(r_0,\theta_0)$ referred to as \emph{initial condition} (IC). Furthermore, the map is a recursive rule that is easily implemented in numerical simulations.

\subsubsection*{Equilibrium configuration $\mathbf{B}^0$}
Let us now consider a typical equilibrium magnetic configuration in a tokamak, essentially given by the following setup: Strong magnetic coils are assembled over the plasma chamber producing the toroidal magnetic field $B^0_\phi = B_0$; A toroidal electric current, known as the plasma current, is induced by non-static electric fields, giving rise to a poloidal magnetic field $B^0_\theta=B^0_\theta(r)$ that can be obtained via Ampere's law while considering a suitable electric current density function for typical tokamak discharges; The radial component of the magnetic field $B_r = 0$ is said to be null. The superposition of these components forms the typical magnetic configuration at equilibrium $ \mathbf{B}^0=(B^0_r = 0, B^0_\theta = B^0_\theta(r), B^0_\phi = B_0)$. 

In this case, the field lines equations shown in Eqs.\ (\ref{eq:dr/db}) are written as 

\begin{equation}
 \begin{aligned}
  \frac{\mathrm{d}r}{\mathrm{d}\phi}&=0~,\\
  \frac{\mathrm{d}\theta}{\mathrm{d}\phi}&=\frac{R_0B^0_\theta(r)}{rB^0_\phi}~,
 \end{aligned}
 \label{eq:tokamak_dr/db}
\end{equation}where $R_0$ is the tokamak's larger radius.  

Here we introduce an important concept from the dynamical systems theory known as the \emph{rotation number} $\iota$. Intuitively, the rotation number is the average angular rotation of a given orbit, that in our case can be defined as 

\begin{equation}
 \iota\equiv\left\langle\frac{\mathrm{d}\theta}{\mathrm{d}\phi}\right\rangle=\frac{1}{2\pi}\int^{2\pi}_0\left(\frac{\mathrm{d}\theta}{\mathrm{d}\phi}\right)\mathrm{d}\phi'~,
\label{eq:rotation_numer}
\end{equation}consisting, thereby, of the average poloidal displacement over toroidal turns. The normalisation factor $1/2\pi$ makes complete turns equal unity.

Rewriting Eq.\ (\ref{eq:tokamak_dr/db}) in terms of $\iota$ follows that 

\begin{equation}
 \begin{aligned}
  \frac{\mathrm{d}r}{\mathrm{d}\phi}&=0~,\\
  \frac{\mathrm{d}\theta}{\mathrm{d}\phi}&=\frac{R_0B^0_\theta(r)}{rB^0_\phi}=\iota(r)~,
 \end{aligned}
 \label{eq:tokamak_iota_dl}
\end{equation}which integrating from $\phi = c$ to $\phi = c+2n\pi$, where $c$ is constant and $n \in \mathbb{Z^+}$ i.\ e.\ between successive intersections with the poloidal section, is equivalent to the mapping $T^0$ given by the following dynamical equations 

\begin{equation}
T^0:\left\{\begin{array}{ll}
  r_{n+1}=r_n\\
  \theta_{n+1}=\theta_n+2\pi\iota(r_{n+1})~\mathrm{mod} 2\pi
  \end{array}
  \right..
  \label{eq:tokamak_mapping}
\end{equation}

The equilibrium map presented in Eq.\ (\ref{eq:tokamak_mapping}) is, by construction, a symplectic map. Hence, let us confirm the aforementioned features {\it (i)} and {\it (ii)}: 

\begin{itemize}
    \item[{\it (i)}] Let $G^0(r_{n+1},\theta_n)$ be a second-order generating function as follows 

    \begin{equation}
        G(r_{n+1},\theta_n)=r_{n+1}\theta_n+2\pi\int^{r_{n+1}}_0\iota(r')\mathrm{d}r'~.
        \label{eq:generating}
    \end{equation}\noindent The map should be derived from it by the relations 

    \begin{equation}
        r_n=\frac{\partial G}{\partial \theta_n}~,~~\theta_{n+1}=\frac{\partial G}{\partial r_{n+1}}~;\\
    \label{eq:relations}
    \end{equation}

    \item[{\it (ii)}] Let $J_{T^0}$ be the Jacobian matrix of the map as follows

    \begin{equation*}
        J_{T^0} =
        \begin{bmatrix}
        \dfrac{\partial r_{n+1}}{\partial r_n} & \dfrac{\partial r_{n+1}}{\partial \theta_n} \\
        \dfrac{\partial \theta_{n+1}}{\partial r_n} & \dfrac{\partial \theta_{n+1}}{\partial \theta_n} \\
        \end{bmatrix}=
        \begin{bmatrix}
        1 & 0 \\
        2\pi \dfrac{\partial \iota(r_{n+1})}{\partial r_n}  & 1\\
        \end{bmatrix}~.
    \end{equation*}The determinant $\mathrm{det} (J_{T^0})$ should be equal unity.
\end{itemize}
These features assured, the map $T^0$ is, indeed, a symplectic map. 

\subsubsection*{Safety factor}
The aforementioned concept of rotation number is strictly connected to the \emph{safety factor} $q = q(r)$, an important quantity that essentially measures the pitch of the helical magnetic field lines in tokamaks. At equilibrium configuration $q(r) = q^0(r)$ can be calculated considering Eqs.\ (\ref{eq:tokamak_dr/db}) as 

\begin{equation}
q^{0}(r)=\frac{rB^0_\phi}{R_0B^0_{\theta}(r)}~.
\label{eq:q0}
\end{equation}Note that it is, thereby, the inverse of the rotation number $\iota = 1/q$. This means that the safety factor may be interpreted as the ratio between the number of toroidal cycles to the poloidal ones for any field line on a given magnetic surface. Hence, a magnetic surface possesses a given value of $q = a / b$. If the ratio $a / b$ is a rational number, $a, b \in \mathbb{N}$ and a field line that lay on this surface closes on itself after $a$ toroidal turns and $b$ poloidal turns. In this case, $a$ and $b$ are said to be \emph{commensurable}. 

However, there will be the case where the ratio $a / b$ assumes irrational values, meaning that the field line on
the corresponding surface will never be able to close itself, densely filling said surface. This is the case when we have an \emph{incommensurable} number of toroidal and poloidal cycles.

Therefore, on one hand, we have rational magnetic surfaces constructed by wrapping periodic field lines. On the other, we have irrational surfaces where one single magnetic field line fills the entire region. Indeed, when we consider natural or forced-induced magnetic perturbations to the system, rational magnetic surfaces can be destroyed and, consequently, the phase space can be layered by \emph{chaotic regions}. To study that, we briefly discuss non-integrable Hamiltonian systems and the KAM theory as follows.

\subsection*{Integrability}
In dynamical systems theory, integrability is a well-studied property with various notions that fit different settings to distinct problems \cite{Tabor1989}. Here we sternly restrain this concept to the idea of an explicit determination of solutions considering specific Hamiltonian systems that can be written in terms of \emph{action-angle} variables $\mathcal{H} = \mathcal{H}(\boldsymbol{I},\boldsymbol{\theta})$. In this special case, the Hamiltonian depends only upon the actions $\mathcal{H}=\mathcal{H}(\boldsymbol{I})$ and the equations of motion shown in Eq.\ (\ref{eq:hamilton}) become as simple as 

\begin{equation}
    \dot{\boldsymbol{I}} = 0 ~, ~~\dot{\boldsymbol {\theta}} = \Omega(\boldsymbol{I}),
    \label{eq:action_angle}
\end{equation}where $\Omega$ can be understood as natural frequency vector; Solutions are $\boldsymbol{\theta}$-angle trajectories moving along the invariant torus at $\boldsymbol{I}=\boldsymbol{I_0}$ with fixed frequency $\Omega$. This simple setup is an example of an integrable Hamiltonian system. 

Now, as a convenient example, let us consider a two-dimensional dynamical system given by the following Hamiltonian

\begin{equation}
    H(I_1,I_2,\theta_1,\theta_2) = H^0(I_1,I_2) + \epsilon H^1(I_1,I_2,\theta_1,\theta_2)~,
    \label{eq:h0+h1}
\end{equation}where $I_i$ and $\theta_i$, $i=1,2$ represents the action and angle variables respectively. Here, $H^0$ corresponds to the \emph{integrable} part and, conversely $H^1$ is related to a \emph{non-integrable} part that is controlled through the parameter $\epsilon$, often referred to as the \emph{control parameter}.

It is also worth mentioning that, similar to what was presented in Eq.\ (\ref{eq:action_angle}), solutions of Eq.\ (\ref{eq:h0+h1}) for $\epsilon \neq 0$ possess two natural frequencies $\Omega^0$ and $\Omega^1$. Depending on the ratio $\Omega^0 / \Omega^1$ of these frequencies, the system may present \emph{resonances} that modify the phase space configuration, interfering also with the trajectories' stability. In addition, the number of resonances that can be created on the phase space depends on the value of the control parameter $\epsilon$ \cite{Leonel2015}.  

Furthermore, since Eq.\ (\ref{eq:h0+h1}) is an autonomous Hamiltonian i.\ e.\ does not explicitly depends on time, the energy of the system is constant and we reduce one of the four $(I_1,I_2,\theta_1,\theta_2)$ in terms of it. Therefore, the system is now described by three variables that may possess solutions in the three-dimensional space intercepting an arbitrary Poincaré section at $\theta_2$ constant for instance. It follows that the Hamiltonian flux can be mapped by a return map over the section at the plane $I_1 \times \theta_1$ with $\theta_2$ constant. 

In that sense, a generic mapping $T^{\epsilon}$ is proposed to map the points of these trajectories while crossing the aforementioned section. $T^{\epsilon}$ is written by the following dynamical equations 

\begin{equation}
    T^{\epsilon}:\left\{\begin{array}{ll}
    I_{n+1} = I_n + \epsilon f(\theta_n,I_{n+1})\\
    \theta_{n+1}=\theta_n + h(I_{n+1}) + \epsilon l(\theta_n,I_{n+1})
    \end{array}
    \right.,
\label{eq:genDyn}
\end{equation}where $f = f(\theta_n,I_{n+1})$, $h = h(I_{n+1})$ and $l = l(\theta_n,I_{n+1})$ are adjustable functions. Indeed, with suitable change of coordinates and a specific function $h(I_{n+1})$ it is possible to show that, for $\epsilon = 0$, the generic map on Eq.\ (\ref{eq:genDyn}) is reduced to $T^0$ previously show in Eq.\ (\ref{eq:tokamak_mapping}). For $\epsilon \neq 0$, Eq.\ (\ref{eq:genDyn}) describes a family of Hamiltonian mappings \cite{Leonel2015} found throughout the literature on non-linear and chaotic dynamics. Moreover, since we reduced a Hamiltonian flow to a map $T^{\epsilon}$, we can verify the area-preserving property via $\mathrm{det} (J_{T^{\epsilon}})$, imposing the following relation  

\begin{equation}
    \frac{\partial l(\theta_n,I_{n+1})}{\partial \theta_n} + \frac{\partial f(\theta_n,I_{n+1})}{\partial I_{n+1}} = 0~.
    \label{eq:function_relations}
\end{equation}

In practical terms, the Hamiltonian system from Eq.\ (\ref{eq:h0+h1}) and, consequently the map in Eq. (\ref{eq:genDyn}) are integrable systems if $\epsilon = 0$, and non-integrable if $\epsilon \neq 0$. Interestingly, the case $0 < \epsilon \ll 1$ gives rise to \emph{near-integragle Hamiltonian systems}, or also referred to as weakly perturbed Hamiltonian systems, an important class of system that are the subject of the KAM theory \cite{Broer2004}. 

For our intents and purposes, the KAM theory states that a small volume-preserving perturbation applied to an integrable Hamiltonian system would still preserve a finite fraction of regular trajectories conforming to defined KAM tori. The remaining fraction of trajectories would exhibit chaotic motion characterised by the well-known sensitivity to initial conditions. Regular trajectories, when examined in a surface of section, manifest as invariant circles\footnote{Or invariant sets of points corresponding to the periodic orbits.} which are often referred to as \emph{KAM islands} (or \emph{magnetic islands} in our context of application); Chaotic trajectories densely fill a finite region of the same surface, a region commonly known as the \emph{chaotic sea}. This constitutes the fundamental dynamical scenario of a near-integrable Hamiltonian system, where its phase space is said to be of \emph{mixed} type due to the coexistence of regions of stability or regular motion, along with chaotic areas \cite{Lichtenberg1992}.

Finally, let us shift our focus to a pragmatic example of a non-linear symplectic map namely the standard map, which offers a suitable "numerical laboratory" for exploring fundamental properties of mixed phase spaces. Ultimately, this brief investigation leads us to the significant notion of chaotic transport in our research.

\section{The Standard map}
\label{sec:standard_map}
Proposed by Boris Chirikov in 1969, the Standard map (SM) \cite{Chirikov1969} was initially conceived for applications in plasma dynamics. Due to its rich dynamical scenarios, the map describes a universal behaviour of area-preserving maps with mixed phase space. Its physical model is related to the motion of a particle constrained to a movement on a ring while being kicked periodically by an external field. 

The map derives from the Hamiltonian
\begin{eqnarray}
H\left(q,p,t\right) & = & \frac{p^2}{2mr^2}+K\cos\left(q\right)\sum_{n=-\infty}^{\infty}\delta\left(\frac{t}{T}-n\right)~, 
\label{eq:ham_stdmap}
\end{eqnarray}where $\delta$ is the Dirac delta function, $q$ is the angular coordinate and $p$ is its conjugate momentum. It is worth remarking that the particle's mass $m$, the ring's radius $r$, and the period of kicks $T$ are all considered to be one for simplicity.

Although the Hamiltonian in Eq.\ (\ref{eq:ham_stdmap}) is sufficient to analyse the dynamics, it is possible to define a symplectic non-linear discrete map $T_{\text{SM}}$ to investigate the dynamics via extensive numerical simulations since iterating the symplectic map is much faster than solving the equations of motion. 

The mapping $T_{\text{SM}}\left(p_{n},q_{n}\right)=\left(p_{n+1},q_{n+1}\right)$ gives the position and momentum for the $(n+1)^{th}$ iteration by the following equations

\begin{equation}
T_{\text{SM}}:\left\{\begin{array}{ll}
    p_{n+1} = p_n +k\sin(q_n) ~ \mod(2\pi)\\
    q_{n+1} = q_n + p_{n+1} ~ \mod(2\pi)
\end{array}
\right.,
\label{eq:stdmap}
\end{equation}where the parameter $k$ controls the intensity of the non-linearity brought by the sine function. Indeed, the standard map is equivalent to the generic map $T^{\epsilon}$ presented in Eq.\ (\ref{eq:general_map}) while considering $f(\theta_n,I_{n+1}) = f(\theta_n) = \sin(\theta_n)$ periodic in $2\pi$, $h(I_{n+1}) = I_{n+1}$ and $l(\theta_n,I_{n+1}) = 0$.

\begin{figure}[h]
\centering
\includegraphics[scale=0.62]{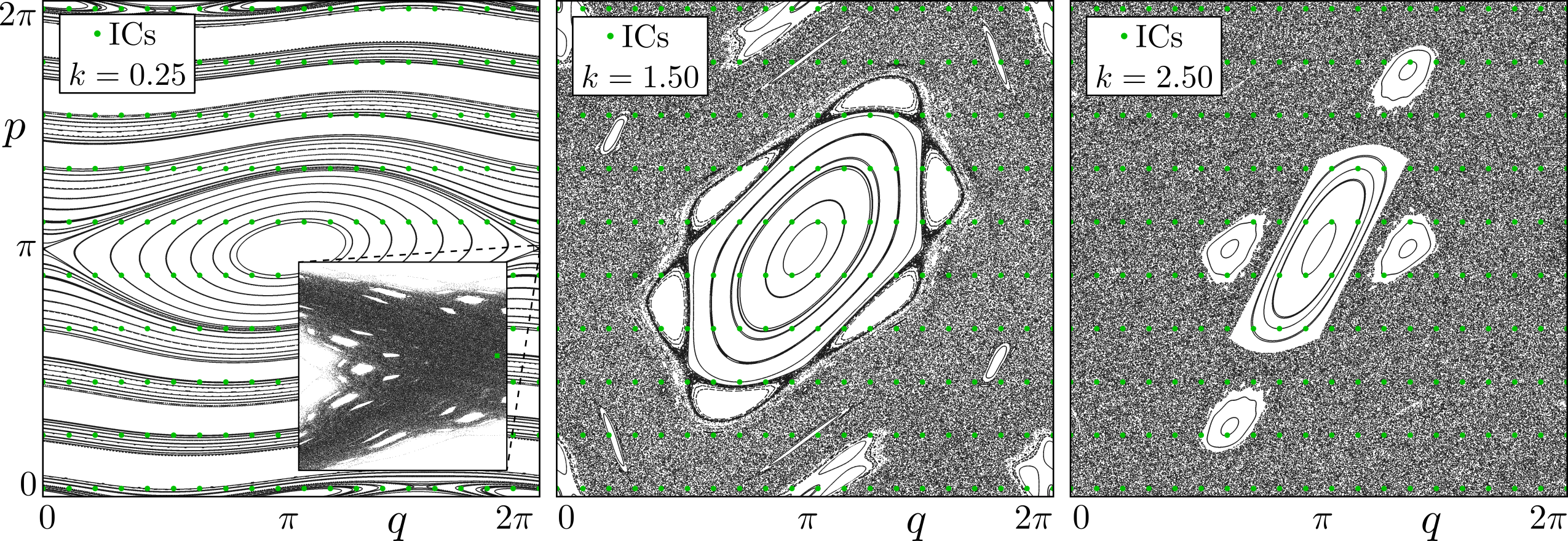}
\caption[Phase spaces of the Standard map (SM)]{Characteristic phase spaces of the SM considering three values of the parameter $k$. On the left $k=0.25$ and the inset shows the chaotic region around the unstable point at $(2\pi,\pi)$. On the central panel $k=1.50$ and on the right $k=2.50$. This phase space construction considers $+\pi$ at the $q$-equation so that the main island is always centred at $(\pi,\pi)$. $20 \times 10$ ICs, represented by the green points, were evolved up to $N = 10^3$ iterations.}
\label{fig:STD_3ps}
\end{figure}

Figure \ref{fig:STD_3ps} presents the characteristic phase spaces\footnote{A technical note: All phase spaces of the SM are constructed by adding $\pi$ at the $q$-equation of the mapping so that the central island, around the stable fixed point at $(\pi,\pi)$, is centralised in all figures.} of the SM considering three different values of the control parameter $k$. First, on the left panel, the space is mostly composed of \emph{invariant spanning curves} that are: {\it invariant} with respect to iterations i.\ e.\ an IC placed on the curve will evolve to a trajectory that lies only in this curve, no matter how long you iterate it and; {\it spanning} because it spans through the hole $q$-axis from $0$ to $2\pi$. Nevertheless, still on the first panel, $k=0.25$ already introduces a chaotic region around the unstable point located at $(2\pi,\pi)$, or $(0,\pi)$ due to modulated equations, that is displayed on the inset.

The large chaotic seas shown in the middle and right panels of Fig. \ref{fig:STD_3ps} are due to the relatively high values of the control parameter, namely $k = 1.5$ and $k = 2.5$, respectively. These chaotic regions contain all chaotic trajectories in the system. In our context, a chaotic trajectory $\varphi$ is an orbit evolved from an IC $(q_0,p_0)$ in the chaotic sea that presents an unpredictable and complex evolution, exhibiting various dynamical behaviours until reaching the maximum iteration time $N$. It is important to note that the chaotic evolution $\varphi(q_0,p_0)$ of a specific initial condition is deterministic, but the system is highly \emph{sensitive to initial conditions}. Therefore, even two initial conditions that are very close to each other can lead to completely different evolution, a well-known characteristic of chaotic dynamics \cite{Strogatz2000}. 

\subsection*{Sensitivity to initial conditions and Stickiness}
\begin{figure}[h]
\centering
\includegraphics[scale=0.72]{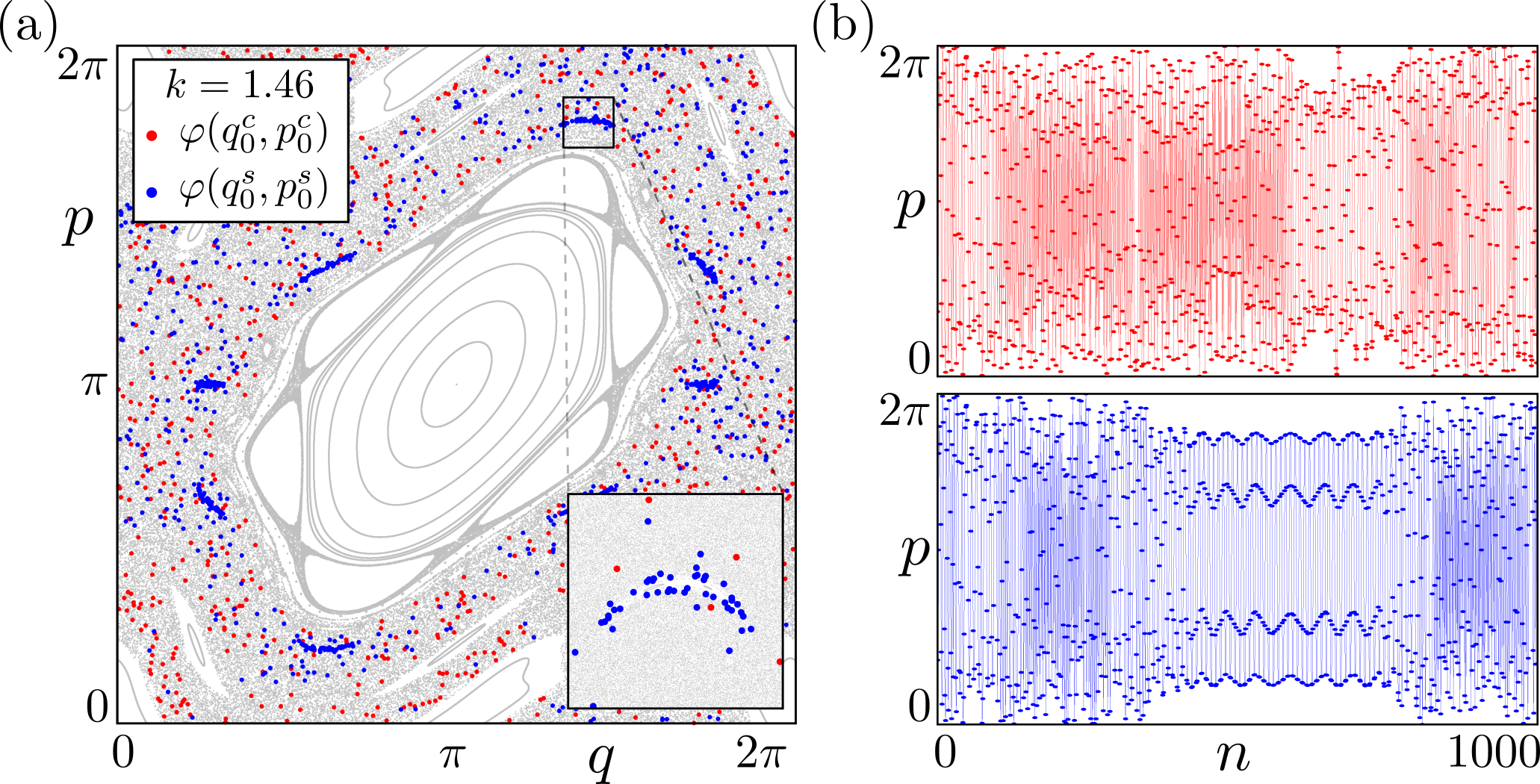}
\caption[Example of the sensitivity to ICs in the SM]{(a) Phase space of the SM for $k=1.46$ (grey on the background) along with trajectories $\varphi(q_0^c,p_0^c) = (0.51234567899870,\pi)$ in red and $\varphi(q_0^s,p_0^s) = (0.51234567899871,\pi)$ in blue. The inset shows an amplification of the stickiness region; (b) Behaviour of the generalised momentum $p$ in function of the iterations $n$. }
\label{fig:STD_IC_sensi}
\end{figure}

Let us consider an specific example in the SM to explore this idea of sensitivity to ICs and the stickiness phenomenon. For $k=1.46$, we select an IC located in the chaotic sea at $(q_0^c,p_0^c) = (0.51234567899870,\pi)$ and the evolved trajectory $\varphi(q_0^c,p_0^c)$ is analysed. Additionally, we select $(q_0^s,p_0^s) = (q_0^c+1\times 10^{-14},p_0^c)$ i.\ e.\ a very close IC that changes $q_0$ only by $10^{-14}$, and the trajectory $\varphi(q_0^s,p_0^s)$ is also analysed. Figure\ \ref{fig:STD_IC_sensi} show the phase space of the SM for $k=1.46$ with the aforementioned trajectories and the evolution of $p(n)$ for each one.

The two selected chaotic trajectories on the phase space shown in Fig.\ \ref{fig:STD_IC_sensi} (a) have distinct evolution despite the proximity of their ICs. Trajectory $\varphi(q_0^c,p_0^c)$, depicted in red, explores a larger region of the available chaotic sea throughout its evolution, a fact that is also verified by its $p(n)$ behaviour in (b). Conversely, $\varphi(q_0^s,p_0^s)$ begins its chaotic evolution similarly to $\varphi(q_0^c,p_0^c)$, however around $450$ iterations, the trajectory is temporally confined to specific regions of the phase space, which are clearly observed from the concentrated blue clusters in Fig.\ \ref{fig:STD_IC_sensi} (a). The inset displays one of these regions amplified, where we note the blue trajectory in a close chaotic vicinity around a small periodic island found in the chaotic sea. At this moment in its evolution, the chaotic orbit sustains a \emph{quasi-periodic} behaviour until approximately 750 iterations, also evidenced by its $p$ time-series in (b). After this trapping time, the orbit is again free to explore other regions of the chaotic sea.  

The intermittent dynamical evolution experienced by the trajectory $\varphi(q_0^s,p_0^s)$ shown in Fig.\ \ref{fig:STD_IC_sensi} is one example of the \emph{stickiness phenomena} \cite{Contopoulos2010,Altmann2006}. Due to mixed phase spaces, near-integrable Hamiltonian systems may present chaotic orbits that spend a considerable amount of time experiencing successive dynamical traps. The trajectory that was once free to explore all chaotic regions of the phase space, can be temporally confined in a peculiar quasi-periodic motion in the nearest chaotic vicinity around the stability islands. Stickiness is a well-studied phenomenon that strongly affects the transport and statistical properties of chaotic orbits \cite{Zaslavsky2002}.

\subsection*{Chaotic transport}
The notion of chaotic transport refers to the irregular motion of particles or fluid elements in systems that can be described by nonlinear dynamics. We showed that, in such systems, small perturbations in the ICs can lead to vastly different outcomes, making long-term predictions very difficult. Chaotic transport is a common phenomenon in many physical systems, including atmospheric and oceanic flows \cite{Wiggins2005}, chemical reactions \cite{Bringer2004}, and plasma confinement \cite{Bickerton1997}.

One approach to understanding the transport of chaotic trajectories in phase spaces is to use techniques from statistical mechanics, mainly studying diffusion. Indeed, for fully chaotic phase spaces, it is possible to show that an ensemble of chaotic orbits diffuses accordingly to the Heat equation, being a process of \emph{normal diffusion} \cite{Leonel2004,Palmero2020}. In this case, the average behaviour of all chaotic trajectories is equivalent to particles moving randomly in a defined space, analogously to a Brownian motion.

However, in mixed phase spaces, the chaotic area coexists with regions of periodic motion, such as the KAM islands shown in phase spaces of the SM. In these cases, chaotic trajectories may exhibit a complicated diffusion process often referred to as \emph{anomalous diffusion} \cite{Shlesinger1993,Meroz2015}. Taking the previous situation on the SM as an example, the trajectory $\varphi(q_0^s,p_0^s)$ begins its diffusion on the chaotic sea, but the stickiness forces a temporarily constrained diffusion around the small KAM island, highly affecting the orbit's transport. In that sense, investigations on anomalous effects in chaotic transport are needed to advance our understanding of these particular situations, improving also the general knowledge on how and why these effects may interfere in practical situations such as in the magnetic confinement of fusion plasma.     

\begin{figure}[b!]
\centering
\includegraphics[scale=0.77]{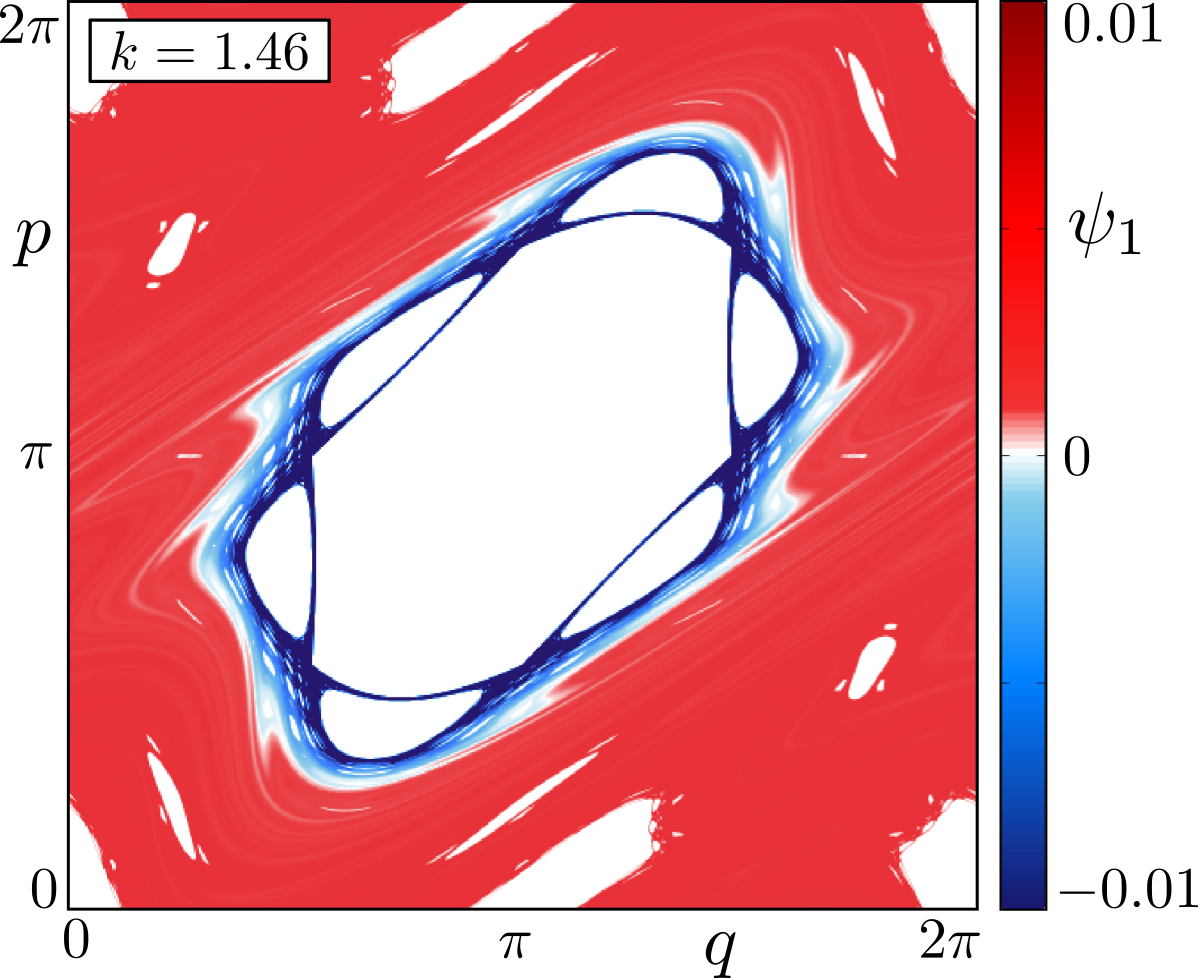}
\caption[First diffusion mode in the phase space of the SM]{First diffusion mode, given by the first eigenvector $\psi_1$ of the Perron-Frobenius-like operator, on the phase space of the SM for the selected control parameter $k=1.46$.}
\label{fig:STD_diff}
\end{figure}

Another statistical approach for studying chaotic transport is based on probability distributions that can be used to estimate the probability of finding a trajectory in a certain region of phase space. These distributions can be constructed using techniques such as the Perron-Frobenius operator \cite{Klus2016} or Markov partitions \cite{Vollmer2002}. In particular, we developed an approach to investigate the chaotic transport in the SM considering trajectories evolved from specific ICs. Our original contribution presented at {\it Sub-diffusive behavior in the Standard Map}, by Matheus S. Palmero, Gabriel I. D\'iaz, Iber\^e L. Caldas and Igor M. Sokolov, Eur.\ Phys.\ J.\ Spec.\ Top.\ {\bf 230}, published in June 2021, showed that it is indeed possible to find sub-diffusive behaviour in the SM \cite{Palmero2021}.   

In summary, our approach is based on chaotic, yet sticky, trajectories that can be described by the Continuous Time Random Walk (CTRW) model. Since CTRW is a classic example of anomalous diffusion, we studied some consequences of the Fractional Diffusion Equation (FDE) and how to connect it to a numerical method that approximates the Perron-Frobenius operator for the SM. With that, we established a relation between the eigenvalues of a Perron-Frobenius-like operator and the solution of FDE, providing an approximated value for the anomalous diffusion exponent $\alpha$ for the SM considering $k=1.46$. The found value of $\alpha\approx0.25$ shows that the evolution of trajectories in this particular scenario is indeed associated with anomalous diffusion, distinctively a sub-diffusive behaviour. 

\begin{figure}[b!]
\centering
\includegraphics[scale=0.8]{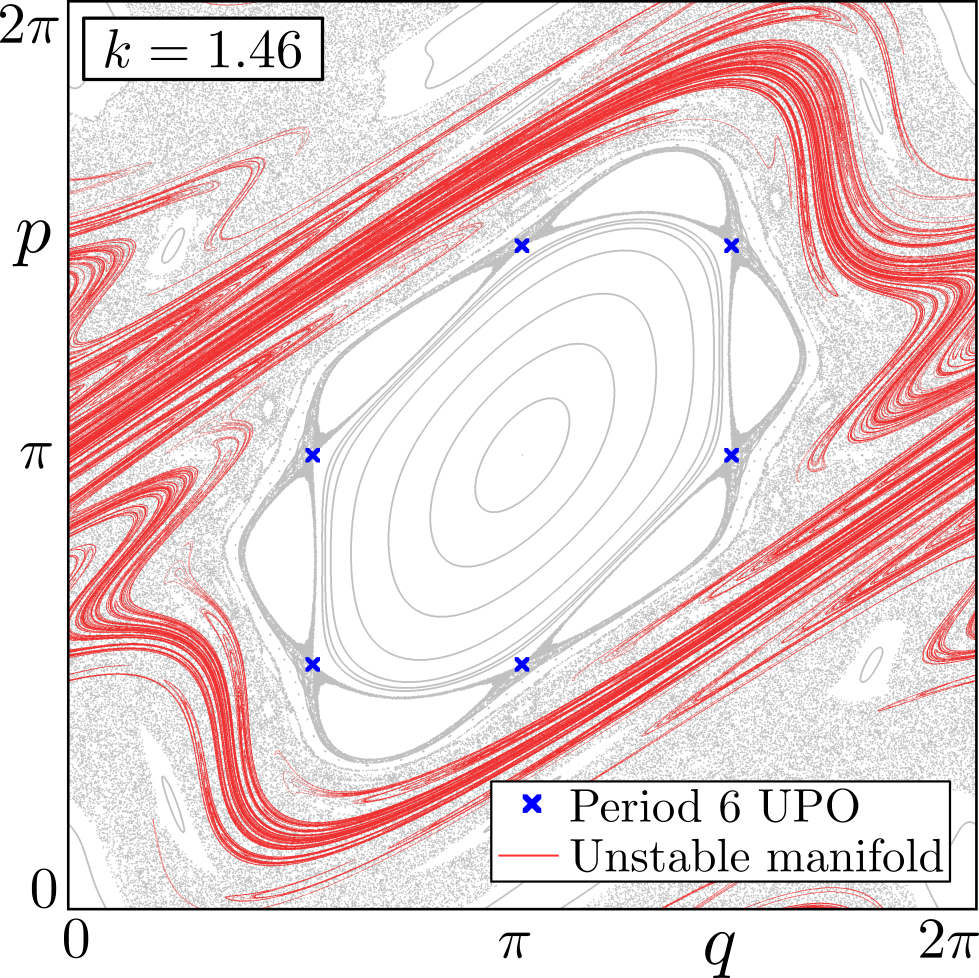}
\caption[Period 6 UPO and unstable manifold in the phase space of the SM]{Phase space of the SM for $k=1.46$ (grey on the background) along with the period 6 UPO (blue crosses) and the unstable manifold associated with $(q^{\star}, p^{\star}) = (0, \pi)$ (red line).}
\label{fig:STD_upo_mani}
\end{figure}

One of the main results of the aforementioned approach is shown in Fig.\ \ref{fig:STD_diff}, where we display the numerical result for the first diffusion mode $\psi_1$, eigenvector of the calculated Perron-Frobenius-like operator, on the phase space of the SM for $k=1.46$. It is clear from the colour map that the phase space is divided into two distinct regions. This division is properly characterised by the change of signs of the first diffusion mode $\psi_1$. Negative values of $\psi_1$ are attained only at fine chaotic regions around the main island and the other resonant ones. Yet, positive values of $\psi_1$ depict the outer chaotic sea. 

Additionally, Fig.\ \ref{fig:STD_diff} also displays fine regions within the chaotic sea where $\psi_1=0$, depicted in white, are of fundamental importance for understanding the main features of chaotic transport. These regions are related to underlying structures on phase space known as \emph{invariant manifolds}. In practical sense for our intents and purpose, manifolds are geometrical structures that are regarded as the skeleton of all possible dynamical behaviours of the system \cite{Alligood2012}. For that reason, determining how these geometrical structures are spatially organised is essential to comprehend how different regions of a given phase space are linked or severed. Invariant manifolds often act as transport barriers, partial transport barriers and transport channels \cite{Wiggins2013}. In Chapter \ref{chap4:transient} we provide a further detailed mathematical definitions of the invariant manifolds.

The construction of the results shown in Fig.\ \ref{fig:STD_diff} is explained in detail in \cite{Palmero2021}. Nevertheless, essentially we selected a specific IC and evolved it for $10^{10}$ iterations while computing the Perron-Frobenius-like operator of the SM to gain a better understanding of its chaotic evolution. The IC of the considered trajectory was at a period 6 Unstable Periodic Orbit (UPO) that surrounds the main island as secondary resonances caused by the non-linearity. UPOs are special points within the chaotic sea that are the unstable counterparts of the stable fixed point inside (at the centre) the KAM islands. In this special case, the selected UPO is located deep within the fine chaotic region that surrounds the main island of the SM for $k=1.46$, making it a suitable place for strong stickiness effects. In Chapter \ref{chap5:recurrence} we further address stickiness by analysing large ensembles of ICs near UPOs of interest.

In order to illustrate and exemplify both aforementioned elements, we show in Fig.\ \ref{fig:STD_upo_mani} the phase space of the SM with the period 6 UPO, depicted by the blue crosses, and the unstable invariant manifold associated with the unstable fixed point at $(q^{\star}, p^{\star}) = (0, \pi)$ depicted by the red line. As mentioned, the selection of ICs in close vicinity of UPOs, and the spatial organisation of invariant manifolds in the phase space are both important concepts for upcoming analysis in the next chapters.    

Finally, it is worth noting that we determined the exact location of the UPO and the computed unstable manifold using the method proposed by Ciro, outlined in \cite{Ciro2018}. In Chapter \ref{chap4:transient} and Chapter \ref{chap5:recurrence}, we use the same method for computing and tracing invariant manifolds, and determining suitable locations around UPOs for ICs to study stickiness, all considering the models of our interest in application to magnetic confinement. The following chapter introduces the main models of our research for magnetic field lines in tokamaks under different configurations.

\chapter{Models and Escape analysis}
\label{chap3:models}

In this chapter, we present the \emph{models} and the adopted methodology for our numerical investigations brought by an \emph{escape analysis}. The two selected symplectic maps model the magnetic field lines of tokamaks under two different configurations; The \emph{Single-null divertor map}, a phenomenological model that describes the magnetic configuration of a tokamak equipped with a poloidal divertor and; The \emph{Ergodic magnetic limiter map}, a freely adjustable map, derived from a suitable generating function, that describes the magnetic field lines of a tokamak assembled with an ergodic magnetic limiter. These two models are discussed in detail in the first and second sections respectively.  

The methodology also discussed here is based on an important detail present in both models. Due to the perturbations introduced by the divertor, or the limiter, there will be \emph{escaping magnetic field lines} that might be unfavourable for the tokamak purpose. In that sense, we introduce in the third section the adopted methodology for our numerical investigations while considering a wide range of perturbation strengths for the two models. The escaping analysis as a methodology is an original approach that prepares the ground for the upcoming phase space analyses in chapters four and five.  

\section{Single-null divertor map}
\label{sec:chap3_boozer}

\begin{figure}[h]
\centering
\includegraphics[scale=0.8]{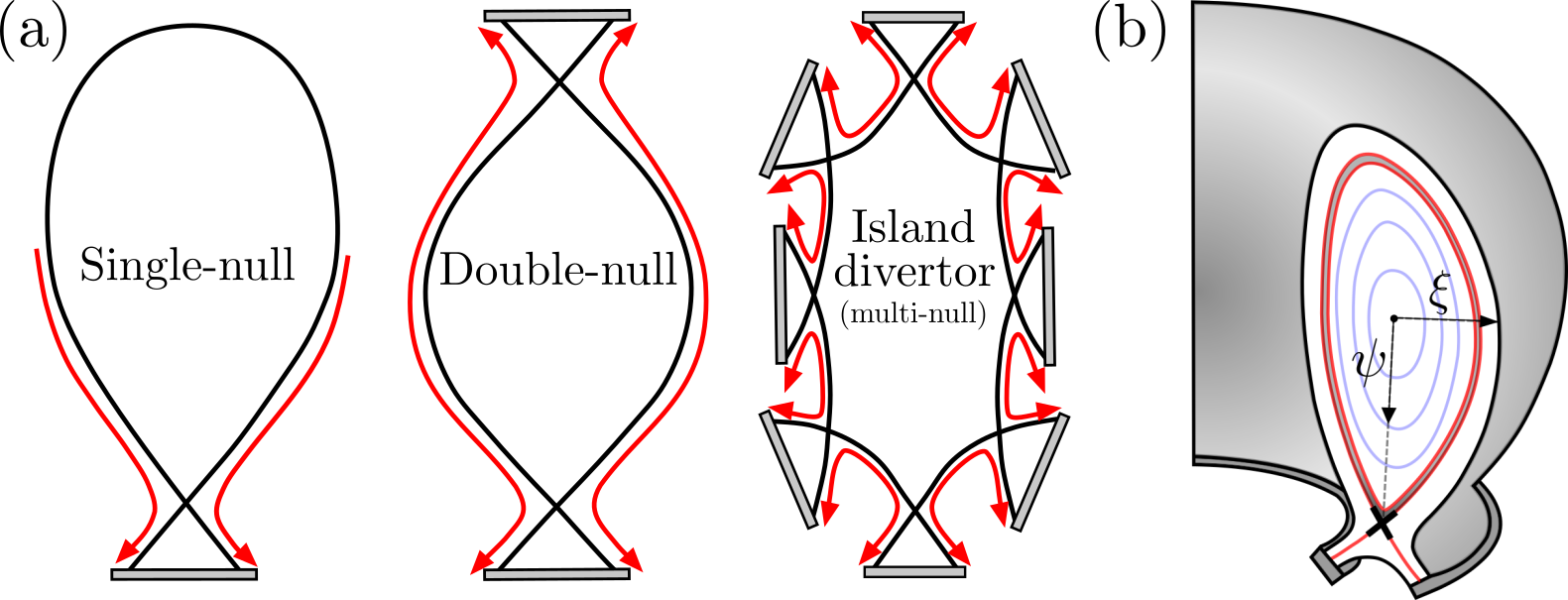}
\caption[Divertor configurations and schematics of magnetic separatrix]{(a) Three different types of configurations for divertors; (b) Schematics of a poloidal section of a divertor tokamak, showing the closed magnetic field lines (light blue), the region around the magnetic separatrix (grey region between the red lines), magnetic saddle (black cross) and the general rectangular coordinates $(\xi,\psi)$.}
\label{fig:BM_schematics}
\end{figure}

The single-null divertor map, also known as the \emph{Boozer map} (BM), was proposed by Punjabi, Verma and Boozer \cite{Punjabi1992} as a phenomenological model for the magnetic field lines of a tokamak equipped with poloidal divertors. Divertors are external devices placed at a poloidal section of the tokamak, designed to exhaust unwanted particles from the plasma to maintain the fusion reaction inside the tokamak. The divertor cassette is composed, essentially, of a steel body that encloses a poloidal section, and divertor targets that are placed inside the tokamak chamber, next to the plasma edge. The targets are specifically constructed to receive and rapidly extract the thermal load caused by the high heat flux from the fusion plasma.   

Technically, the divertor induces a magnetic configuration with a saddle point (x-point) known as the \emph{magnetic saddle}. This configuration allows particles to follow the field lines towards exit points precisely placed near the divertor targets. Due to perturbations in the magnetic field, a chaotic layer is formed around the saddle, allowing the open field lines to escape through the x-points, striking the divertor target. The striking points are commonly referred to as \emph{magnetic footprints}, which form specific patterns that will be further addressed in this section.

There are a few different types of configurations for poloidal divertors, each one inducing a certain magnetic topology in the tokamak. Figure \ \ref{fig:BM_schematics}(a) depicts three different configurations for the divertors: The single-null, which induces only one magnetic saddle; The double-null, with upper and lower x-points and; The island divertor, or multi-null, often used for the construction of the stellarator \cite{Spitzer1958}, another toroidal shaped plasma device that is well-known for its complex design intended to highly enhance magnetic confinement.

Considering the single-null configuration, it is possible to study the behaviour of the magnetic field lines around the magnetic saddle via the symplectic separatrix map $T_{\text{BM}}$ given by the following equations
\begin{equation}
T_{\text{BM}}:\left\{\begin{array}{ll}
\xi_{n+1}=\xi_n-k\psi_n(1-\psi_n)\\
\psi_{n+1}=\psi_n+k\xi_{n+1}
\end{array}
\right.~,
\label{eq:bm_map}
\end{equation}where the pair $(\xi, \psi)$ are generic rectangular coordinates over a poloidal section surface, as depicted in Fig.~\ref{fig:BM_schematics}(b); Note that $\psi$ is positive from the centre of the poloidal section in the direction of the x-point. The control parameter $k$ is related to the amplitude of toroidal asymmetries that perturb the magnetic field configuration \cite{Punjabi1997}.  

In regards to the parameter $k$, which is the main parameter of the model, there are a few works in the literature that relates numerical values of $k$ to the safety factor calculated at the plasma edge. Accordingly to these works,  $k \approx 0.6$ is a fair value to simulate the diverted magnetic field configuration while considering, specifically, large tokamaks like the ITER \cite{Punjabi1997}. 

\begin{figure}[h]
\centering
\includegraphics[scale=0.75]{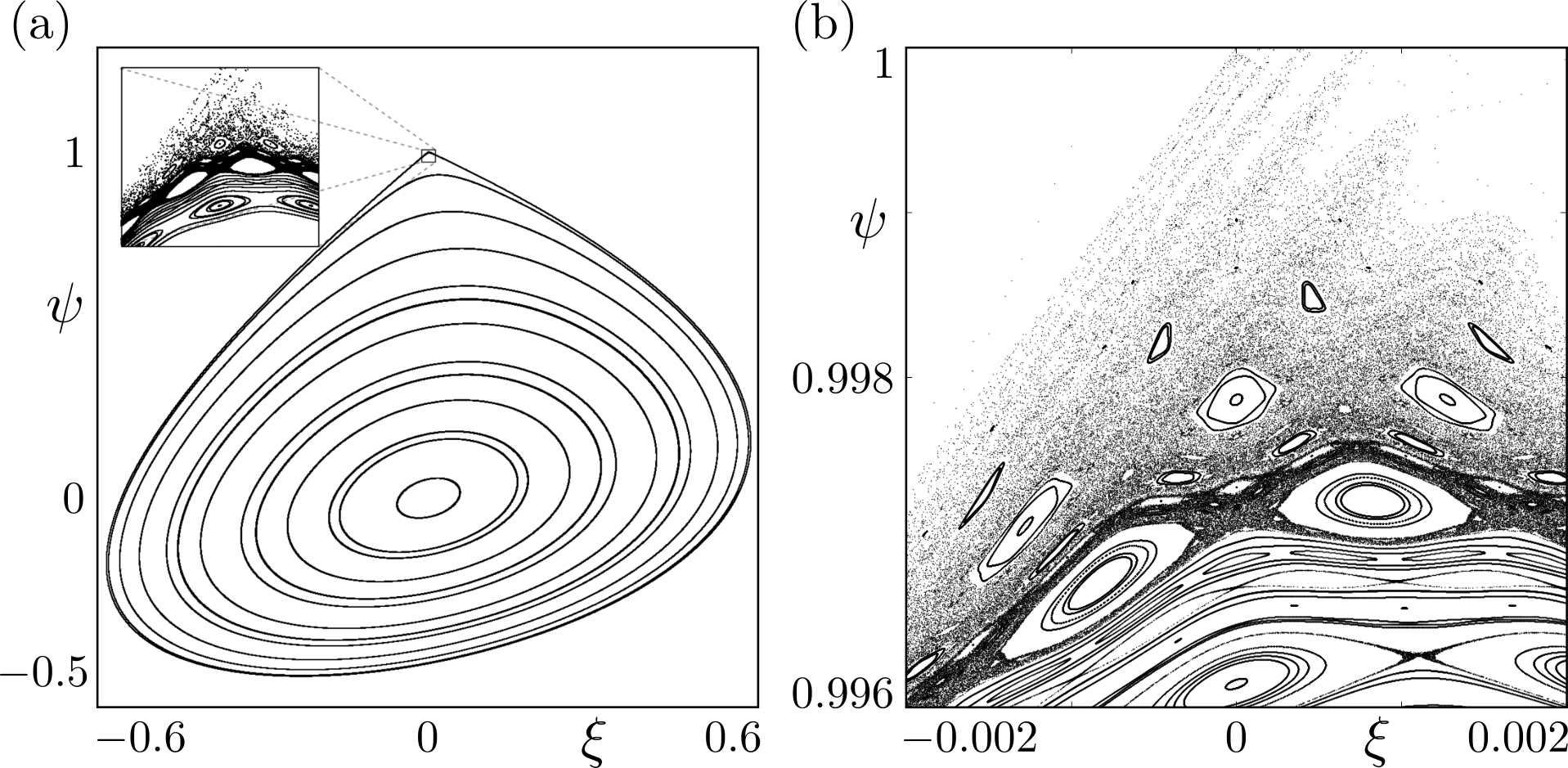}
\caption[Phase space of the Single-null divertor map (BM)]{Characteristic phase space of the BM considering $k = 0.6$. (a) Full phase space with the inset showing the region around the magnetic saddle; (b) The inset amplified, showing the mixed chaotic-periodic states present in this region.}
\label{fig:BM_phase_space}
\end{figure}

The characteristic phase space of the model is shown in Fig.\ \ref{fig:BM_phase_space}. In particular, panel (b) shows the amplified region around the magnetic saddle located at $(\xi^{\star},\psi^{\star}) = (0,1)$, where the separatrix chaotic layer, embedded with several highly-periodic island chains, is clearly visible. The chaotic layer is essentially composed of open field lines that will, eventually, escape through the x-point, hitting the divertor target. Punjabi {\it et. al.} also provides a procedure to numerically obtain the collision points between the open magnetic field lines and the divertor target. The procedure is essentially the following: For a given set of ICs and a fixed value of $\psi_{\text{target}}$, in which the divertor target is located, we iterate the map $T_{\text{BM}}$ until the escape condition $\psi_n \leq \psi_{\text{target}} \leq \psi_{n+1}$ is satisfied. Then, a toroidal coordinate $\phi_s$ is introduced to calculate the striking points using the equations of the map, Eq.\ \ref{eq:bm_map}, as follows 

\begin{equation}
    \phi_s = \frac{\psi_{\text{target}} - \psi_n}{k[\xi_n - k\psi_n(1-\psi_n)]}~,
    \label{eq:bm_psi_s}
\end{equation}where, by definition, $0 \leq \phi_s \leq 1$. Analogously, the abscissa of the striking points $\xi_s$ is given by

\begin{equation}
    \xi_s = \xi_n - k\phi_s\psi_n(1 - \psi_n)~.
    \label{eq:bm_xi_s}
\end{equation}

\begin{figure}[t]
\centering
\includegraphics[scale=0.65]{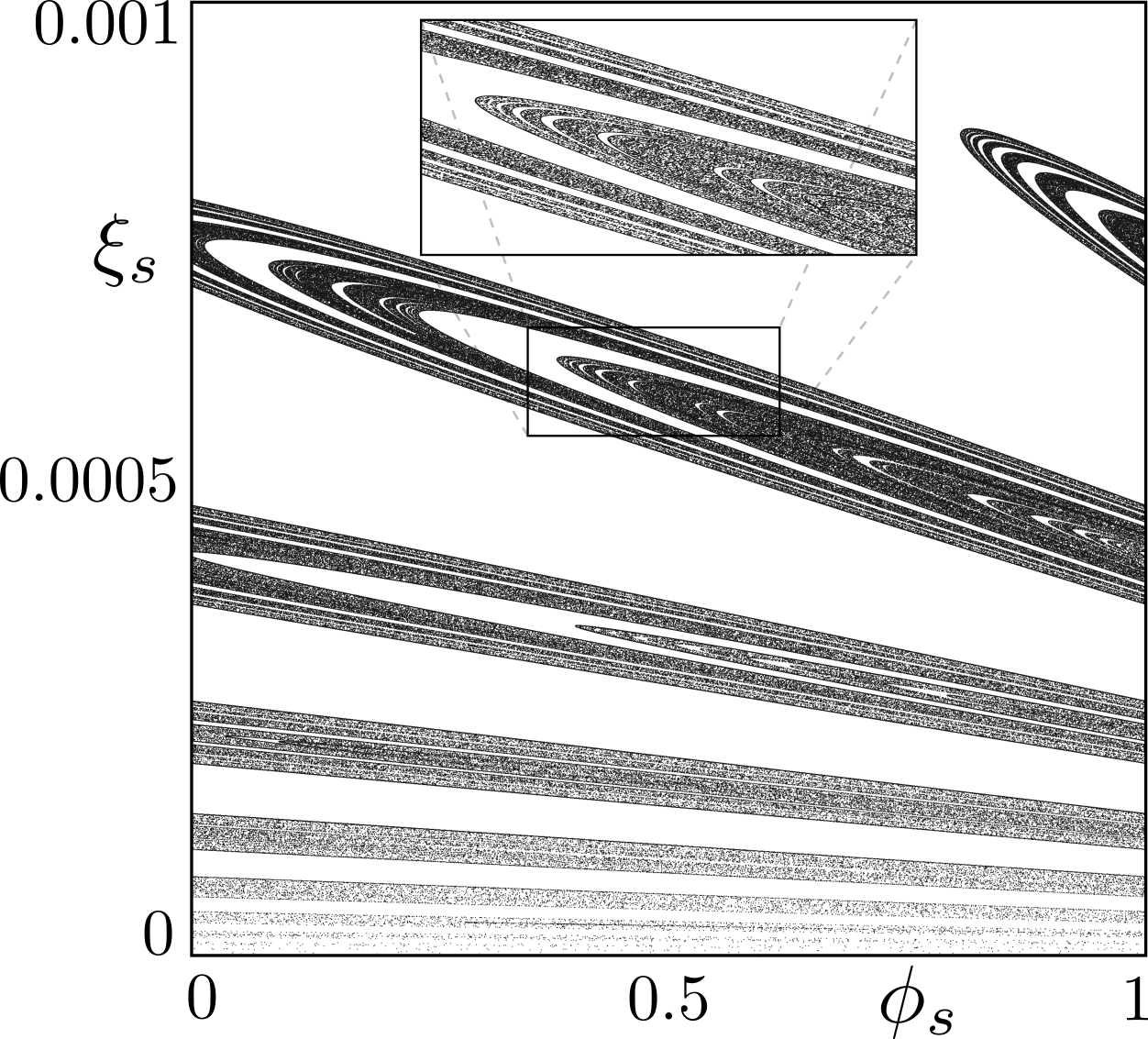}
\caption[Magnetic footprints]{Magnetic footprints formed by the open field lines that strike the divertor target. The highlighted square shows the respective region amplified. $10^6$ ICs ($\xi_0 \in [0.0, 0.001]$ and $\psi_0 = 0.999$) were evolved up to $10^4$ iterations of the map, considering $k = 0.6$.}
\label{fig:BM_mag_footprint}
\end{figure}

Considering $\psi_{\text{target}} = 1.0$, we recover a noted result from the literature for the aforementioned magnetic footprints shown in Fig.\ \ref{fig:BM_mag_footprint}. This result shows the fractal pattern created by the escaping field lines while striking the target of the poloidal divertor. Indeed, experimental evidence \cite{Ciro2016} based on heat patterns at the divertor target, suggests that ions from the plasma may be following open magnetic field lines, striking the target and forming specific heat signatures that closely follow the expectation from the magnetic footprints.

From the theoretical point of view, specifically from the theory of weakly perturbed Hamiltonian systems, it is known that this fractal pattern is strictly related to the homoclinic tangles formed by the infinite intersections between the unstable and stable invariant manifolds associated with the magnetic saddle \cite{Ott2002}.    

Although simple, the BM is a very suitable model to perform extensive numerical simulations that might improve our understanding of the complex behaviour of the magnetic field lines around magnetic saddles in divertor tokamaks. 

\section{Ergodic magnetic limiter map}
\label{sec:chap3_ullmann}

The ergodic magnetic limiter map, or the \emph{Ullmann map} (UM), was proposed as a symplectic two-dimensional non-linear map that models the magnetic field lines of a tokamak assembled with an ergodic limiter \cite{Ullmann2000}. Inside the tokamak, in the plasma core, the magnetic field is strong and stable enough for the duration of a typical discharge. However, on the plasma edge, closer to the inner walls of the machine, the magnetic field lines are often perturbed, forming regions of strong instabilities. In many cases, to either control or change the magnetic configuration in this outer region, the tokamak is assembled with devices placed at the border of the machine. This is the case of the ergodic magnetic limiter which is, basically, an outer ring composed of several helical coils that periodically perturb the field lines at the plasma edge.  

\begin{figure}[h]
\centering
\includegraphics[scale=0.69]{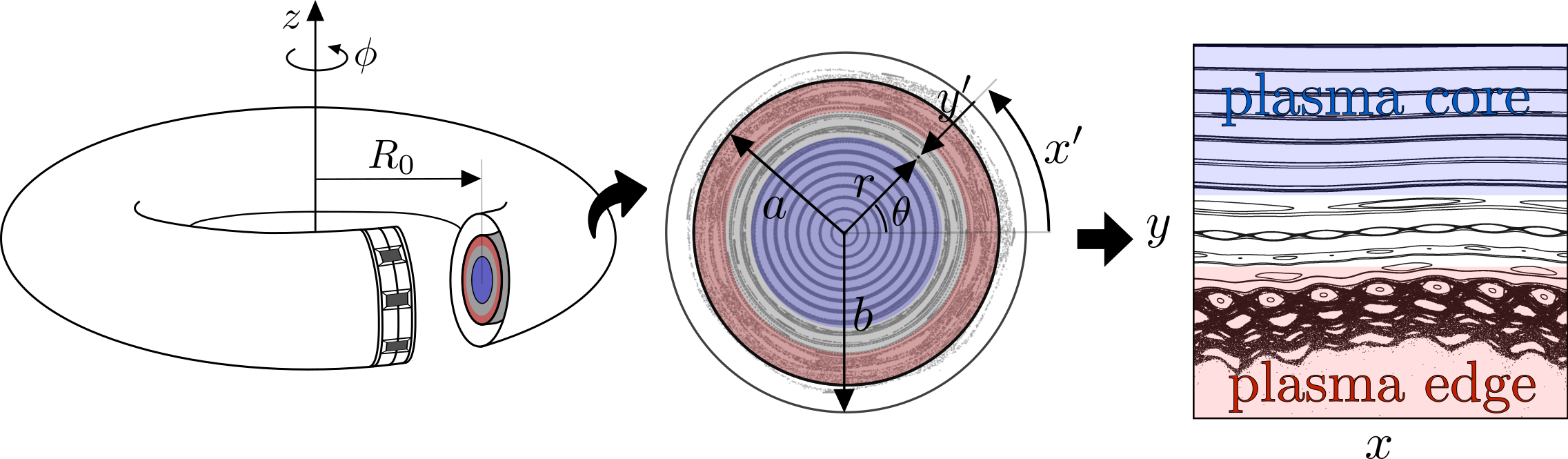}
\caption[Modelling process for the Ergodic magnetic limiter map]{Schematic example of the modelling process that begins at the torus (representing the tokamak), passing by the poloidal section near the limiter (ring placed in a section of the torus), showing also the calculated magnetic field lines (grey in the background), arriving at the rectangular coordinates $(x,y)$ drawing the full phase space of the model. The colour blue represents regions closer to the plasma core and the colour red regions around the plasma edge.}
\label{fig:UM_scheme}
\end{figure}

In the practical sense, there are a few features that make the UM a fitting model for our analysis: {\it (i)} Since it is a symplectic map, it can be derived from suitable generating functions that include the appropriate periodic perturbation; {\it (ii)} The profile of the safety factor $q(r)$ is freely adjustable for a given tokamak discharge, allowing also an analysis considering non-monotonic profiles \cite{Caldas2012}; {\it(iii)} The parameters of the model are directly linked to experimental parameters of a tokamak, such as the intensity of the toroidal magnetic field $B_0$, large radius $R_0$, small radius $b$, and radius of the plasma column $a$ \cite{Ullmann2000}. Figure \ref{fig:UM_scheme} displays a schematic example of the modelling process, along with the aforementioned parameters.

The pair of dimensionless coordinates $(x,y)$, that draws the characteristic phase space of the model shown in the last panel of Fig.\ \ref{fig:UM_scheme}, is calculated considering a relative distance $y^{\prime} = b - r$ (in relation to the minor radius $b$) and a modulated angle $x^{\prime} = b \theta$, thereby defining $x = x^{\prime}/2\pi b = \theta / 2\pi$ and $y = y^{\prime}/b = 1 - r /b$.  

The complete model is a composition of two maps $T^0_{\text{UM}} \circ T^1_{\text{UM}} (x_n,y_n) = (x_{n+1},y_{n+1})$. The first part $T^0_{\text{UM}}$ is the equilibrium dynamics with a toroidal correction to the field line equations proposed by Ullmann. For that, it is assumed that the equilibrium magnetic field is $\mathbf{B}^0 = (B^0_r = 0, B^0_\theta = B^0_\theta(r), B^0_\phi = B_0)$ and one can introduce the generating function  

\begin{equation}
G^0_{\text{tor}}(r_{n+1},\theta_n)=G^0(r_{n+1},\theta_n)+\sum^{\infty}_{i=1}a_i\left(\frac{r_{n+1}}{R_0}\right)^{i}\cos^{i}(\theta_n)~,
\label{eq:um_g_tor}
\end{equation}in which, considering just the first term $a_1 = -0.04$\footnote{This toroidal correction is introduced to the system to take into consideration a small outward radial displacement of the centre of flux surfaces in the tokamak, usually caused by a phenomenon known as the Shafranov shift.} and $a_i = 0$ for $i > 1$, and following the same procedure described in chapter two for $G^0(r_{n+1},\theta_n)$, the first part of the map is obtained as follows

\begin{equation}
T^0_{\text{UM}}:\left\{\begin{array}{ll}
y_{n+1}^{*}=1-\frac{(1 - y_n)}{1-a_1\sin(x_n)}\\
x_{n+1}^{*}=x_n+\frac{2\pi}{q^{0}(y_{n+1}^{*})}+a_1\cos(x_n)\\
\end{array}
\right.~,
\label{eq:um_eq}
\end{equation}where $q^{0}(y)$ is the safety factor, obtained from the poloidal magnetic field $B^0_\theta(r)$, calculated at the new dimensionless coordinate $y$.

In order to model the periodic perturbation caused by the ergodic magnetic limiter, it is assumed that the limiter is sufficiently narrow in relation to the torus, ensuring a pinpoint perturbation where would be defined the poloidal section of interest. That way, the perturbed magnetic field $\mathbf{B}^1=(B^1_r = B^1_r(r,\theta,\phi), B^1_\theta = B^1_\theta(r,\theta,\phi),B^1_{\phi}=B_0)$ is written considering    

\begin{equation}
\begin{aligned}
B^1_r(r,\theta,\phi)&=-B^1(r)\sin(m\theta)\sum^{\infty}_{j=-\infty}\delta(\phi-2\pi j)~,\\
B^1_\theta(r,\theta,\phi)&=-B^1(r)\cos(m\theta)\sum^{\infty}_{j=-\infty}\delta(\phi-2\pi j)~,
\end{aligned}
\label{eq:um_bperturb}
\end{equation}where $m$ and $I_h$ are, respectively, the number of coils (also known as the perturbation mode) and the induced helical current in the limiter. The amplitude of the perturbation is explicitly given by

\begin{equation}
B^1(r) = \frac{\mu_0mI_h}{\pi b}\left(\frac{r}{b}\right)^{m-1}~.
\label{eq:um_b1}
\end{equation}Once defined the perturbed magnetic field, one can introduce the following generating function

\begin{equation}
G^1(r_{n+1}, \theta_n)=r_{n+1}\theta_n-\frac{C}{m-1}\left(\frac{r_{n+1}}{b}\right)^{m-1}\cos(m\theta_n)~,
\label{eq:um_g_pert}
\end{equation}and the second part of the map is obtained as follows

\begin{equation}
T^1_{\text{UM}}:\left\{\begin{array}{ll}
y_{n}=y_{n+1}^{*}+\frac{m}{m-1}C(1-y_n)^{m-1}\sin(m x_n^{*})\\
x_{n+1}=x_n^{*}-C(1-y_{n})^{m-2}\cos(m x_n^{*})\\
\end{array}
\right.~,
\label{eq:um_pert}
\end{equation}where the dimensionless constant $C$ arrange all main parameters of the model, relating them in the following way

\begin{equation}
    C = \frac{2 \pi}{q^0(a)}\left(\frac{b}{a}\right)^{m-2} \frac{B^1(a)}{B^0_\theta(a)} = \frac{4 m \pi}{q^0(a)} \left(\frac{a}{b}\right)^2 \frac{I_h}{I_p}~.
    \label{eq:um_c}
\end{equation}Defining $\delta B = B^1(a) / B^0_\theta(a)$ and $\delta I = I_h / I_p$, we finally reach the proper control parameters of the model: $\delta B$ defined as the relative perturbation of the poloidal magnetic field or; $\delta I$ as the relative current factor. In practical terms, $\delta B = 1.0\%$ means that the intensity of the magnetic field caused by the limiter is $1.0\%$ of the intensity of the poloidal magnetic field at the plasma edge or, $\delta I = 0.5\%$ means that the electric current of the limiter is $0.5\%$ of the plasma current.  

\begin{figure}[h]
\centering
\includegraphics[scale=0.8]{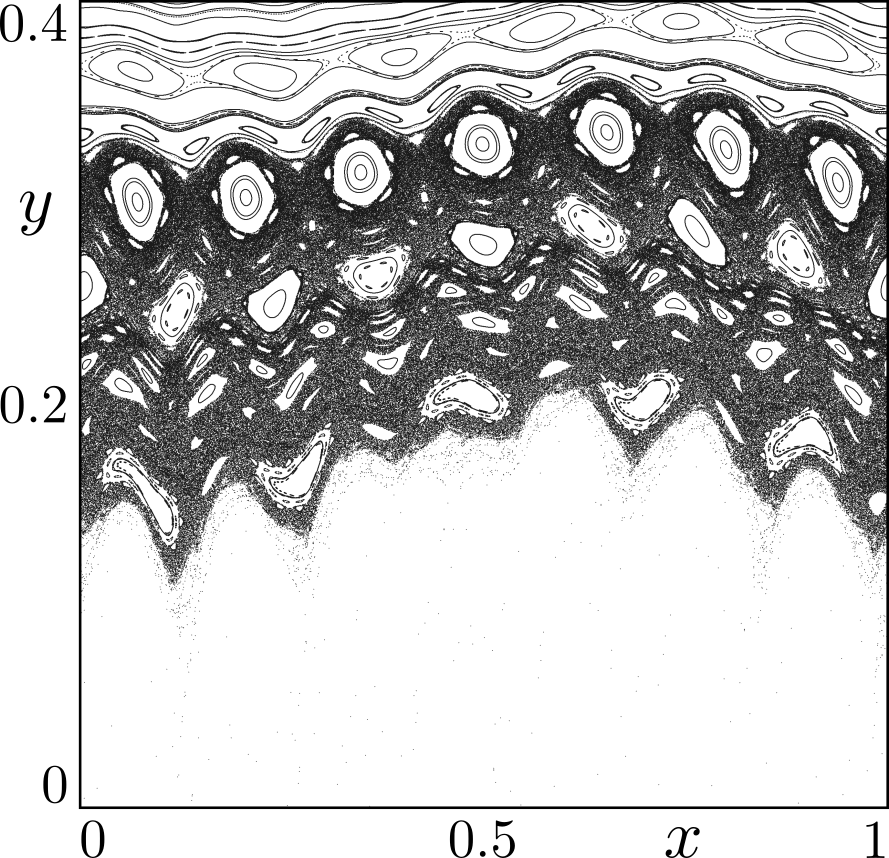}
\caption[Phase space of the Ergodic magnetic limiter map (UM)]{Characteristic phase space of the UM considering $m = 7$, $q^0(a) = 5.0$ and $\delta B = 1.5 \%$ ($\delta I \approx 0.45\%$).}
\label{fig:UM_phase_space}
\end{figure}

With that, all parameters of the model can be set for a specific tokamak and, the iteration of the maps $T^0_{\text{UM}} \circ T^1_{\text{UM}}$ provide different magnetic field configurations considering different values of $\delta B$ or $\delta I$. In our numerical simulations, we use the parameters of the TCABR, the tokamak of the Physics Institute, University of S\~ao Paulo, given by the following table:   

\begin{table}[h]
   \caption{TCABR - Main parameters \cite{Elton2002}} 
   \label{tcabr}
   \small 
   \centering 
   \begin{tabular}{lccr} 
   \hline\hline   
   \textbf{Parameter} & \textbf{Symbol} & \textbf{Value} & \\ 
   \hline
   Larger radius & $R_0$ & $0,615$ m \\
   Minor radius & $b$ & $0,21$ m \\
   Plasma column radius & $a$ & $0,18$ m \\
   Toroidal magnetic field & $B_0$ & $1,07$ T \\
   Plasma current (equilibrium) & $I_p$ & < $100$ kA \\
   Safety factor (equilibrium at $r=a$) & $q^0(a)$ & $\approx 5.0$ \\
   \hline
   \end{tabular}
\end{table}The full phase space of the model was presented in the last panel of Fig.\ \ref{fig:UM_scheme} however, from now on, we focus only on the region of the plasma edge $0.0 < y < 0.4$. Figure \ref{fig:UM_phase_space} displays the characteristic phase space of the UM considering the perturbation mode $m = 7$, the safety factor $q^0(a) = 5.0$ and $\delta B = 1.5 \%$ ($\delta I \approx 0.45\%$).

The chaotic region, depicted by the extended connected black sea around the periodic islands in white, present in phase space drawn in Fig.\ \ref{fig:UM_phase_space} will be an essential framework for all phase space analyses in the upcoming chapters. It's worth noting that the open field lines, that compose the aforementioned chaotic sea, can eventually escape hitting the inner wall of the tokamak at $y=0$. Therefore, the model's escape condition is satisfied when $y_n < 0 < y_{n+1}$.

Escaping field lines, featured in both BM and UM models, are the main focus of the methodology proposed in the next section.

\section{Escape analysis}
\label{sec:chap3_escape}

The methodology outlined here was developed to address one fundamental practical question: Which values of control parameters should be selected for all numerical simulations of the models?

Our approach to addressing this question is based on the Escape Rate ($ER$), which is defined as the proportion of escaping trajectories ${\bf x}^e$ that correspond to the escaping magnetic field lines, relative to the total number $M$ of initial conditions (ICs) provided to the models; $ER = {\bf x}^e / M$. By defining $ER$, we can analyse it as a function of the control parameters, $k$ (for the BM) and $\delta B$ (for the UM). To do so, we establish a suitable range of parameter values that preserves the key characteristic features of the phase spaces. The parameter ranges are defined as follows:

\begin{equation}
     \begin{aligned}
    k\in\frac{[0.53, 0.63]}{L}~\text{and}~
    \delta B \in \frac{[1.2\%, 2.2\%]}{L}~,
    \end{aligned}
    \label{eq:method_ranges}
\end{equation}where $L$ is the number of parameters between the considered range. 

\begin{figure}[h]
\centering
\includegraphics[scale=0.60]{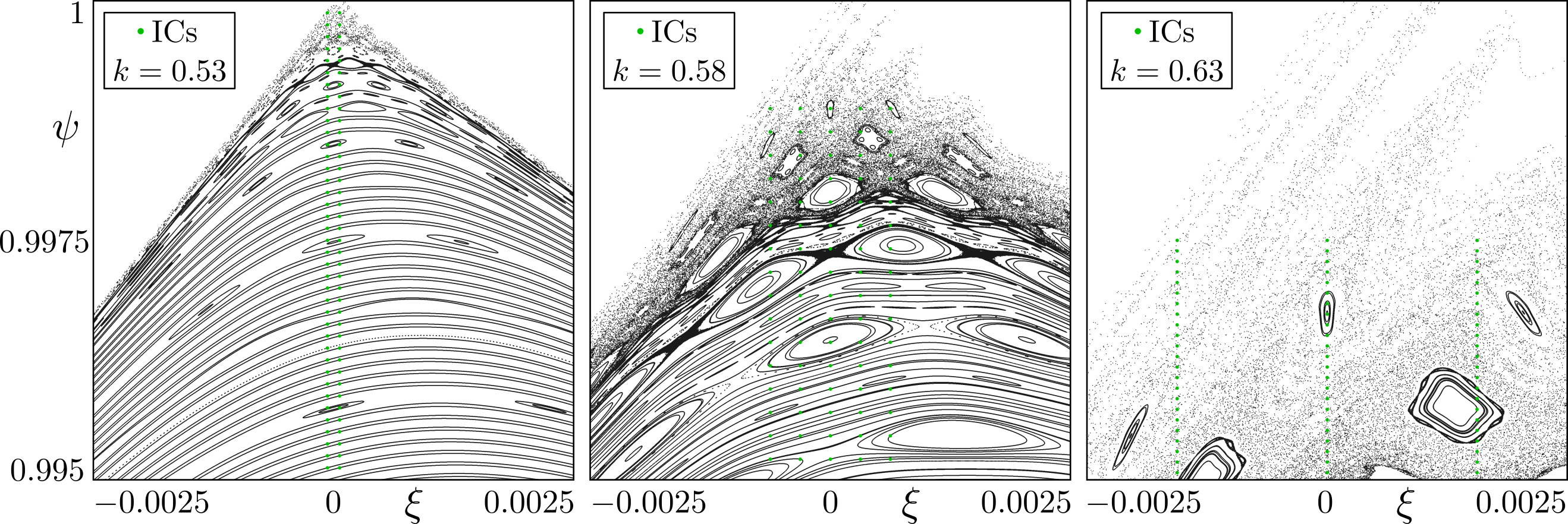}
\caption[Phase spaces of the BM increasing the perturbation strength]{Phase spaces of the BM considering $k=0.53$ (left), $k=0.58$ (centre) and $k=0.63$ (right). The selected ICs are represented by the green points and their evolution was up to $5 \times 10^5$ iterations of the map $T_{\text{BM}}$. All three phase spaces are drawn considering only the region $\xi \in [-0.0025, 0.0025]$ and $\psi \in [0.995, 1.0]$.}
\label{fig:BM_3ps}
\end{figure}

\begin{figure}[t]
\centering
\includegraphics[scale=0.61]{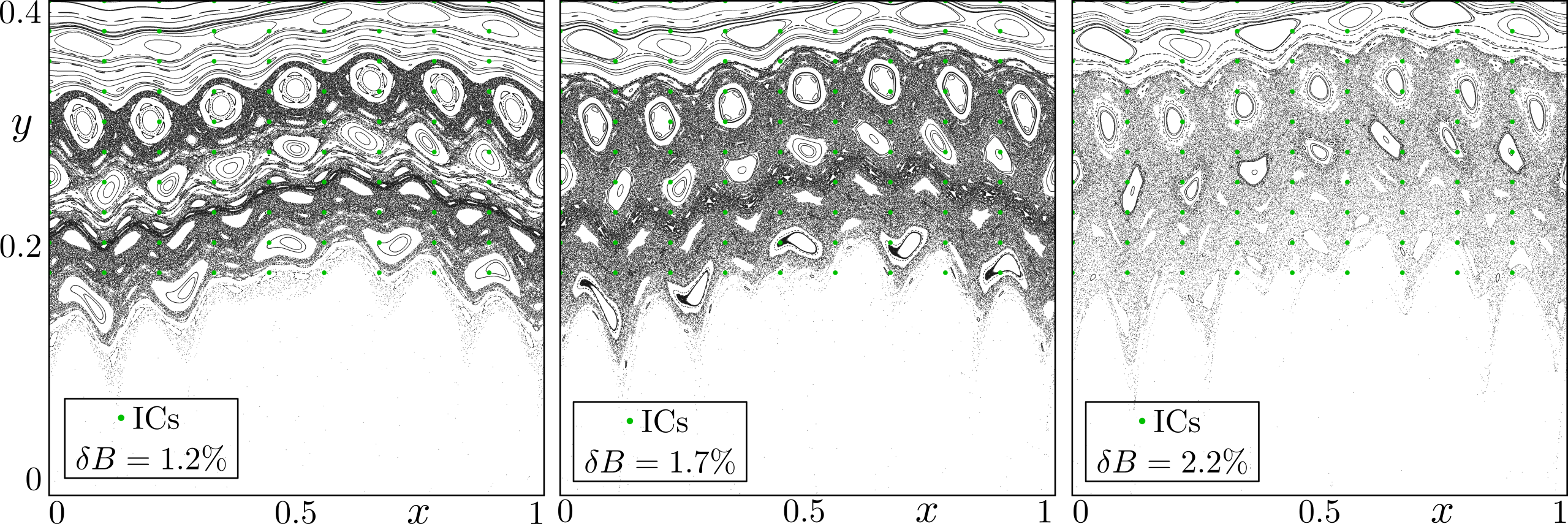}
\caption[Phase spaces of the UM increasing the perturbation strength]{Phase spaces of the UM considering $\delta B=1.2\%$ (left), $\delta B = 1.7\%$ (centre) and $\delta B = 2.2\%$ (right). The selected ICs are represented by the green points and their evolution was up to $2 \times 10^4$ iterations of the composed map $T^0_{\text{UM}} \circ T^1_{\text{UM}}$. All three phase spaces are drawn considering the region $x \in [0,1]$ and $y \in [0, 0.4]$.}
\label{fig:UM_3ps}
\end{figure}

Antes de analisar o $ER$ como uma função dos parâmetros de controle, é necessário garantir que as condições de fuga possam ser realmente satisfeitas pelas trajetórias evoluídas. Como apenas as trajetórias caóticas podem escapar, uma região caótica adequada deve ser encontrada nos espaços de fase. Para isso, as Figuras \ref{fig:BM_3ps} e \ref{fig:UM_3ps} mostram os espaços de fase para o BM e o UM, respectivamente, considerando três valores diferentes de parâmetros. Esses valores correspondem aos extremos (maior e menor) das faixas definidas na Eq. \ref{eq:method_ranges} e a um valor intermediário. É importante observar como a configuração do espaço de fase muda ao aumentar o valor do parâmetro de controle.

Once the ranges of parameters of interest are established and the phase spaces are drawn, it is possible to identify a region where all given ICs can be placed within the chaotic sea. However, this specific chaotic region must be maintained throughout the entire parameter range\footnote{Identifying this robust chaotic region is a meticulous procedure specifically for the BM. As evidenced by Fig.\ \ref{fig:BM_3ps}, both the topology and the size of the phase space are highly sensitive to the parameter change.}. These conditions assured, only chaotic trajectories are evolved and, therefore, the escape conditions can be satisfied.

\begin{figure}[h]
\centering
\includegraphics[scale=0.55]{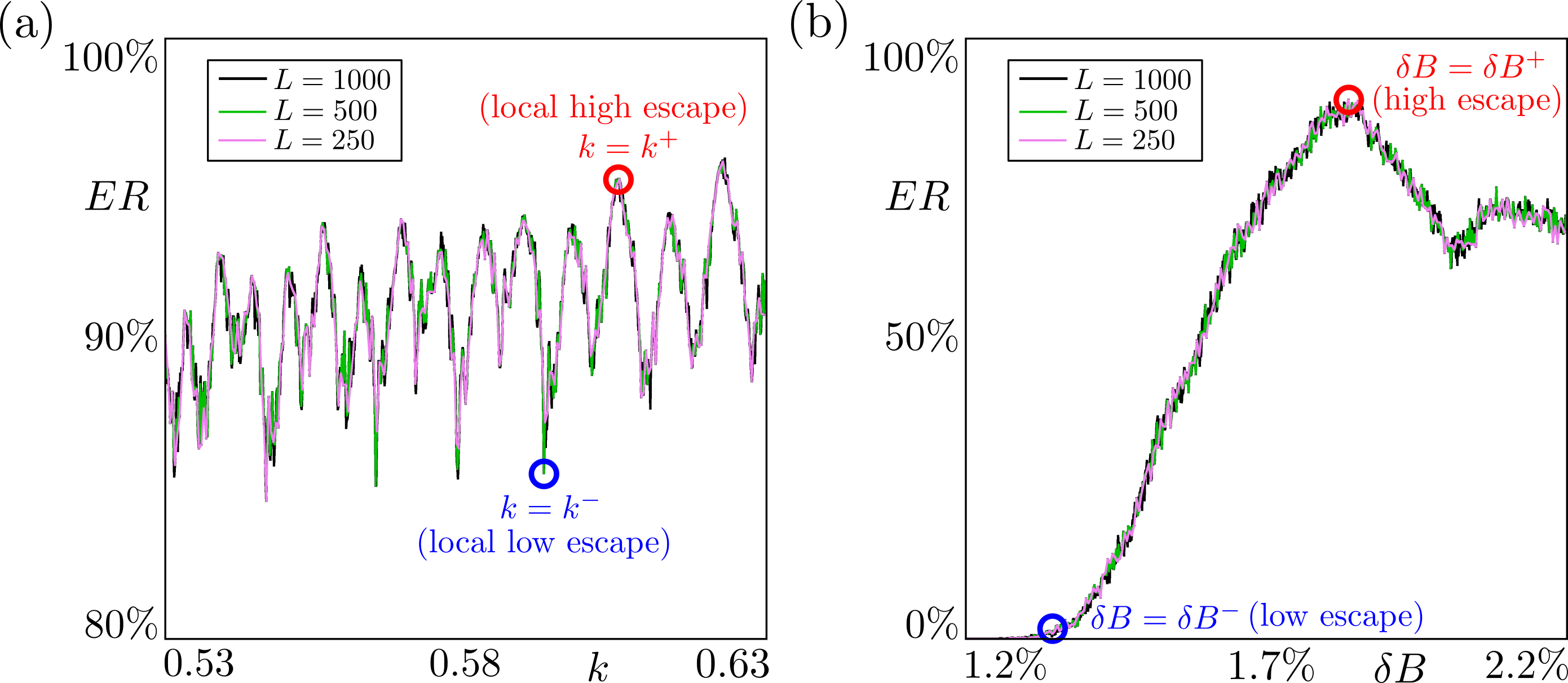}
\caption[Escape rate as a function of the perturbation strengths]{Behaviour of the escape rate as a function of the control parameters for both the BM (a) and UM (b) models. Their defined ranges were divided into $L$ different parameters and, the colour lines show the behaviour of $1000$ (black), $500$ (green) and $250$ (violet) parameters.}
\label{fig:MTD_ER_parameter}
\end{figure}

First, for the BM, $M = 10^5$ ICs placed at $\xi_0 = 0$ and $\psi_0 \in [1-10^{-9},1-10^{-10}]$, were evolved up to $10^3$ iterations of the map. The computed $ER = ER(k)$, considering $L = 250$, $500$ and $1000$ parameters distributed at $k \in [0.53, 0.63]$, is shown in Fig.\ \ref{fig:MTD_ER_parameter}(a). It is worth noting that the general behaviour of $ER(k)$ is the same for the tree values of $L$.

For the UM, $M = 10^3$ ICs where placed at $x_0 \in [10^{-6},10^{-5}]$ and $y_0 = 0.3$ and evolved up to $10^5$ iterations. The computed $ER = ER(\delta B)$, considering $L = 250$, $500$ and $1000$ parameters distributed at $\delta B \in [1.2\%, 2.2\%]$, is shown in Fig.\ \ref{fig:MTD_ER_parameter}(b). It is also worth noting that the general behaviour of $ER(\delta B)$ is the same for the tree selected values of $L$.

In general terms, the escape analysis illustrated in Fig.\ \ref{fig:MTD_ER_parameter} yields precise values for the control parameters that indicate a configuration which enhances or restrains the escape. For the BM, $k = k^{-} = 0.5930$ and $k = k^{+} = 0.6056$ are the parameters of interest: $ER(k^{+})$ is the local high around $k\approx 0.6$, which not only provides better visualisation of the phase space but also is an adequate value as discussed in Sec.\ \ref{sec:chap3_boozer} and, $ER(k^{-})$ the local low. For the UM, $\delta B = \delta B^{-} = 1.334\%$ and $\delta B = \delta B^{+} = 1.836\%$ are the parameters of interest: $ER(\delta B^{+})$ is the global high and, $ER(k^{-})$ is the global low that also indicates the first parameter value that escapes occurs.  

Furthermore, it is worth remarking that the general behaviour displayed in both panels of Fig.\ \ref{fig:MTD_ER_parameter} provides additional relevant interpretations. On one hand, the analysis of the BM shows a curious oscillatory behaviour that might be related to intrinsic dynamical structures, such as homoclinic tangles that arises and vanishes, around the saddle point. On the other hand, the analysis of the UM yields an expected growth while increasing the parameter value, however, there is a maximum followed by an arguably puzzling decay. Nevertheless, these interesting features would be further and properly investigated in other opportunities. 

Finally, once defined the values of interest $k^+$ and $k^-$ for the BM and, analogously $\delta B^{+}$ $\delta B^{-}$ for the UM, it is possible to thoroughly investigate their respective phase spaces via two of our original methods. Chapters four and five are devoted to detailing the analyses and presenting the results.



\chapter{Transient motion}
\label{chap4:transient}

This chapter presents the first phase space analysis developed to improve our understanding of the field lines' behaviour. The \emph{transient motion} analysis introduces a numerical method to visually illustrate the transient dynamics in Hamiltonian systems. In our context, Hamiltonian systems that are either open, leaking, or contain holes in the phase space possess solutions that eventually escape the system's domain and, in that sense, the motion described by such escape orbits before crossing the escape threshold can be understood as a transient behaviour. 

The method is based on the \emph{transient measure}, a finite-time version of the natural measure often used in analyses of dissipative systems for calculating dimensions of chaotic attractors \cite{Ott2002}. Once the profile of this new measure is computed throughout the phase space, preferable paths taken by the escape trajectories are outlined. The detailed knowledge of these possible paths, namely what a given orbit may experience in the transient dynamics before escaping, is important and may have further implications for the analysed systems. Here we focus on both models (BM and UM), considering the values of perturbations strengths given by the escape analysis presented in Chapter \ref{chap3:models}, Sec.\ \ref{sec:chap3_escape}.  

It is important to mention that this is an original analysis of general Hamiltonian systems; Contribution presented at {\it Measure, dimension, and complexity of the transient motion in Hamiltonian systems} by Vitor M. de Oliveira, Matheus S. Palmero, Iberê L. Caldas, Physica D {\bf 431}, 133126, published in March 2022 \cite{deOliveira2022}.

\section{Mathematical framework}
\label{sec:chap4_math}

Before presenting the numerical results of the method considering both models described in the last chapter, it is important to introduce the mathematical framework related to this analysis. In the first subsection, we define the transient measure and consequently related quantities; The second subsection is devoted to general definitions of unstable and stable invariant manifolds that are relevant to the upcoming numerical results. 

\subsection*{Mean transient measure}
\label{subsec:chap4_measure}

In general terms, let ${\boldsymbol{\varphi}}_{t}(\boldsymbol{x}_0)$ be a solution of our dynamical system in the $D$-dimensional phase space with initial condition ${\boldsymbol{x}}_0$ and at time $t$, and let us cover the region of the phase space that we are interested in by a grid of $D$-dimensional boxes of side-length $\varepsilon$.

We call $\eta(B_i,{\boldsymbol{\varphi}}_{t}({\boldsymbol{x}}_0),T)$ the total time spent by the solution ${\boldsymbol{\varphi}}_{t}({\boldsymbol{x}}_0)$ inside the box $B_i$ in the time interval $t\in[0,T]$. If $\eta$ is the same for almost every ${\boldsymbol{x}}_0$, the \emph{natural measure} for each box $B_i$ can be defined as \cite{Ott2002}

\begin{equation}
    \mu_i=\lim_{T\to\infty}\dfrac{\eta(B_i,{\boldsymbol{\varphi}}_{t}({\boldsymbol{x}}_0),T)}{T},
    \label{eq:natural_measure_def}
\end{equation}if the limit exists. It follows that $\sum_{i=1}^{N} \mu_i = 1$, where $N$ is the number of boxes in the grid, which depends on the box side-length $\varepsilon$.

The natural measure is defined in the asymptotic limit $T\to\infty$ and is usually associated with the dynamics of an orbit on a chaotic attractor. We are interested here, however, in the transient dynamics of escape orbits in Hamiltonian systems. Then, we propose a finite-time version of Eq.\ (\ref{eq:natural_measure_def}), which we call the \emph{transient measure},

\begin{equation}
    \nu_i=\dfrac{\eta(B_i,{\boldsymbol{\varphi}}_{t}({\boldsymbol{x}}_0),T^e)}{T^e},
    \label{eq:escape_measure_def}
\end{equation}where $T^e$ is the escape time, i.e., the time it takes for the orbit that starts at ${\boldsymbol{x}}_0$ to reach a predefined escape region. Here, $\eta$ is the total time spent by the orbit inside the box $B_i$ before leaving the system. It is important to note that $\sum_{i=1}^{N} \nu_i = 1$.

If we consider an orbit in the chaotic sea, the transient measure reflects the path followed by the orbit up until exiting the system. Hence, this measure is able to depict the transient dynamics of an escape orbit, including effects such as stickiness.

Generally, a single chaotic orbit will visit only a small portion of the available chaotic area before escaping, making it hard to describe the behaviour of escape orbits in the phase space by looking solely at $\nu$. Therefore, we now define the \emph{mean transient measure}, the average of the transient measure on an ensemble $E$ composed by $M$ initial conditions,

\begin{equation}
    \langle\nu_i\rangle=\dfrac{1}{M}\sum^{M}_{j=1}\nu_{i,j},
    \label{eq:mean_escape_measure_def}
\end{equation}where $\nu_{i,j}=\eta(B_i,{\boldsymbol{\varphi}}_{t}({\boldsymbol{x}}_{0,j}),T^e_j)/T^e_j$ is the transient measure for the $j$-th initial condition ${\boldsymbol{x}}_{0,j}$ and box $B_i$. As was the case for the transient measure, we have that $\sum_{i=1}^{N}\langle\nu_i\rangle=1$. 

Equation\ (\ref{eq:mean_escape_measure_def}) is well-defined for any discrete ensemble. For us, $E$ has a small volume and a high number of elements $M$ which are uniformly distributed on a grid. The ensemble is centred at an initial condition of interest, and the mean transient measure, therefore, describes the transient dynamics associated with a small neighbourhood of the said point. In practice, $M$ is chosen to be large enough so that the orbits do visit a sufficient number of boxes and clearly depict the transient behaviour throughout the phase space.

Apart from the finite-time aspect, another difference of Eq.\ (\ref{eq:escape_measure_def}) from Eq.\ (\ref{eq:natural_measure_def}) is that we do not demand it holds for almost every ${\boldsymbol{x}}_0$. With that, the mean transient measure, Eq.\ (\ref{eq:mean_escape_measure_def}), is, in fact, a function of the ensemble of ICs $\langle\nu_i\rangle={\langle\nu_i\rangle}_E$. Hence, there may be different transient behaviours depending on the ICs. Indeed, we investigate in \cite{deOliveira2022} several different behaviours, considering two distinct Hamiltonian systems, depending on the correspondingly chosen location for the ensemble $E$.

Here, however, we use the method for a slightly different approach; The profiles of the mean transient measure, computed throughout the phase spaces, can visually distinguish the field lines' behaviours while considering specifically perturbations strengths that enhance or restrain the escape, previously discussed in Chapter \ref{chap3:models}, Sec.\ \ref{sec:chap3_escape}. These behaviours are strictly connected to underlying geometrical structures in the phase space known as invariant manifolds.  

\subsection*{Stable and unstable invariant manifolds}
\label{subsec:chap4_manifolds}

First, since our models of interest are two-dimensional non-linear maps, we restrict this brief general discussion to only this type of planar system. Hence, let $T$ be a two-dimensional invertible map, with both $T$ and $T^{-1}$ differentiable, and let $\alpha=\{\boldsymbol{o}_1, \boldsymbol{o}_2,\dots, \boldsymbol{o}_m\}$ be an unstable periodic orbit (UPO) of period $m$ of the mapping $T$. The \emph{stable manifold} $W^s(\alpha)$ and the \emph{unstable manifold} $W^u(\alpha)$ associated with $\alpha$ are given by \cite{Alligood2012}

\begin{equation}
    \begin{aligned}
    W^{s}(\alpha) &= \{\boldsymbol{x} \in U \subset \mathbb{R}^{2} ~|~ T^{n}(\boldsymbol{x}) \rightarrow T^{n}(\boldsymbol{o_i}) ~\text{as}~ n \rightarrow \infty, ~ i = 1, ..., m\}~,\\
    W^{u}(\alpha) &= \{\boldsymbol{x} \in U \subset \mathbb{R}^{2} ~|~ T^{-n}(\boldsymbol{x}) \rightarrow T^{-n}(\boldsymbol{o_i}) ~\text{as}~ n \rightarrow \infty, ~ i = 1, ..., m\}~.
    \end{aligned}
    \label{eq:mani_defs}
\end{equation}In practice, $\boldsymbol{x}$ is the pair of the defined coordinates and $U$ can be any suitable region of the system's phase space.  

The manifolds $W^{s,u}(\alpha)$ are, indeed, one-dimensional curves which are locally tangent to the respective subspaces \cite{Wiggins2003}. The stable manifold is tangent to the eigenvector in the stable subspace, while the unstable manifold is tangent to the unstable eigenvector. They are additionally, by definition, invariant under the action of the dynamical system and formed by infinite sets that, in non-integrable cases, can intercept. The crossings between $W^{s}(\alpha)$ and $W^{u}(\alpha)$ compose what is known as \emph{homoclinic orbtis}. Moreover, the crossings between $W^{s}(\alpha_1)$ and $W^{u}(\alpha_2)$, with $\alpha_1 \neq \alpha_2$, are known as \emph{heteroclinic orbtis}. 

Tracing manifolds is a difficult task that can only be overcome using fitting numerical tools. In our results, we use the efficient method proposed by Ciro \cite{Ciro2018} that, essentially, chooses an appropriate segment in a linear approximation for the dynamics and evolves said segment under the mapping and under its inverse. The method also proved to be useful for calculating homo/heteroclinic intersections in planar maps \cite{deOliveira2020}. 

\section{Numerical results}
\label{sec:chap4_results}

The transient motion analysis offers a visual aid for the hidden transient behaviour of escaping trajectories. In that sense, we employ the method on extensive numerical simulations of both BM and UM models. The results are shown separately in the subsections below.

\subsection*{Single-null divertor map}
As described in Chapter \ref{chap3:models}, Sec.\ \ref{sec:chap3_boozer}, the topology of the magnetic configuration induced by the single-null poloidal divertor, represented via the characteristic phase space of the BM shown in Fig.\ \ref{fig:BM_phase_space}, presents a saddle point at $(\xi^{\star}, \psi^{\star}) = (0,1)$. Accordingly, the position of the divertor target is considered $\psi_{\text{target}} = 1.0$ to be the nearest to the saddle point, imposing the escape condition $\psi_n < 1.0 < \psi_{n+1}$ to the map equations. 

\begin{figure}[h]
\centering
\includegraphics[scale=0.80]{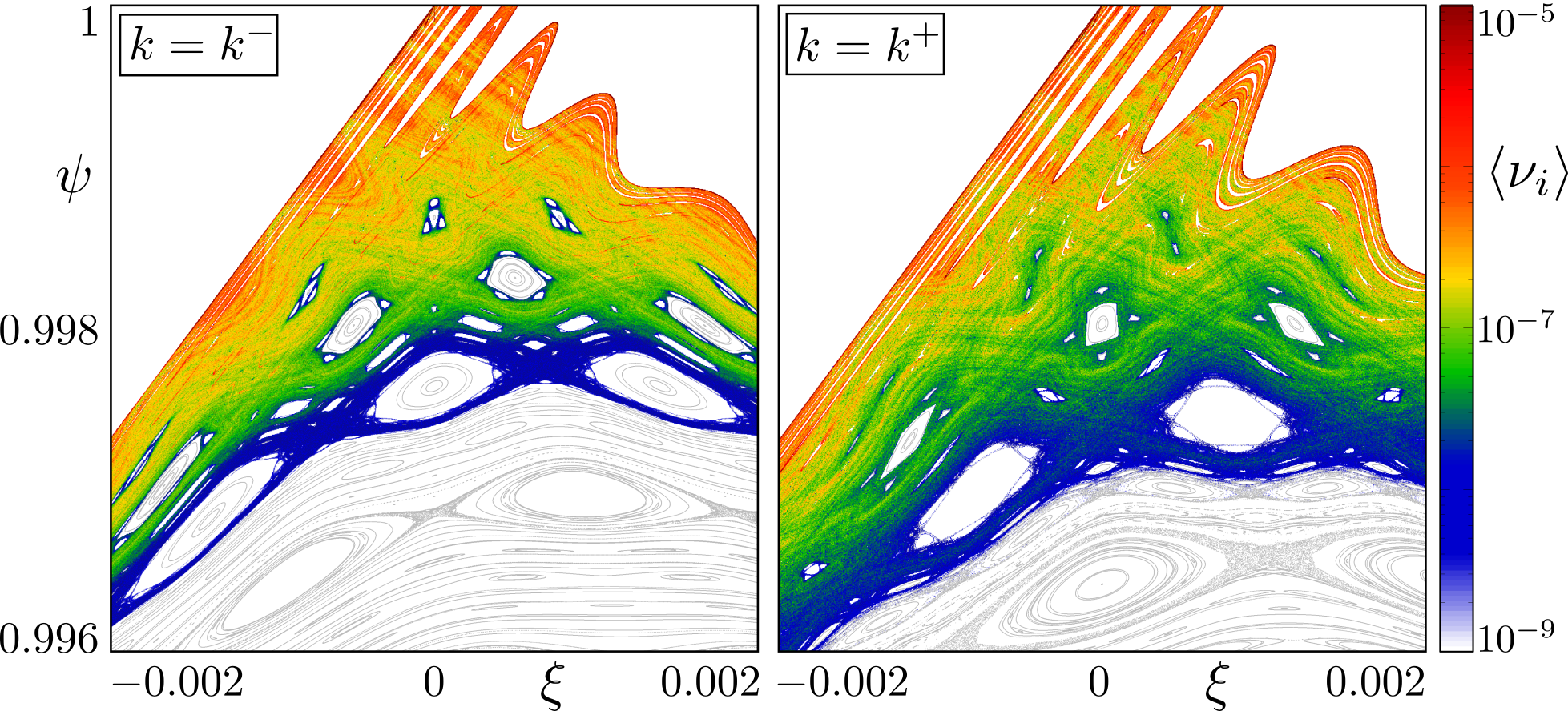}
\caption[Mean transient measure profile for the BM]{Profiles of the mean transient measure $\langle \nu_i \rangle$ in logarithmic scale for the BM, calculated on a $1024\times1024$ grid, considering $k=k^{-}=0.5930$ (left) and $k=k^{+}=0.6056$ (right). Their respective phase spaces are depicted in grey on the background. The ensemble of ICs and the colour range for the mean transient measure were kept the same for both cases.}
\label{fig:TMA_bm_profiles}
\end{figure}

Once defined the escape condition, we select an ensemble of $M = 10^5$ ICs, uniformly distributed in a dense small line positioned at $\xi_0 = 0$, $\psi_0 \in [1 - 1\times10^{-4}, 1 - 1\times10^{-6}]$ evolved up to $10^7$ iterations of the map. For the computation of the transient measure, it was considered $\varepsilon = 1024$ as the side-length of the boxes that compose a grid over the region of interest $\xi \in [-0.0025, 0.0025]$ and $\psi \in [0.995,1.0]$. Then, it is finally possible to compute the profile of the mean transient measure $\langle \nu_i \rangle$ throughout the phase space of the BM, considering the two special values of the control parameter: $k = k^{-}$, the perturbation strength that gives a low escape rate and; $k = k^{+}$, the perturbation strength that gives a high escape rate. Figure \ref{fig:TMA_bm_profiles} displays the results.  

From Fig.\ \ref{fig:TMA_bm_profiles} we readily observe that the $\langle \nu_i \rangle$ profile, depicted by the logarithmic colour scale, highlights the differences between both cases. As expected, since $k^{+} > k^{-}$, the available chaotic portion of the phase space for $k=k^{+}$ is larger than the phase space for $k=k^{-}$. However, the gradient from the colour scales provides better insights into how the chaotic orbits are experiencing these chaotic regions before escaping the system. On one hand, for $k^{-}$ escape orbits frequently visits regions around upper smaller islands, as depicted by the surrounding blue colours, strongly indicating the presence of stickiness. On the other, $k^{+}$ presents a relatively thinner profile, where most of the previous islands are already destroyed and the gradient from yellow to dark red reveals more visible gaps (in white) farther from the saddle. 

\begin{figure}[h]
\centering
\includegraphics[scale=1.0]{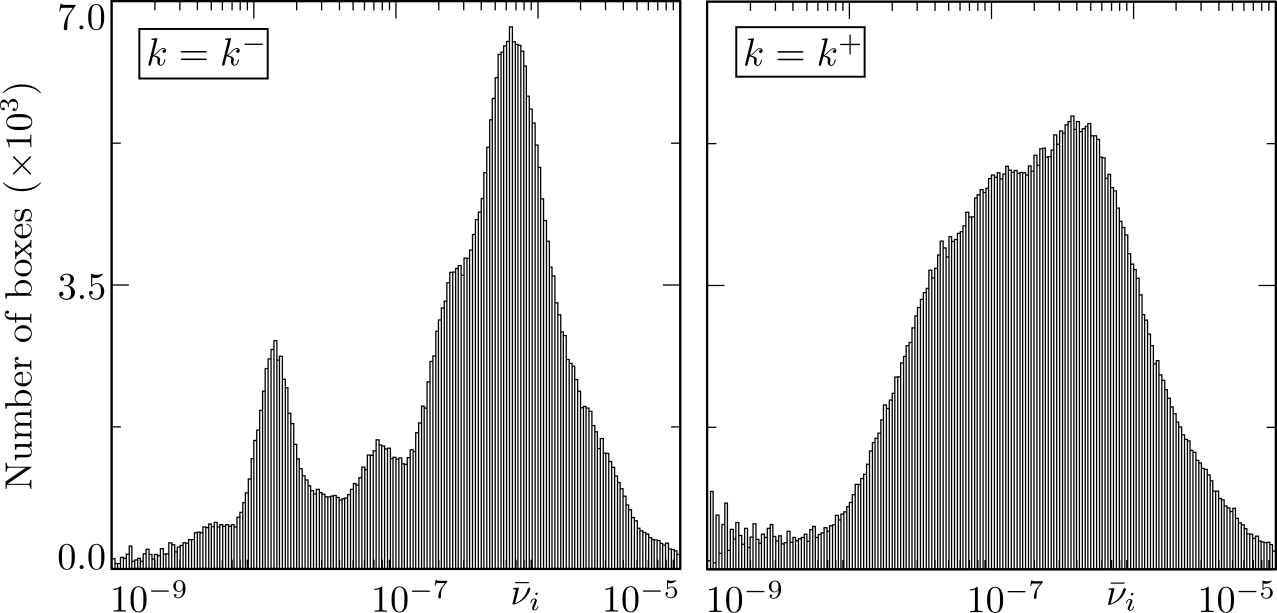}
\caption[Histograms of the mean transient measure for the BM]{Histogram distributions of the mean transient measure $\langle \nu_i \rangle$ for the escape trajectories of the BM, considering both $k=k^{-}=0.5930$ (left) and $k=k^{+}=0.6056$ (right). All parameters were kept the same as for Fig.\ \ref{fig:TMA_bm_profiles}.}
\label{fig:TMA_bm_histograms}
\end{figure}

Before diving into the comparison between the complicated structures uncovered by the profile and the correspondingly invariant manifolds, it is possible to statistically investigate the mean transient measure profiles by computing their histograms. Considering only the boxes visited at least once by the simulated dynamics for both cases, the result is shown in Fig.\ \ref{fig:TMA_bm_histograms}.  

The calculated histogram distributions stress the different transient behaviours that emerge from the complex dynamical scenario of the system, especially while comparing different perturbation strengths. Confronting both panels of Fig.\ \ref{fig:TMA_bm_histograms} to their respective profiles on the phase space, shown in Fig.\ \ref{fig:TMA_bm_profiles}, we note that, indeed, orbits within $k=k^{-}$ frequently visit regions that are not available in $k=k^{+}$. The smaller peak around $\langle \nu_i \rangle \approx 10^{-8}$ indicates a relatively high concentration for all possible paths in areas coloured blue (respective colour for $\langle \nu_i \rangle \approx 10^{-8}$) in Fig.\ \ref{fig:TMA_bm_profiles}. Still in $k=k^{-}$, the higher peak, around $\langle \nu_i \rangle \approx 10^{-6}$ is sharper compared to the broader distribution for $k=k^{+}$. This suggests a somewhat contra-intuitive realisation that although the phase space of $k^{-}$ presents a smaller chaotic region compared to $k^{+}$, escaping chaotic trajectories from $k^{-}$ are experiencing thoroughly most of the available regions and, in that sense, enabling stickiness phenomena, in comparison to a more erratic visitation of the orbits in $k^{+}$. 

\begin{figure}[h]
\centering
\includegraphics[scale=1]{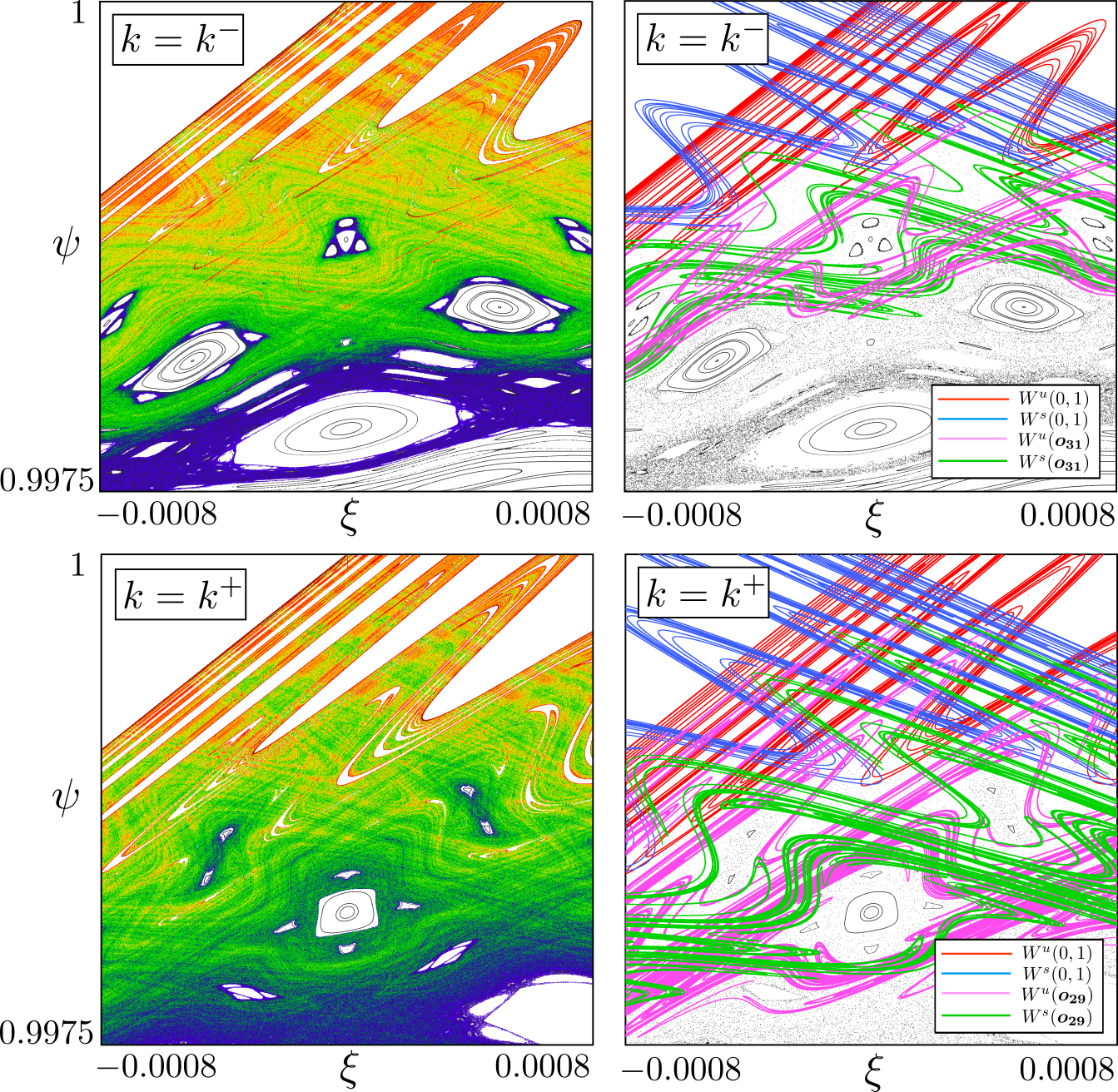}
\caption[Visual comparison between profiles and invariant manifolds for the BM]{Comparison between the $\langle \nu_i \rangle$ profile (left panels) and invariant manifolds $W^{s,u}$ (right panels) associated with the saddle point and nearest UPOs of interest. Their respective phase spaces are depicted in grey on the background. All parameters and the colour scale (omitted) are the same as in Fig.\ \ref{fig:TMA_bm_profiles}.}
\label{fig:TMA_bm_mani}
\end{figure}

Finally, as a last visual investigation for the BM, we compare the uncovered structural details shown in Fig.\ \ref{fig:TMA_bm_profiles} to the relevant invariant manifolds present in both phase spaces. To further improve the visualisation, we consider an amplified region $\xi \in [-0.0008, 0.0008]$ and $\psi \in [0.9975,1.0]$ where the respective details are more evident. Figure \ref{fig:TMA_bm_mani} shows, on the left panels, the computed profile throughout this region of the phase space and, on the right, the same region covered by stable and unstable invariant manifolds associated with the saddle $(\xi^{\star}, \psi^{\star}) = (0,1)$ and the nearest UPOs from the last visible chain of islands. For $k^{-}$, the nearest UPO is associated with a period 31 chain of islands and, for $k^{+}$ it is a period 29 UPO.

The colour gradient of $\langle \nu_i \rangle$ depicts, in both cases, complex geometrical structures, formed by seemingly erratic curves, that are embedded in their respective phase spaces. These structures accurately agree with the invariant manifolds shown in the right panels of Fig.\ \ref{fig:TMA_bm_mani}. Curiously, the structures revealed by the mean transient measure are consistent not only with the unstable manifolds but also with the stable ones. Moreover, while comparing both cases of $k$, it is possible to note that the traced manifolds for $k^+$ seem more interconnected\footnote{As a technical remark; It is known that, since these invariant manifolds are composed by infinite sets, both homo/heteroclinic intersections are, by theoretical definition, also invariant and infinite sets. That being said, we discuss the interconnectivity of these manifolds focusing only on what was possible to display in Fig.\ \ref{fig:TMA_bm_mani}.} than the ones presented in $k^-$. Of course, knowing that the manifolds are often related to transport channels, the strong interconnected structures behind the phase space for $k^{+}$ might explain why it is a configuration that enhances the escape, differently from what is observed for $k^{-}$.       

\subsection*{Ergodic magnetic limiter map}
Analogously to what was presented for the BM, Chapter \ref{chap3:models}, Sec.\ \ref{sec:chap3_ullmann} was devoted to the discussion that the phase space of the UM, composed by both Eqs.\ (\ref{eq:um_eq}) and (\ref{eq:um_pert}), describes the configuration of the magnetic field lines of a tokamak with an ergodic magnetic limiter. Depending on the values for the perturbation strength, provided either by $\delta B$ or $\delta I$, it was shown that escape field lines may be found, considering the escape condition $y_n < 0 < y_{n+1}$, meaning that a field line crossed the inner wall at $y = 0$. 

\begin{figure}[h]
\centering
\includegraphics[scale=0.77]{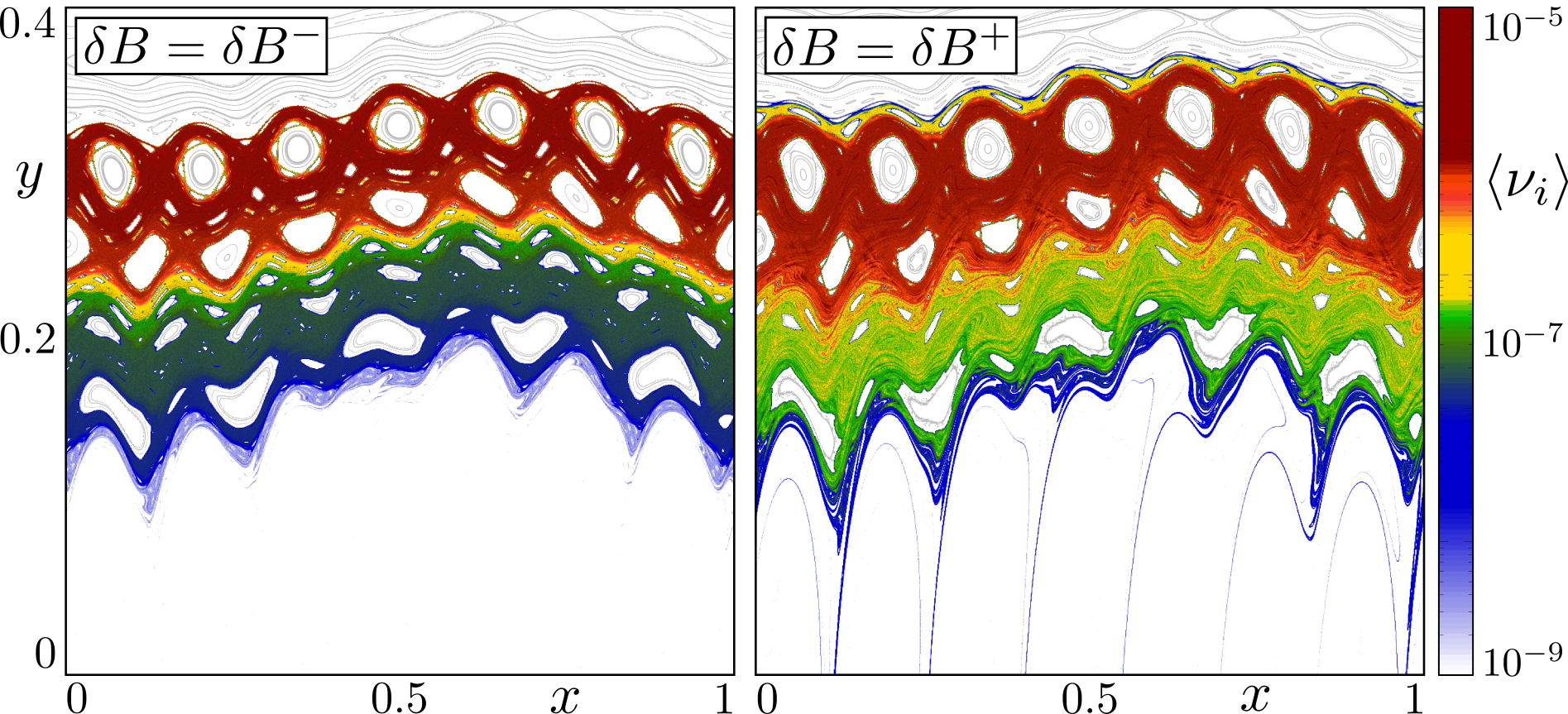}
\caption[Mean transient measure profile for the UM]{Profiles of the mean transient measure $\langle \nu_i \rangle$ in logarithmic scale for the UM, calculated on a $1024\times1024$ grid, considering $\delta B=\delta B^{-}=1.334\%$ (left) and $\delta B=\delta B^{+}=1.836\%$ (right). Their respective phase spaces are depicted in grey on the background. The ensemble of ICs and the colour range for the mean transient measure were kept the same for both cases.}
\label{fig:TMA_um_profiles}
\end{figure}

To numerically calculate the transient measure and the respective mean transient profile, we select an ensemble of $M = 10^4$ ICs, uniformly distributed in a dense small square in $x_0 \in [-1 \times 10^{-7}, 1 \times 10^{-7}]$ and $y_0 \in [0.3 - 1 \times10^{-7}, 0.3 + 1\times10^{-7}]$ evolved up to $10^5$ iterations. Only the lower region $y < 0.4$ with $x \in [0,1]$ was selected for the analysis, considering $\varepsilon = 1024$ as the boxes' side-length of the grid in this region. Then, the profile of the mean transient measure $\langle \nu_i \rangle$ for the UM was computed for the two special values: $\delta B = \delta B^{-}$, the perturbation strength that gives a low escape rate and; $\delta B = \delta B^{+}$, the perturbation strength that gives a high escape rate. Figure \ref{fig:TMA_um_profiles} displays the results.  

It is possible to readily observe that the $\langle \nu_i \rangle$ profile, depicted by the logarithmic colour scale in Fig.\ \ref{fig:TMA_um_profiles}, highlights the differences between both cases similarly to results for the BM. However, one main difference between the simulation of the UM and the BM is the selected location for the ensemble of ICs. For the UM, ICs are located at $y \approx 0.3$, far from the escape condition at $y =0$, whereas the ensemble of ICs for the BM was set much closer to its respective escape condition. This discrepancy is due to the fact that the standard phase space configuration for the UM is more robust to changes in the perturbation strength, compared to the phase spaces for the BM. Nevertheless, investigating the transient behaviour of escape field lines originating from a region closer to the highly confined magnetic fields (modelled by $y > 0.5$ on the UM) may have important implications for further understanding of plasma-wall interactions in tokamaks.

\begin{figure}[h]
\centering
\includegraphics[scale=1.0]{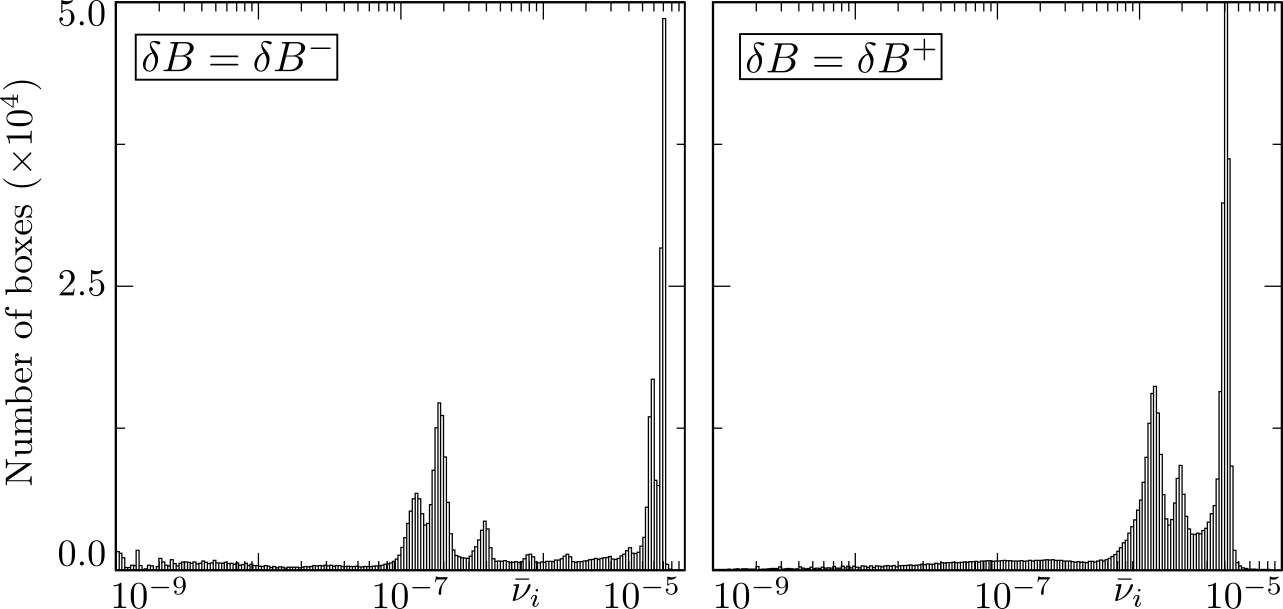}
\caption[Histograms of the mean transient measure for the UM]{Histogram distributions of the mean transient measure $\langle \nu_i \rangle$ for the escape trajectories of the UM, considering both $\delta B=\delta B^{-}=1.334\%$ (left) and $\delta B=\delta B^{+}=1.836\%$ (right). All parameters were kept the same as for Fig.\ \ref{fig:TMA_um_profiles}.}
\label{fig:TMA_um_histograms}
\end{figure}

As a complementary statistical result to Fig.\ \ref{fig:TMA_um_profiles}, we show in Fig.\ \ref{fig:TMA_um_histograms} the computed histogram distributions associated with the $\langle \nu_i \rangle$ profiles.

Figure \ref{fig:TMA_um_histograms} displays rather different distributions in comparison to Fig.\ \ref{fig:TMA_bm_histograms} for the BM. This is expected since for the UM we analyse ICs placed far from the escape condition. We note that the average behaviour of escaping trajectories has similar high peaks around $10^{-5}$, depicted by dark red colours in both panels of Fig.\ \ref{fig:TMA_um_profiles} that are related to the area nearest to the ensemble of ICs. However, other minor peaks are in different positions on the $\langle \nu_i \rangle$ range; While $\delta B^{-}$ presents secondary peaks far from the first one, around $10^{-7}$, the secondary peaks of the distribution for $\delta B^{+}$ are closer to the primary. One way to interpret this result is that the average behaviour of the escaping orbits for $\delta B^{-}$ is more influenced by regions depicted in dark green ($\langle \nu_i \rangle \approx 10^{-7}$), while for $\delta B^{+}$ the regions of influence are closer to the area where the ensemble of ICs was set, namely the regions depicted in red and yellow on the right panel of Fig.\ \ref{fig:TMA_um_profiles}. 

\begin{figure}[h]
\centering
\includegraphics[scale=1.0]{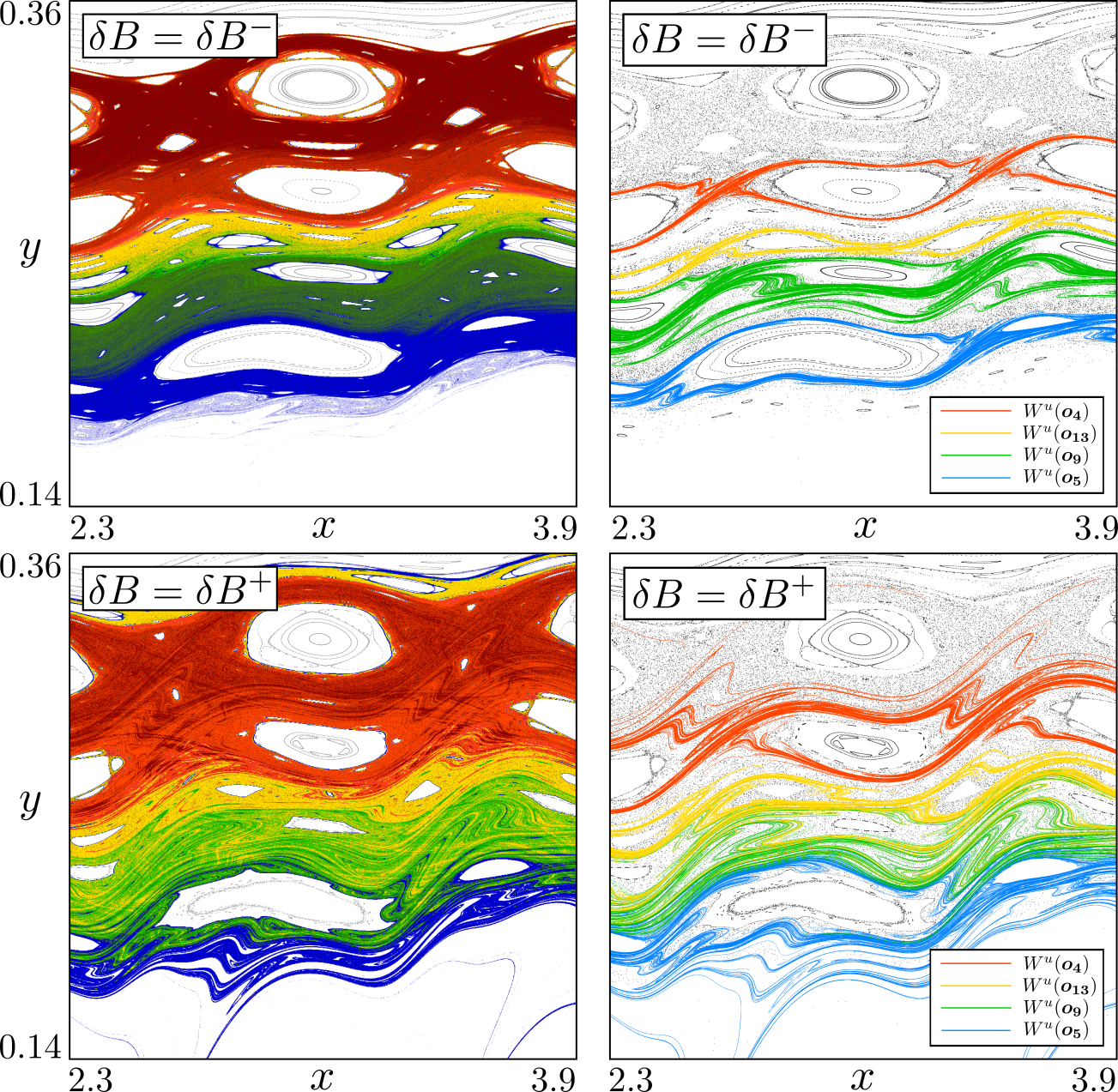}
\caption[Visual comparison between profiles and invariant manifolds for the UM]{Comparison between the $\langle \nu_i \rangle$ profile (left panels) and unstable invariant manifolds $W^{u}$ (right panels) associated with four different chains of islands. Their respective phase spaces are depicted in grey on the background. All parameters and the colour scale (omitted) are the same as in Fig.\ \ref{fig:TMA_um_profiles}.}
\label{fig:TMA_um_mani}
\end{figure}

Moreover, Fig.\ \ref{fig:TMA_um_mani} portrays the last result related to the transient analysis for the UM. We present, in a similar fashion to what was shown for the BM, the comparison between the uncovered structural details and the relevant invariant manifolds over an amplified region of the phase space. It was set $x \in [2.3, 3.9]$ and $y \in [0,14,0.36]$ to make the highlighted structures more evident and, differently from the analysis for the BM, we trace only the unstable manifolds associated with four distinct chains of islands present in the phase space of the UM. Tracing the correspondingly stable manifolds altogether would impair the visualisation, undermining the proposed visual comparison. 

Figure \ref{fig:TMA_um_mani} stress the straightforward link between the average behaviour of all escaping orbits and the underlining invariant manifolds associated with the UM dynamics. Initially, for both values of $\delta B$, the upper region is heavily occupied, as expected for areas closer to the ICs. However, due to the different phase space configurations, the path experienced by all escaping orbits differs in each case; For $\delta B^-$ regions depicted by the colour gradient are well-defined and separated, while for $\delta B^+$ we easily observe a stronger mixing of colours, especially for green and yellow ($\langle \nu_i \rangle \approx 10^7 ~ \text{to} ~ 10^6$). This difference is also revealed by the shapes of all traced unstable manifolds that, for $\delta B^-$, are significantly more restrained in comparison to $\delta B^+$. The strong mixing of the erratic unstable manifolds outlined in the last panel of Fig.\ \ref{fig:TMA_um_mani} might be a fitting explanation of why $\delta B^+$ is a value for the perturbation strength that enhances the escape of this system. 
    
As a final general comment, it is worth paying close attention to all highlighted areas surrounding islands in both systems. The first panel of Fig. \ref{fig:TMA_bm_mani} provides an explicit example for the BM, while the first panel of Fig. \ref{fig:TMA_um_mani} highlights the closer neighbourhood of the upper chain of islands in the UM with a red colour gradient. This observation is closely related to the stickiness phenomena, which is the main issue investigated through the recurrence analysis discussed in the next chapter.

\chapter{Recurrence and Stickiness}
\label{chap5:recurrence}

In this chapter, we describe the second, and final, analysis developed to help our comprehension of the behaviour of the magnetic field lines through the modelling of symplectic maps. The \emph{recurrence and stickiness} analysis consists of a finite-time investigation of several different chaotic trajectories from the dynamics of our models, providing useful prior knowledge of their dynamical behaviour. 

We show that, by defining an ensemble of ICs, evolving them until a given maximum iteration time and computing the \emph{recurrence rate} of each orbit, it is possible to find particular trajectories that widely differ from the average behaviour. On this basis, it is possible to verify that orbits with high recurrence rates are the ones that experience \emph{stickiness}, being dynamically trapped in a strange quasi-periodic motion around specific regions of the phase space, phenomena that strongly affect transport and statistical properties of chaotic trajectories \cite{Altmann2006}. This procedure is proposed as a general method to study the influence of recurrent chaotic orbits of any given phase space.

In that sense, it is worth remarking that this is an additional original analysis suitable for general Hamiltonian systems; Contribution presented at {\it Finite-time recurrence analysis of chaotic trajectories in Hamiltonian systems} by Matheus S. Palmero, Iberê L. Caldas, Igor M. Sokolov, Chaos {\bf 32}, 113144, published in November 2022 \cite{Palmero2022}.

\section{Recurrence analysis}
\label{sec:chap5_rec}

\begin{figure}[b!]
\centering
\includegraphics[scale=0.75]{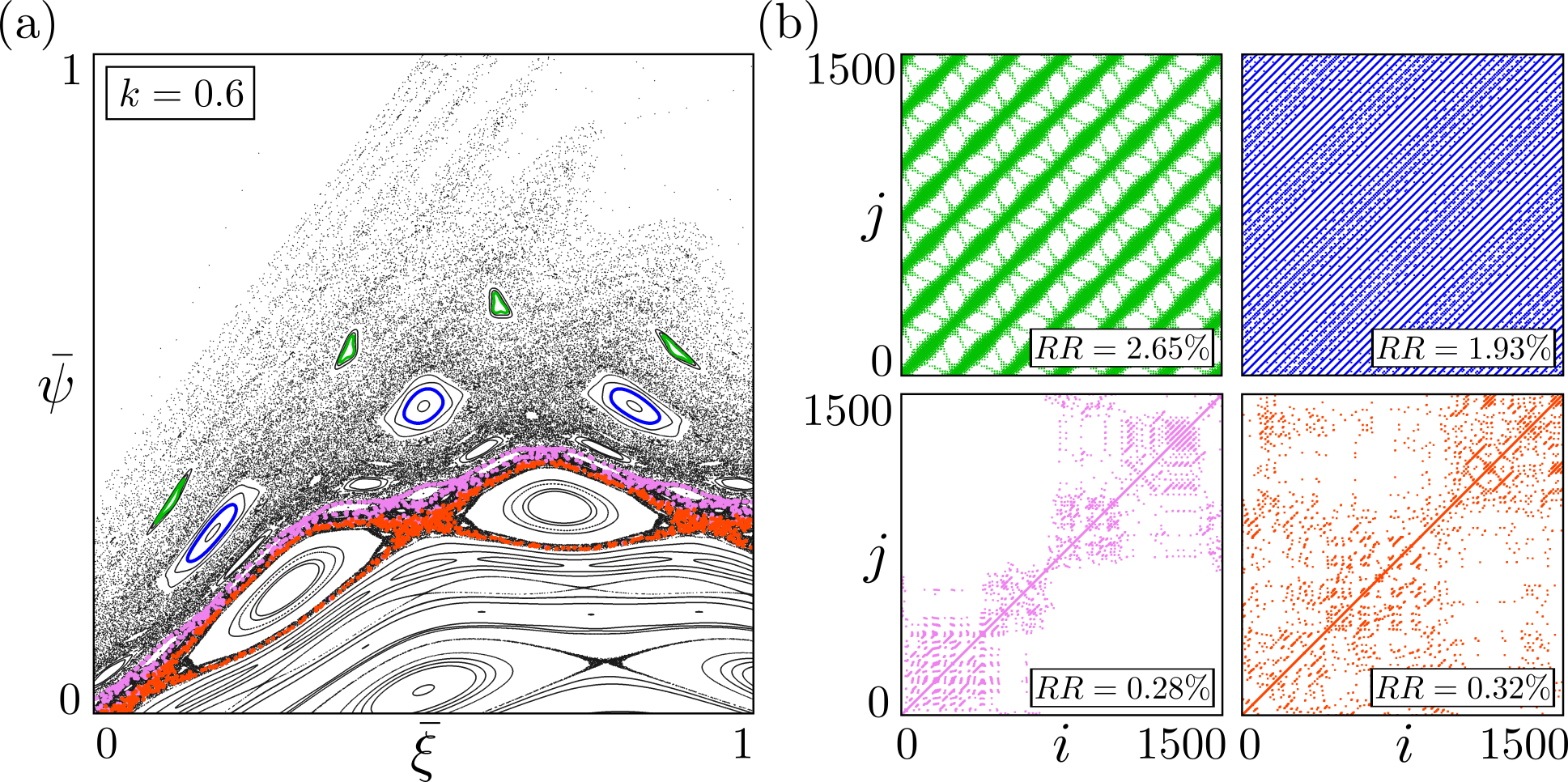}
\caption[Normalised phase pace of the BM and RPs]{(a) Normalised phase space of the BM with $k=0.6$, along with four different trajectories in colours; (b) RPs with threshold distance $\varepsilon = 0.01$ of the selected four trajectories evolved until $N = 1500$ iterations.}
\label{fig:REC_bm_ps_rps}
\end{figure}

Before presenting our numerical observations from this method, it is important to review a few definitions regarding recurrences. It is common to define recurrence if at time $t_j$, a given trajectory $\boldsymbol{x}(t_j)\approx\boldsymbol{x}(t_i)$, with $(t_i < t_j)$ i. e. returns into the dynamical neighbourhood of a previous state at $t_i$. Considering a threshold distance $\varepsilon$, it is possible to write a binary Recurrence Matrix (RM), composed by the elements $R_{i,j}$ defined as

\begin{equation}
    R_{i,j}(\varepsilon)=\left\{\begin{array}{ll}
    1,~\text{if} ~\Vert \boldsymbol{x}(t_i)-\boldsymbol{x}(t_j) \Vert < \varepsilon\\
    0,~\text{otherwise}
    \end{array}
    \right.,
    \label{eq:RM}
\end{equation}where $\Vert\cdot\Vert$ is a suitable norm. Every entry of $1$ in the RM represents a recurrence of the analysed trajectory, meaning that $\boldsymbol{x}(t_i)$ and $\boldsymbol{x}(t_j)$ are dynamically $\varepsilon-$near to each other. Since we focus on a finite-time recurrence analysis, it is worth remarking that $0 < t_i < t_j \leq N$, where $N$ is the trajectory's maximum iteration time. In that regard, RM is size $(N \times N)$.  

The visual representation of the RM is known as the Recurrence Plot (RP). Usually, the RP depicts every null entry of the RM by a white pixel and the $1$s entries by coloured pixels. The RPs can be very different, displaying particular recurrence patterns based on the evolved trajectory, which is inherently determined by its IC in deterministic systems. In addition, one RP may portray different dynamical behaviours for a given trajectory while evolving up to $N$. For instance, a chaotic orbit initiated near periodic regions can produce similar recurrence patterns to a quasi-periodic motion. However, after a sufficient iterated time, this same trajectory can escape towards other chaotic regions of the phase space, producing patterns related to pure chaotic motion. In this case, the RP would show us different recurrence patterns for distinct time windows until the dynamical evolution ends at the maximum iteration time.

Once the RP of a trajectory of interest is computed, many different measures can quantify and differentiate several aspects between different RPs. These are called Recurrence Quantification Analysis (RQA) \cite{Zbilut1992, Marwan2002}. The simplest one is the Recurrence Rate ($RR$), which provides the percentage of recurrence points as follows

\begin{equation}
    RR = \frac{1}{N^2} \sum_{i,j=1}^{N} R_{ij}~,
    \label{eq:RR}
\end{equation}note that $RR = RR(N)$, i. e. the recurrence rate depends on the maximum iteration time considered for the evolution of the given trajectory.

\begin{figure}[h]
\centering
\includegraphics[scale=0.75]{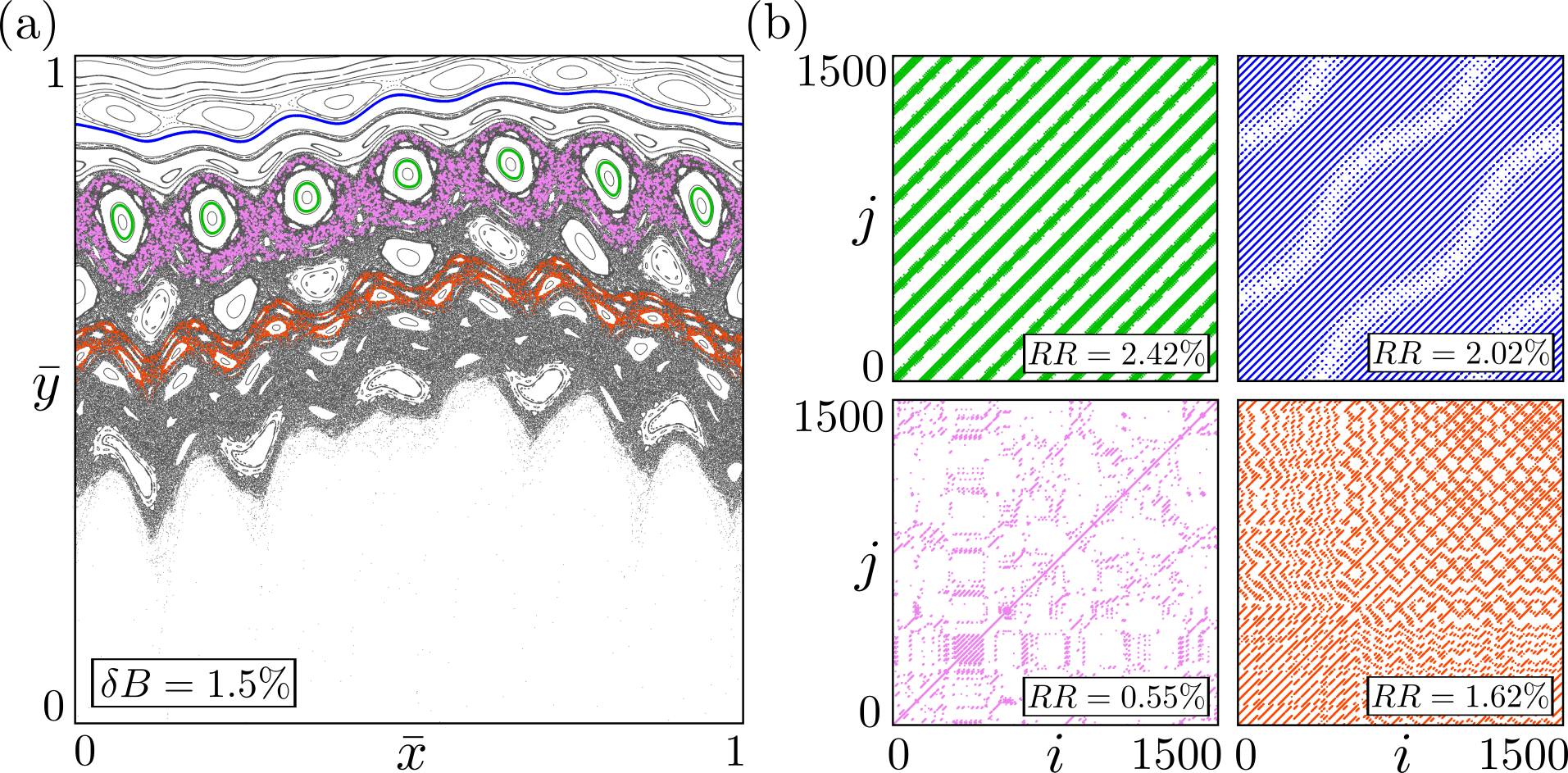}
\caption[Normalised phase pace of the UM and RPs]{(a) Normalised phase space of the UM with $\delta B = 1.5\%$, along with four different trajectories in colours. (b) RPs with threshold distance $\varepsilon = 0.05$ of the selected four trajectories evolved until $N = 1500$ iterations.}
\label{fig:REC_um_ps_rps}
\end{figure}

As initial general examples of RPs, and their respective computed values of $RR$, Figs.\ \ref{fig:REC_bm_ps_rps} and \ref{fig:REC_um_ps_rps} display, on the left panel, the characteristic normalised phase spaces for both BM and UM, along with four different trajectories and, on the right panels, their respective RPs. The normalised phase spaces $(\bar{\xi}, \bar{\psi})$ or $(\bar{x}, \bar{y})$ were constructed considering only the regions of interest for both models: $\xi \in [-0.002, 0.002]$, $\psi \in [0.996, 1.0]$ and $x \in [0,1]$, $y \in [0.0, 0.4]$. In that regard, recurrences are computed accordingly to the threshold distance $\varepsilon$ only if a trajectory is within these regions.

All four right panels in both Figs.\ \ref{fig:REC_bm_ps_rps} and \ref{fig:REC_um_ps_rps} depict distinct dynamical behaviours found in their phase spaces. Particularly for the BM, both upper panels of Fig.\ \ref{fig:REC_bm_ps_rps} (b) show characteristic RPs of pure periodic motion, where orbits depicted in green and blue were placed inside KAM islands embedded within the chaotic separatrix. Orbits in violet and red, however, were selected in chaotic regions between the periodic islands, displaying chaotic RPs marked by different recurrence patterns throughout their evolution. 

Analogously, RPs displayed in Fig.\ \ref{fig:REC_um_ps_rps} (b) show periodic and chaotic behaviours for the UM. Initially in green, we see pure periodic motion from an orbit inside of a period 7 KAM islands. In blue, however, we observe a trajectory placed over an invariant spanning curve, where a different periodic motion is depicted by long diagonal lines wrapped in an oscillatory pattern. In violet, we see distinct patterns produced by a chaotic orbit between the period 7 chain of islands. Furthermore, the last panel of Fig.\ \ref{fig:REC_um_ps_rps} exhibits complex patterns associated with a quasi-periodic behaviour of a chaotic orbit, depicted in red, in a vicinity between many small islands.

Although visually pleasing, recurrence patterns of pure periodic trajectories do not improve our understanding of stickiness, being a phenomenon experienced only by chaotic orbits. In that sense, the upcoming analysis is focused solely on chaotic trajectories and, due to the fundamental nature of the stickiness, as a rare trapped quasi-periodic motion eventually experienced by chaotic orbits, the $RR$ is a suitable measure \cite{Zou2007} that can distinguish purely chaotic dynamics from orbits that, indeed, experience stickiness in the considered time-frame.

\section{Numerical observations}
The analysis proposed here is heavily based on numerical simulations. Essentially, the numerical method to detect stickiness in general planar non-linear symplectic maps via recurrences requires the following steps: 

\begin{enumerate}
    \item[1st.] A well-defined mixed phase space in which is possible to determine coordinates of the chaotic sea or at least a portion of the chaotic region;
    
    \item[2nd.] Knowing the coordinates of ICs $\boldsymbol{x}_0$ that provide chaotic trajectories, is possible to define an ensemble E of $M$ ICs around some particular chaotic region of the phase space;
    
    \item[3rd.] Evolving all $M$ trajectories until a given $N$ maximum number of iterations of the dynamical equations, it is possible to compute the respective $RR(\boldsymbol{x}_0)$ for all trajectories, considering also a given value for the recurrence threshold distance $\varepsilon$;
    
    \item[4th.] Once computed the $RR$ for each trajectory in E, it can be plotted as a function of $\boldsymbol{x}_0$. This forms a distribution of $RR$ over all ICs and can be analysed by changing the ICs' positions in both $x$ and $y$ axis;
    
    \item[5th.] Identifying peaks\footnote{For that, one can calculate the average $\langle RR \rangle$ over the ensemble E along with its standard deviation $\sigma$ and perform a 3-$\sigma$ detection; If a trajectory from E presents a $RR(\boldsymbol{x}_0) > \langle RR \rangle + 3\sigma$, then $\boldsymbol{x}_0=\boldsymbol{x}_0^S$ i.\ e.\ an IC that, when evolved, will produce a highly recurrent trajectory.} at the aforementioned distributions is equivalent to pinpoint the ICs' coordinates $\boldsymbol{x}_0^S$ of highly recurrent trajectories, good candidates for the ones that experience stickiness;   
    
    \item[6th.] Finally, all special $\boldsymbol{x}_0^S$ can be evolved once again and plotted along with the corresponding phase space. Since they are \emph{sticky} trajectories, regions around KAM islands will be highlighted, indicating that, indeed, they experienced stickiness in their finite-time evolution.   
\end{enumerate}

In the upcoming subsections, we employ the method considering both the BM and UM. Consistently with previous results, the perturbations' strengths for the two models were selected by the escape analysis performed in Chapter \ref{chap3:models}, Sec.\ \ref{sec:chap3_escape}. 

\subsection*{Single-null divertor map}

\begin{figure}[h]
\centering
\includegraphics[scale=0.70]{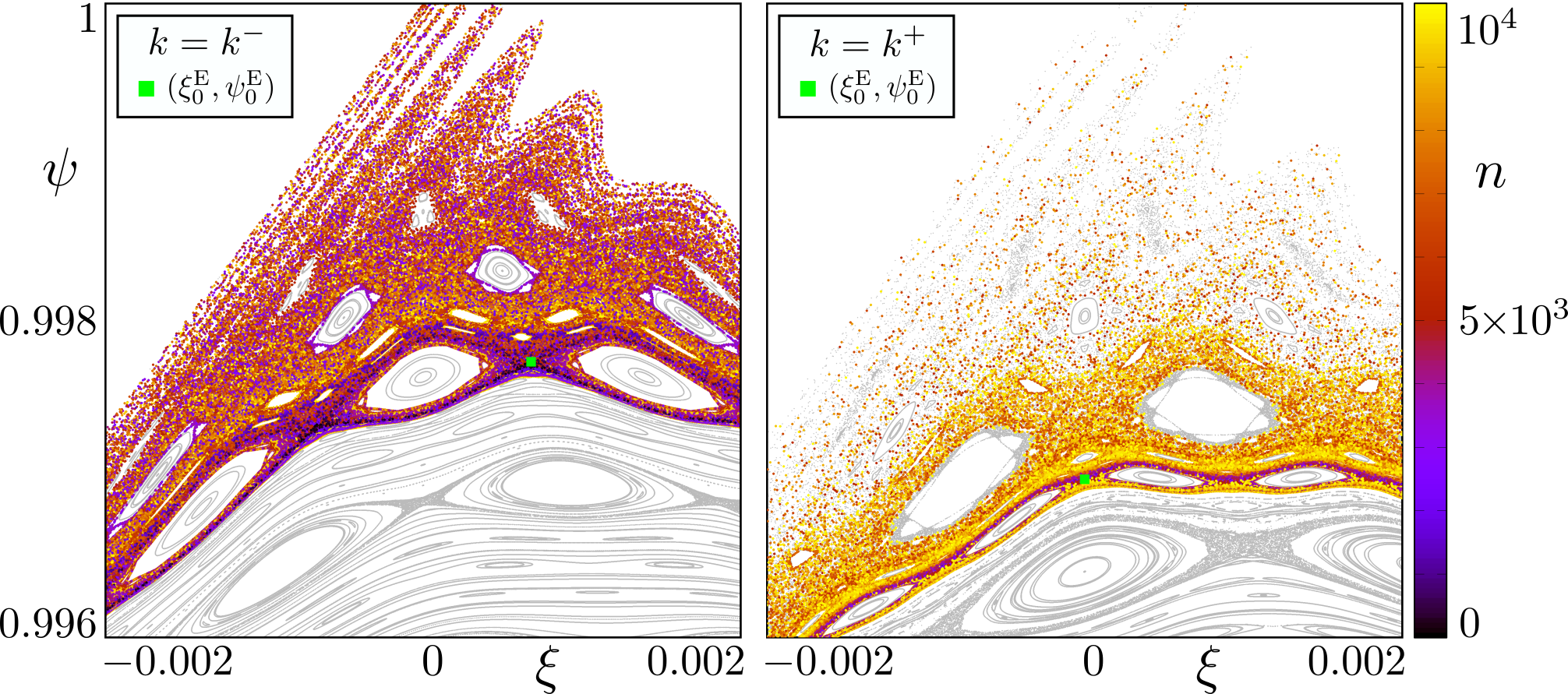}
\caption[Ensemble's evolution on the phase space of the BM]{Phase space of the BM (grey on the background), along with $500$ trajectories from the ensemble E depicted by the green square (out of scale). The colour axis shows the iteration time of the evolved trajectories up to $10^4$ iterations.}
\label{fig:REC_bm_ps_ensembles}
\end{figure}

First, for the BM we selected an ensemble E formed by $M = 4.9 \times 10^7$ ICs uniformly distributed\footnote{A technical observation regarding the total number of ICs is in order: It is actually $M = m_x \times m_y$, with $m_x = m_y = 7 \times 10^3$ ICs; In that sense, the actual coordinates of the ICs are calculated over the intervals $\xi_0 \in [\xi_0^{\text{min}},\xi_0^{\text{max}}]/m_x$ and $\psi_0 \in [\psi_0^{\text{min}},\psi_0^{\text{max}}]/m_y$.} between $\xi_0^{\text{E}} \in [\xi^{u} - 10^{-10}, \xi^{u} + 10^{-10}]$ and $\psi_0^{\text{E}} \in [\psi^{u} - 10^{-10}, \psi^{u} + 10^{-10}]$, where $(\xi^u,\psi^u) = (0.00066855, 0.99774007)$ are the coordinates of the first UPO found within the chaotic portion of the phase space for $k = k^-$. The first panel of Fig.\ \ref{fig:REC_bm_ps_ensembles} depicts the ensemble position by the green square in the phase space, along with some trajectories from E evolved up to $N = 10^4$ iterations. The colour axis shows the iteration time of these trajectories.

Considering the phase space $k = k^+$, an analogous ensemble was placed again in the closest vicinity of the first UPO found within the chaotic separatrix; $\xi_0^{\text{E}} \in [\xi^{u} - 10^{-10}, \xi^{u} + 10^{-10}$ and $\psi_0^{\text{E}} \in [\psi^{u} - 10^{-10}, \psi^{u} + 10^{-10}]$, with $(\xi^u,\psi^u) = (0.00000001, 0.99698287)$. It is possible to note on the second panel of Fig.\ \ref{fig:REC_bm_ps_ensembles} that the ensemble is located in a lower region when compared to the phase space of $k^-$. This is expected since the chaotic region of the separatrix expands while increasing the perturbation strength $k$, making other lower regions accessible by chaotic trajectories.

It is worth mentioning that, in practice, comparing the behaviour of trajectories from ensembles of ICs for different phase space configurations is a meticulous task. In that sense, both ensembles defined above were constructed considering the same number of ICs $M$, placed within small squares with the same width and height, centred around the computed coordinates (up to $10^{-8}$ numerical precision) of the first UPO found in their respective phase spaces. For $k = k^-$, it was found a period 29 UPO and, for $k = k^+$ a period 55 UPO. The UPOs' close vicinity is a convenient location because it warrants only chaotic orbits.

\begin{figure}[h]
\centering
\includegraphics[scale=0.75]{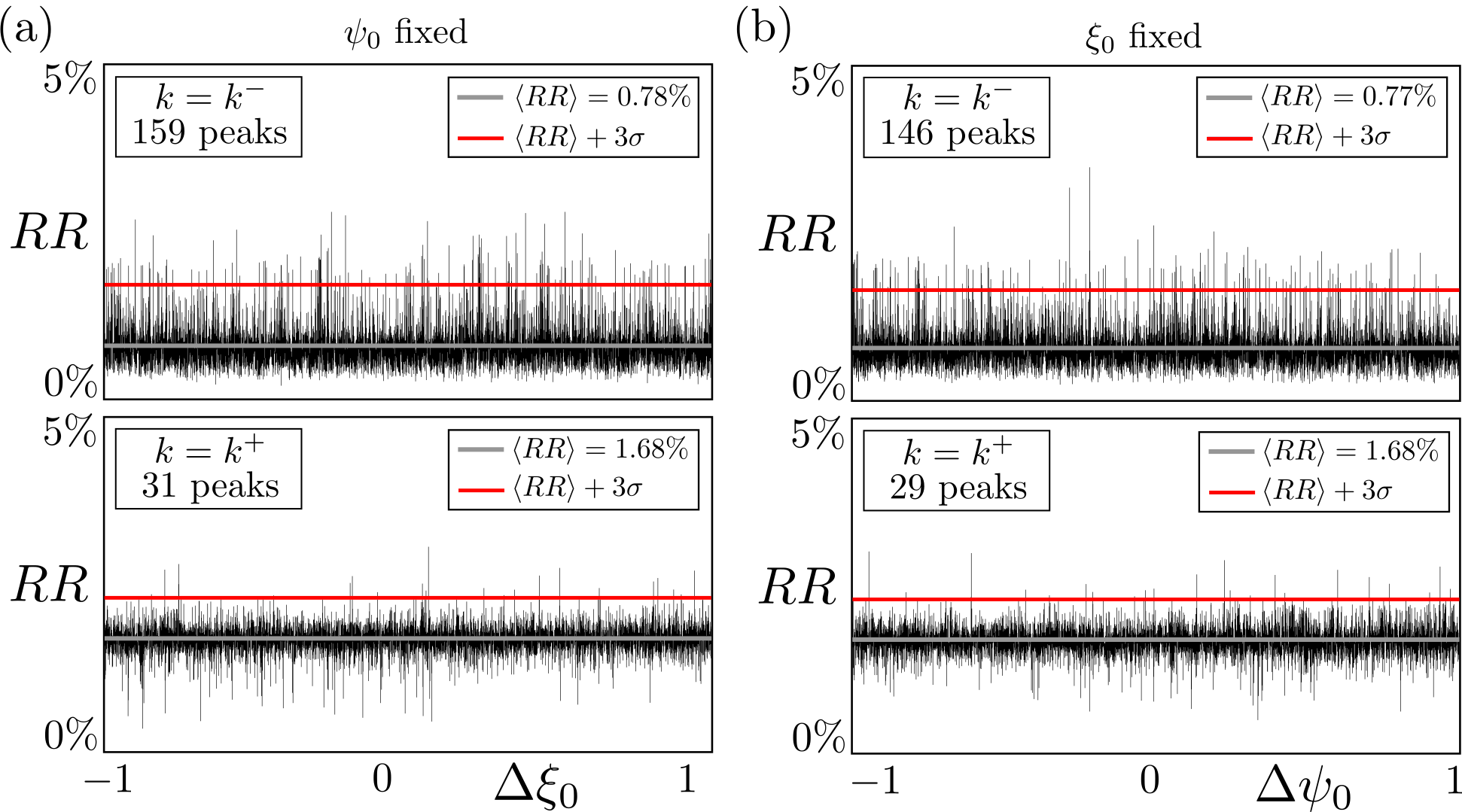}
\caption[Recurrence rate distributions for the BM]{$RR$ distributions as a function of the ICs for the BM. In (a) $\psi_0 = \psi^u$ is fixed, varying $\xi_0$ between the defined interval for both $k-$ (upper panel) and $k^+$ (lower panel). In (b) $\xi_0 = \xi^u$ is fixed, varying $\psi_0$ in the defined interval for $k-$ and $k^+$. Their corresponding values of $\langle RR \rangle$, the upper limit $\langle RR \rangle + 3\sigma$ depicted by the red line and the count of peaks are displayed in all panels.}
\label{fig:REC_bm_rr_ics}
\end{figure}

Once the ensembles are well-defined, it is possible to finally begin the recurrence analysis considering all $M = 4.9 \times 10^7$ ICs, evolved until $N = 10^4$ iterations and the recurrence threshold distance $\varepsilon = 0.01$. Figure \ref{fig:REC_bm_rr_ics} presents the first results of the computed $RR$ for all trajectories in the ensembles.

Since the defined ICs were placed on intervals in both axis, Fig.\ \ref{fig:REC_bm_rr_ics}(a) and (b) separates the analysis fixing $\psi_0$ and $\xi_0$ respectively and varying the other coordinate within the ensemble's size. We define relative distances $\Delta\xi_0$ and $\Delta \psi_0$, taking as a reference the coordinates of their corresponding UPOs $(\xi^u,\psi^u)$, as $\Delta (\xi_0,\psi_0) = [(\xi_0^\text{E}, \psi_0^\text{E}) - (\xi^u,\psi^u)] \times 10^{10}$. Then, the upper and lower panels of Fig.\ \ref{fig:REC_bm_rr_ics} (a) show the recurrence rate distribution over the interval of ICs on $\xi$ for $k^-$ and $k^+$ respectively. Analogously the panels of Fig.\ \ref{fig:REC_bm_rr_ics} (b) display the same distributions now over the interval of ICs on $\psi$.

By defining an upper limit, calculated by $\langle RR \rangle + 3\sigma$ i.\ e.\ the average recurrence rate over the ensemble of ICs plus three times the standard deviation of the average, it is possible to figure out via a 3$\sigma$ detection how many peaks are in the distributions. A peak is associated with a specific value of IC that provides a highly recurrent trajectory.

The foremost important outcome of Fig.\ \ref{fig:REC_bm_rr_ics} is the difference between the number of peaks for $k^-$ and $k^+$. It was established that $k^-$ is the value of the perturbation strength that gives a phase space configuration which restrains escape orbits. In that regard, it is reasonable to expect that this phase space configuration would present many trajectories that experience stickiness and, due to this trapping time, they might not escape the system until the considered evolution time. The opposite is to be expected for $k^+$, being associated with a phase space configuration that enhances escaping trajectories. All four distributions shown in Fig.\ \ref{fig:REC_bm_rr_ics} and their respective peak counts notably agree with these expectations. 

\begin{figure}[h]
\centering
\includegraphics[scale=0.70]{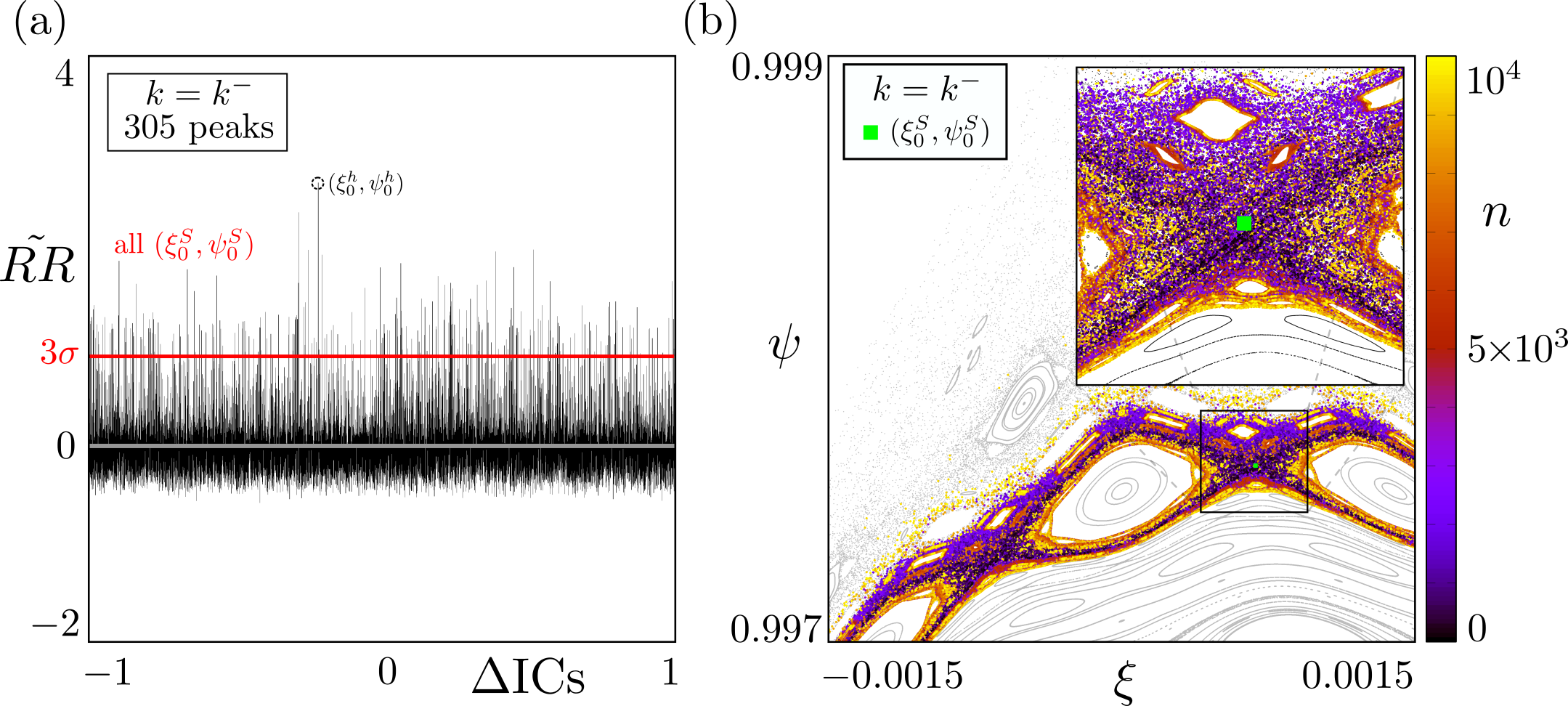}
\caption[Combined distribution and stickiness in the BM for $k=k^-$]{(a) $\tilde{RR}$ distribution as a function of all $M=4.9\times10^7$ ICs evolved until $N=10^4$ iterations of the BM, considering $k=k-$ and the threshold distance $\varepsilon = 0.01$. The IC $(\xi_0^h,\psi_0^h)$ with the highest $RR$ is marked by the dashed circle; (b) 305 trajectories evolved from all $(\xi_0^S,\psi_0^S)$ determined from the 3$\sigma$ limit in (a). The position of the ensemble is depicted by the green square (out of scale) and the inset shows the amplified region inside the black square.} 
\label{fig:REC_bm_rr_ps_k-}
\end{figure}

Furthermore, Fig.\ \ref{fig:REC_bm_rr_ps_k-} (a) combine prior recurrence analyses for $k^-$ on both intervals into one considering $\tilde{RR}$ as a function of $\Delta IC$, where we define $\tilde{RR} = RR - \langle RR \rangle$ as the corrected $RR$ in respect to its ensemble average, and $\Delta IC = \Delta\xi_0 + \Delta \psi_0$ as the combined intervals for the ICs. The upper limit is depicted by the red line at $3\sigma$ since the $\tilde{RR}$ distribution is null at $\langle RR \rangle$ by construction. All ICs that, when evolved up to $N$ iterations, produce trajectories with computed $RR$ larger than the upper limit are suitable candidates to present stickiness, so we label these ICs as $(\xi_0^S,\psi_0^S)$.  

\begin{figure}[h]
\centering
\includegraphics[scale=0.70]{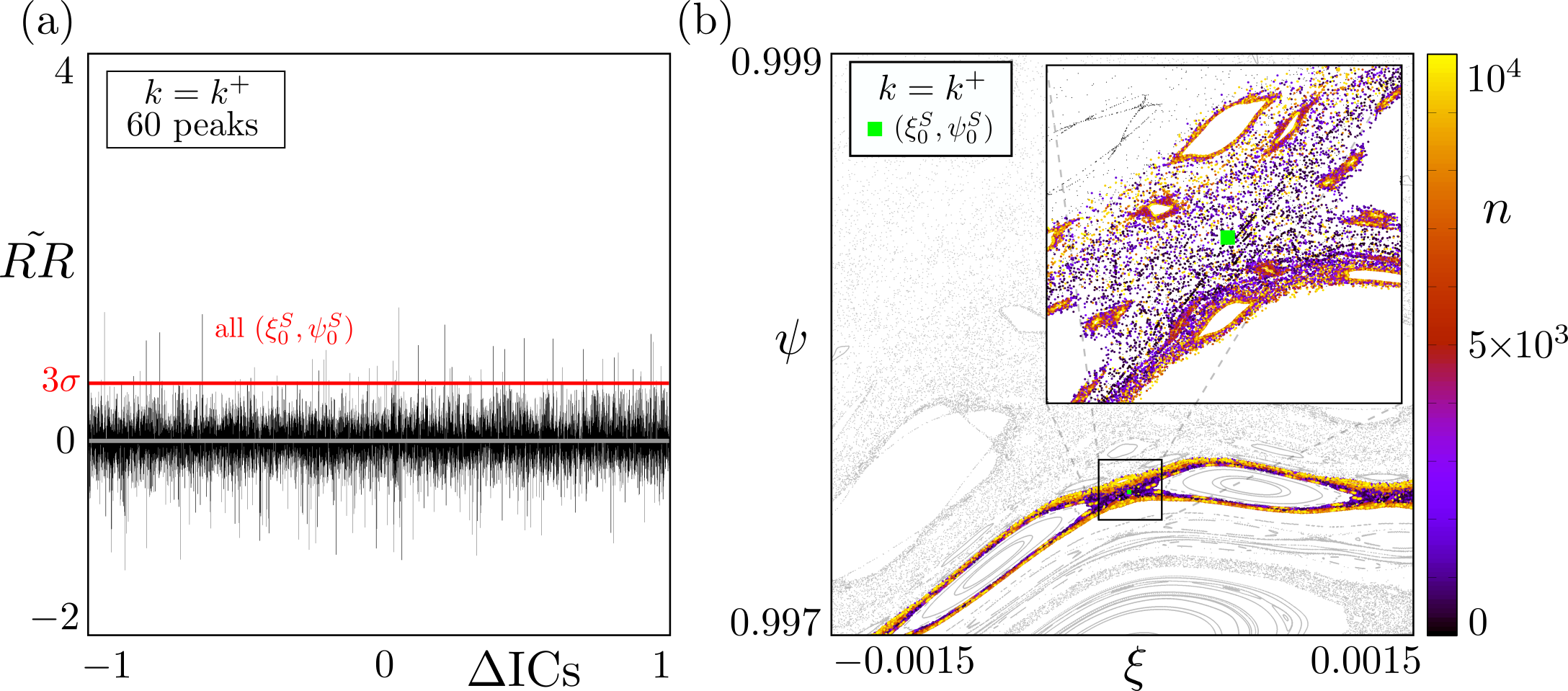}
\caption[Combined distribution and stickiness in the BM for $k=k^+$]{(a) $\tilde{RR}$ distribution as a function of all $M=4.9\times10^7$ ICs evolved until $N=10^4$ iterations of the BM, considering $k=k+$ and the threshold distance $\varepsilon = 0.01$. (b) 60 trajectories evolved from all $(\xi_0^S,\psi_0^S)$ determined in (a). The position of the ensemble is depicted by the green square (out of scale) and the inset shows the amplified region inside the black square.} 
\label{fig:REC_bm_rr_ps_k+}
\end{figure}

\begin{figure}[b!]
\centering
\includegraphics[scale=0.70]{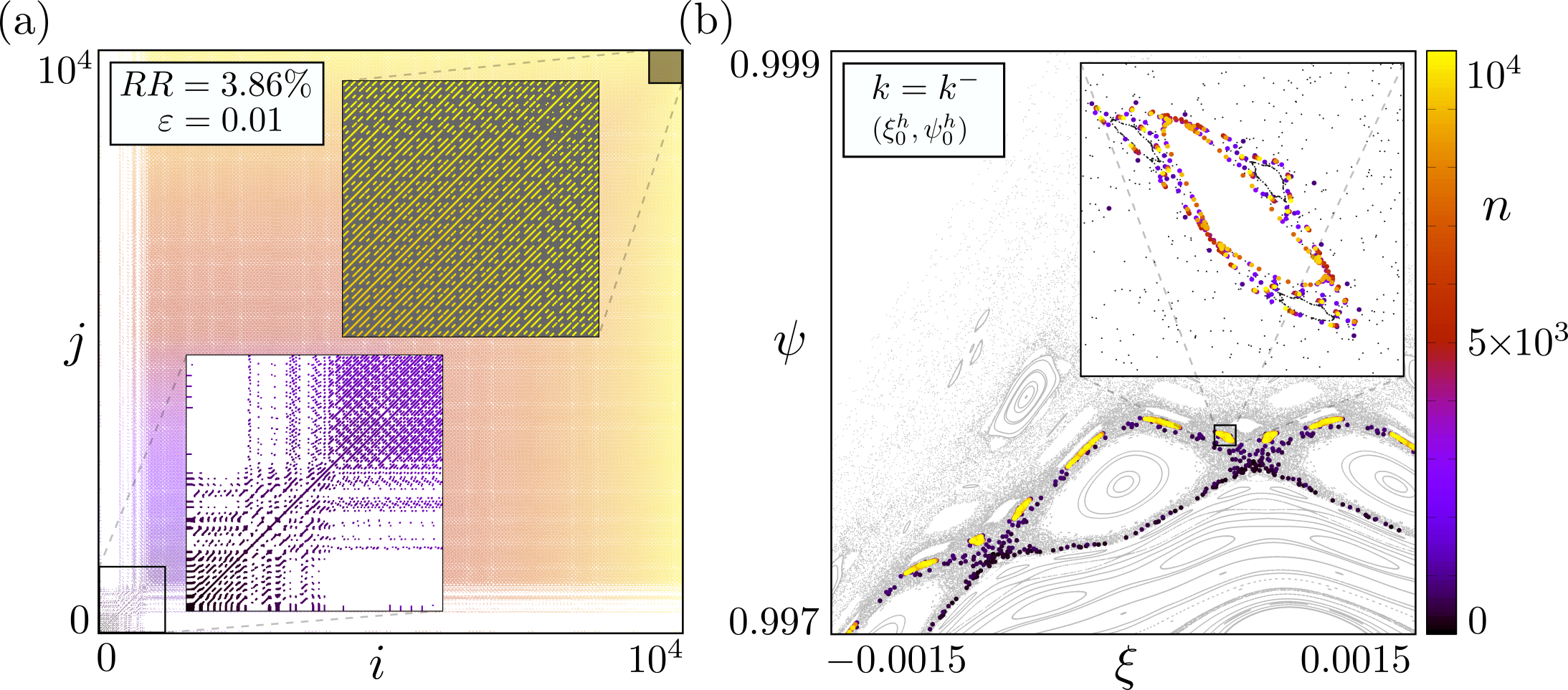}
\caption[RP and the evolution of the most recurrent orbit in the BM]{(a) RP with $\varepsilon = 0.01$ for the trajectory started from $(\xi_0^h,\psi_0^h)$. The insets show amplifications of their respective regions in the black squares. Both axes are set in the same colour gradient of the trajectory's evolution; (b) Phase space for $k = k^-$ (grey on the background) along with the same orbit. The inset shows the region around the small periodic island.} 
\label{fig:REC_bm_highest_rr}
\end{figure}

Essentially, the analysis of the $RR$ distribution over the ICs provides a subset $S$ of the ensemble E, composed solely of ICs that, when evolved, will produce highly recurrent trajectories. For the case of $k = k^-$ the subset $S$ is formed by 305 ICs and their evolution is shown in Fig.\ \ref{fig:REC_bm_rr_ps_k-} (b). We readily note that their evolution is rather different from what was observed in the first panel of Fig.\ \ref{fig:REC_bm_ps_ensembles}. These 305 trajectories do not visit the upper regions of the chaotic separatrix, being dynamically trapped in the chaotic area between the period 29 chain of islands and yet, finishing their evolution by visiting the neighbourhood of all embedded periodic islands. By this numerical observation, these are indeed trajectories that experience stickiness phenomena.  

In the same direction, Fig.\ \ref{fig:REC_bm_rr_ps_k+} (a) presents the results of the distribution $\tilde{RR} \times \Delta IC$ considering $k = k^+$, where the subset $S$ is roughly three times smaller than the previous analysis, composed of 60 ICs. Nevertheless, the evolution of these special trajectories is shown in Fig.\ \ref{fig:REC_bm_rr_ps_k+} (b).  

The evolution of the subset $S$ displayed in Fig.\ \ref{fig:REC_bm_rr_ps_k+} (b) also reflects that different from what was observed in the second panel of Fig.\ \ref{fig:REC_bm_ps_ensembles}, the trajectories stay in the initial chaotic surroundings, never visiting upper regions of the phase space. The inset reveals that these 60 orbits are trapped around many small periodic structures, indicating stickiness once more.

The final numerical observation for the BM is shown in Fig.\ \ref{fig:REC_bm_highest_rr}, where we evolve the IC $(\xi_0^h,\psi_0^h)$ with the highest $RR$ from the distribution in Fig.\ \ref{fig:REC_bm_rr_ps_k-} (a) and we display its RP considering the same $N = 10^4$ and $\varepsilon = 0.01$ established throughout the analysis. The strong quasi-periodic is evident in the RP displayed in Fig.\ \ref{fig:REC_bm_highest_rr} (a). The first inset shows the initial one-thousand iterations, where a chaotic transient-like evolution can be clearly observed before the strong stickiness region. The last inset highlights the periodic patterns within the final five-hundred iterations. While compared to Fig.\ \ref{fig:REC_bm_highest_rr} (b), we observe the exact place in which the orbit gets trapped. The inset shows more clearly this fine chaotic vicinity around the periodic island.   

\subsection*{Ergodic magnetic limiter map}
As the recurrence-based detection approach proved to be effective in identifying stickiness for the BM, we employ a similar analysis for the UM. An ensemble E formed by $M = 2.5 \times 10^7$ ICs is uniformly distributed between small intervals in both $x$ and $y$-axis; $x_0^{\text{E}} \in [x^{u} - 10^{-10}, x^{u} + 10^{-10}]$ and $y_0^{\text{E}} \in [y^{u} - 10^{-10}, y^{u} + 10^{-10}]$, where $(x^u,y^u)$ are the coordinates of the period 7 UPO found in the upper chaotic area of the phase space. Unlikely the BM, the phase space of the UM is more robust to changes in the control parameter of the map, given by the relative perturbation of the magnetic field $\delta B$, making it easier to find a suitable location for positioning E while considering the two different configurations $\delta B^-$ and $\delta B^+$.

\begin{figure}[h]
\centering
\includegraphics[scale=0.77]{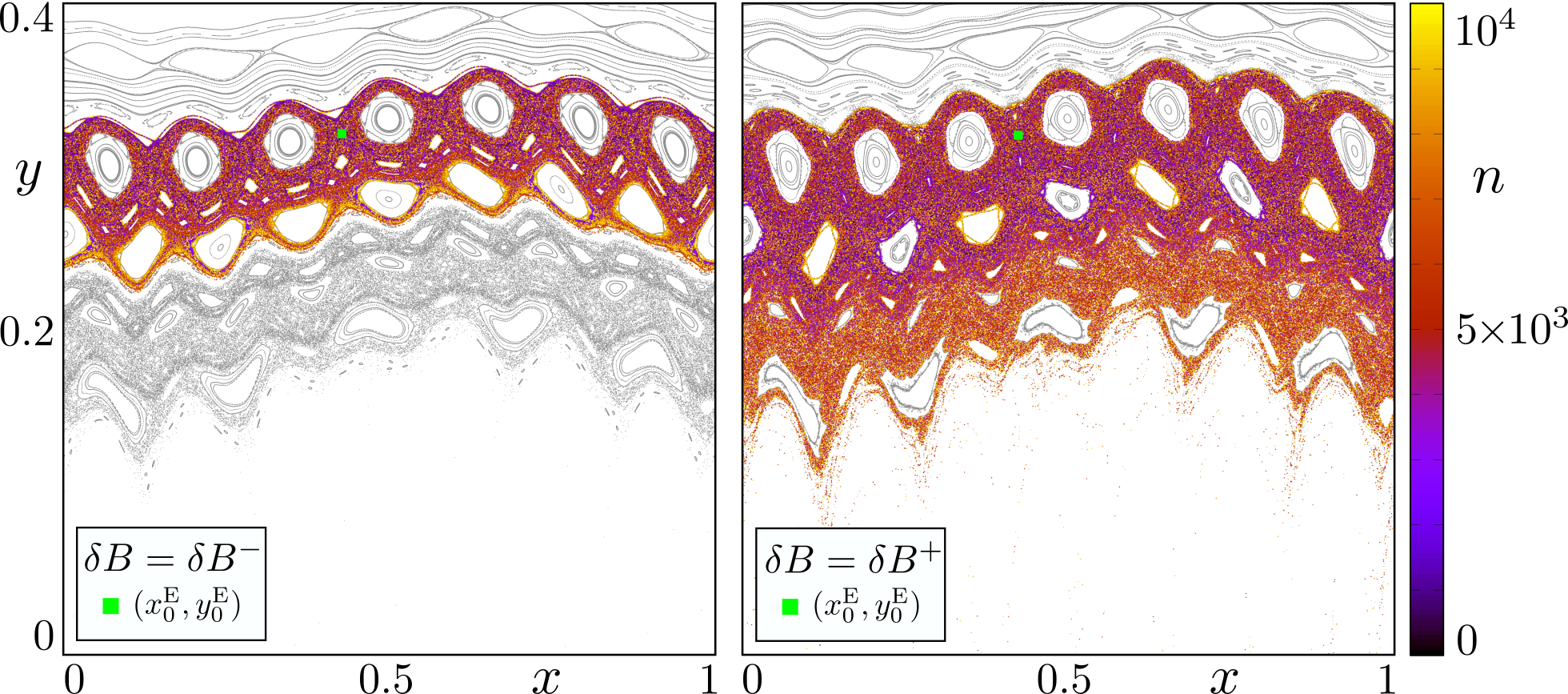}
\caption[Ensemble's evolution on the phase space of the UM]{Phase space of the UM (grey on the background), along with $250$ trajectories from the ensemble E depicted by the green square (out of scale). The colour axis shows the iteration time of the evolved trajectories up to $10^4$ iterations.}
\label{fig:REC_um_ps_ensembles}
\end{figure}

Figure\ \ref{fig:REC_um_ps_ensembles} show the position of the ensembles, depicted by the green squares, in the phase space for $\delta B^-$ in the first panel, and for $\delta B^+$ in the second; E is centred at $(x^u,y^u) = (0.420363188, 0.321734626)$ for $\delta B^-$ and at $(x^u,y^u) = (0.420594827, 0.321787496)$ for $\delta B^+$. Along with their phase spaces, we display the evolution of $250$ ICs from E up to $N = 10^4$ iterations. The colour axis depicts the iteration time.

Now that the ensembles are well-defined and Fig.\ \ref{fig:REC_um_ps_ensembles} shows their typical evolution, we analyse all $M = 2.5 \times 10^7$ ICs, evolved until $N = 10^4$ iterations, considering the recurrence threshold distance $\varepsilon = 0.005$ to compute their corresponding $RR$. We present in Fig.\ \ref{fig:REC_bm_rr_ics} the results for the $RR$ distribution over $x_0$ and $y_0$ intervals separately and the two configurations $\delta B^-$ and $\delta B^+$ of the UM. 

\begin{figure}[h]
\centering
\includegraphics[scale=0.75]{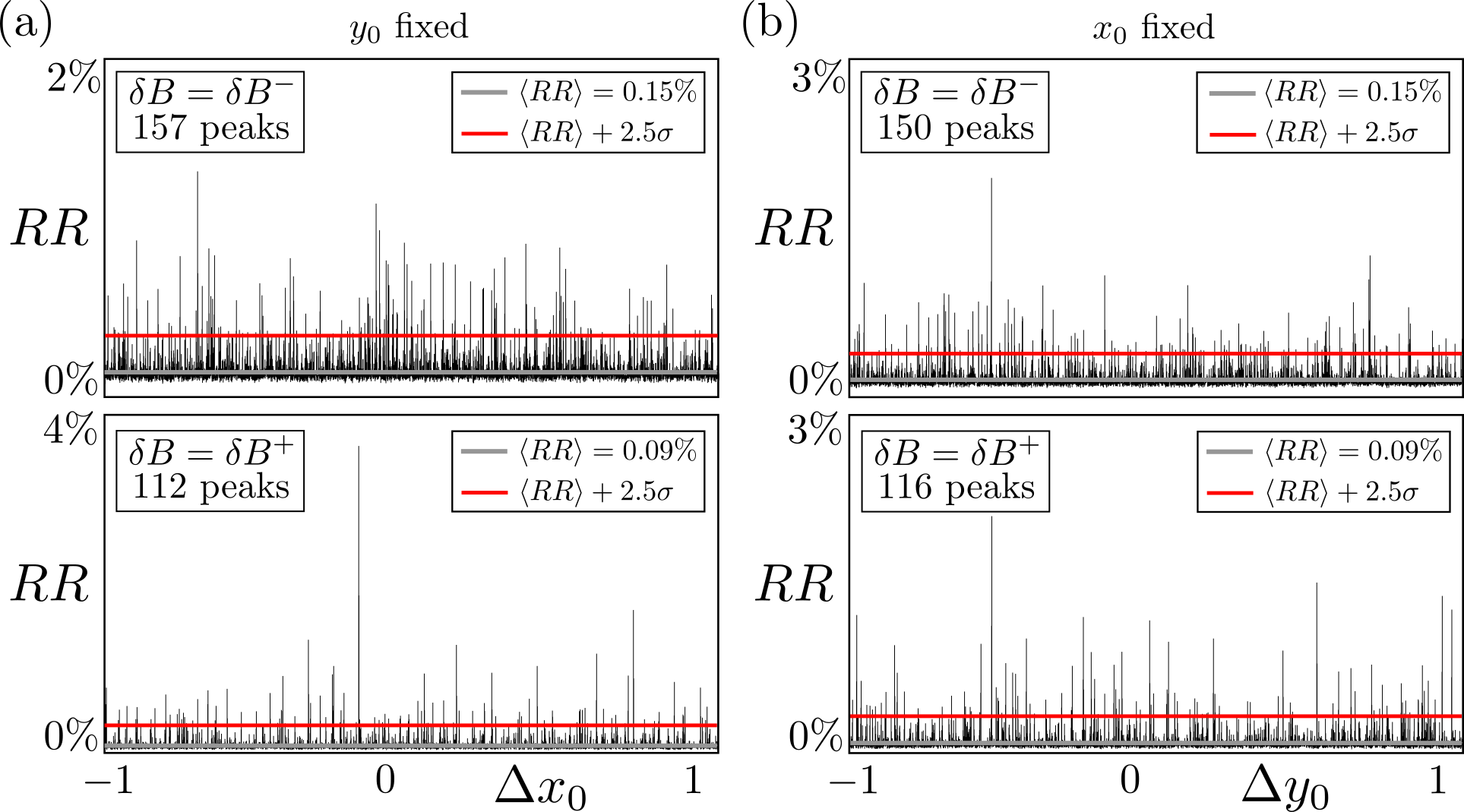}
\caption[Recurrence rate distributions for the UM]{$RR$ distributions as a function of the ICs for the UM. In (a) $y_0 = y^u$ is fixed, varying $x_0$ between the defined interval for both $\delta B^{-}$ (upper panel) and $\delta B^+$ (lower panel). In (b) $x_0 = x^u$ is fixed, varying $y_0$ in the defined interval for $\delta B^{-}$ and $\delta B^+$. Their corresponding values of $\langle RR \rangle$, the upper limit $\langle RR \rangle + 2.5\sigma$ depicted by the red line and the count of peaks are displayed in all panels.}
\label{fig:REC_um_rr_ics}
\end{figure}

In order to further investigate the adaptability of our method, we consider now a different value of the recurrence threshold distance $\varepsilon = 0.005$, set for all analyses of the UM and, we performed a 2.5$\sigma$ detection to pinpoint special initial conditions for stickiness in the UM. 

Figure \ref{fig:REC_um_rr_ics} shows that although the number of peaks for the configuration $\delta B^-$ is higher than for $\delta B^+$ in both $x$ and $y$ distributions, the difference is not as high as for $k^-$ and $k^+$ previously shown in Fig. \ref{fig:REC_bm_rr_ics} for the BM. Nevertheless, the relatively high number of peaks for the configuration $\delta B^+$ indicates that we will find strong stickiness in the corresponding phase space as well.

\begin{figure}[h]
\centering
\includegraphics[scale=0.70]{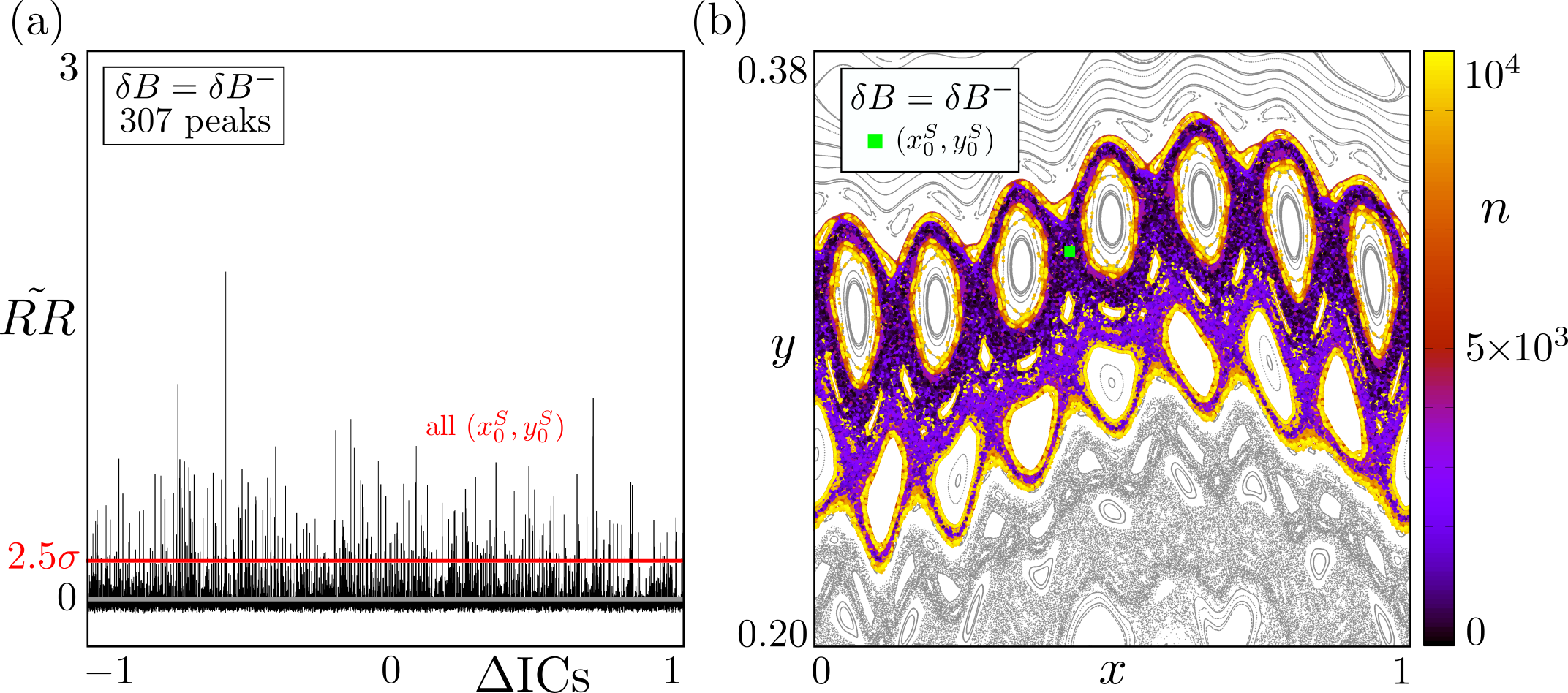}
\caption[Combined distribution and stickiness in the UM for $\delta B = \delta B^-$]{(a) $\tilde{RR}$ distribution as a function of all ICs evolved until $N=10^4$ iterations of the UM, considering $\delta B^-$ and the threshold distance $\varepsilon = 0.005$; (b) 307 trajectories evolved from all $(x_0^S,y^S)$ determined from the 2.5$\sigma$ detection in (a). The position of the ensemble is depicted by the green square (out of scale).} 
\label{fig:REC_um_rr_ps_b-}
\end{figure}

From both upper panels of \ref{fig:REC_um_rr_ics} we found 307 ICs that compose a special subset $S$ of our initial ensemble E. These special ICs are labelled $(x_0^S,y_0^S)$ in Fig.\ \ref{fig:REC_um_rr_ps_b-} (a), where we combine previous distributions on both intervals into one while considering the aforementioned corrected $\tilde{RR}$ recurrence rate. In Fig.\ \ref{fig:REC_um_rr_ps_b-} (b) we display the evolution of $(x_0^S,y_0^S)$ in the corresponding phase space configuration $\delta B^-$, where the colour gradient from the number of iterations reveals that most of these special trajectories spend an expressive amount of time around the KAM islands embedded in this upper region of the phase space. It is worth noting the differences between the typical evolution for ensemble E shown in the first panel of Fig.\ \ref{fig:REC_um_ps_ensembles} and the evolution for the subset $S$ now displayed in Fig.\ \ref{fig:REC_um_rr_ps_b-} (b).  

\begin{figure}[h]
\centering
\includegraphics[scale=0.70]{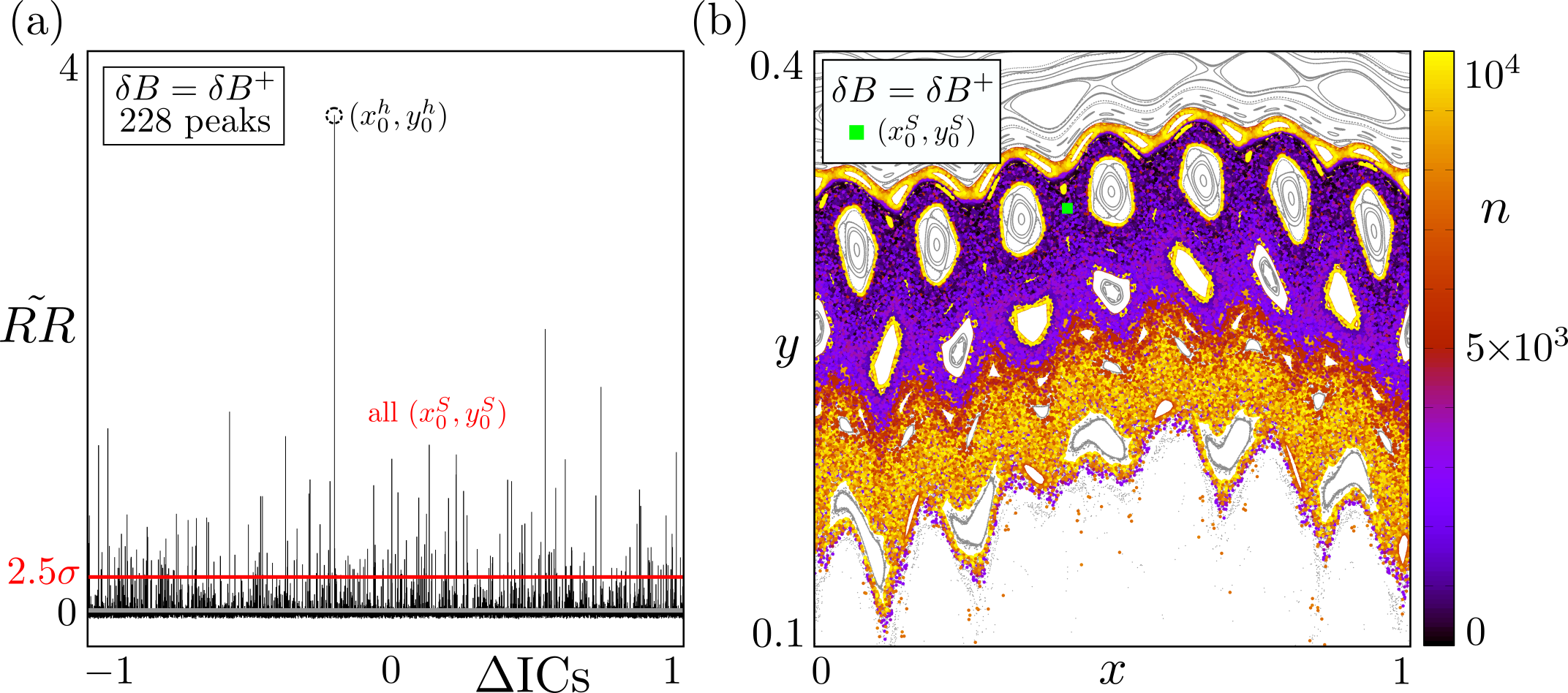}
\caption[Combined distribution and stickiness in the UM for $\delta B = \delta B^+$]{(a) $\tilde{RR}$ distribution as a function of all ICs evolved until $N=10^4$ iterations of the UM, considering $\delta B^+$ and the threshold distance $\varepsilon = 0.005$. The IC $(x_0^h,y_0^h)$ with the highest $RR$ is marked by the dashed circle; (b) 228 trajectories evolved from all $(x_0^S,y^S)$ from the 2.5$\sigma$ detection in (a). The position of the ensemble is depicted by the green square (out of scale).} 
\label{fig:REC_um_rr_ps_b+}
\end{figure}

\begin{figure}[h!]
\centering
\includegraphics[scale=0.70]{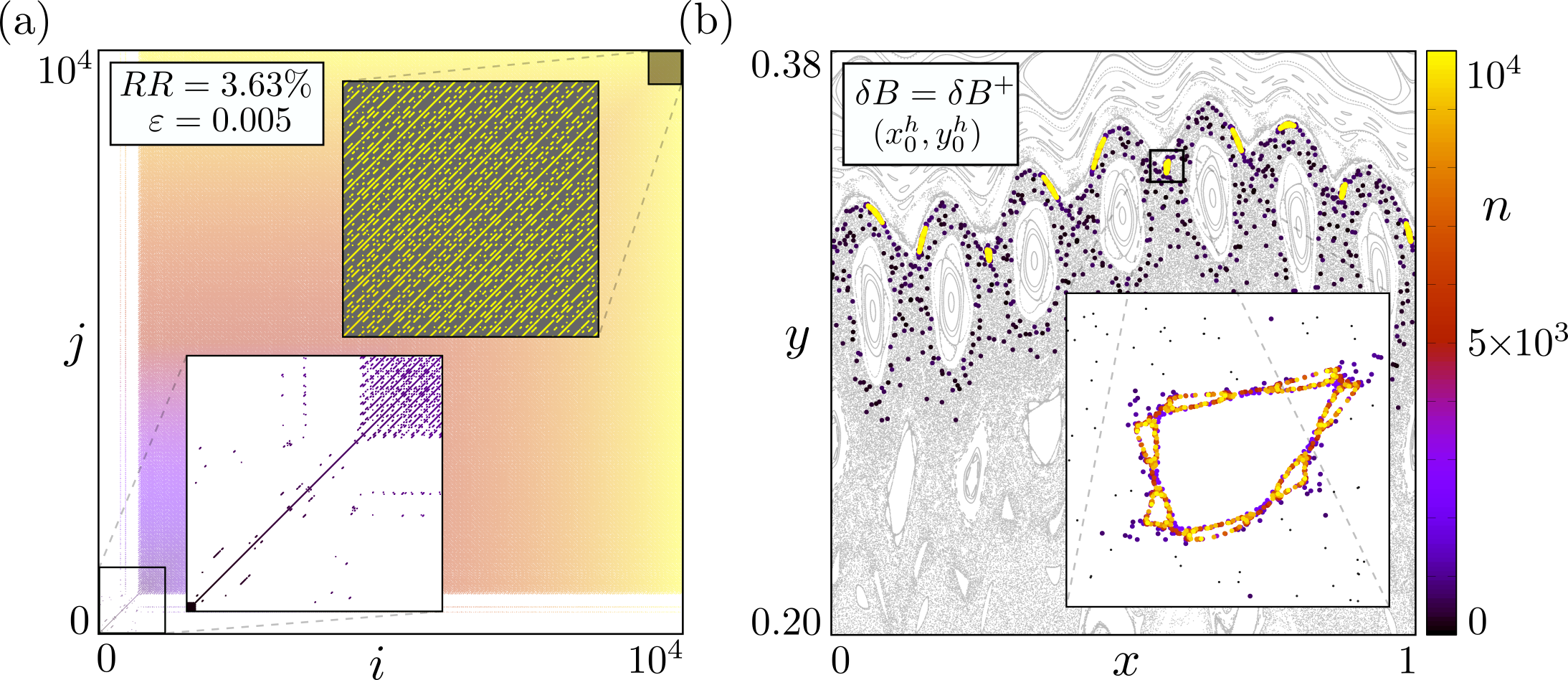}
\caption[RP and the evolution of the most recurrent orbit in the UM]{(a) RP with $\varepsilon = 0.005$ for the trajectory started from $(x_0^h,y_0^h)$. The insets show amplifications of their respective regions in the black squares. Both axes are set in the same colour gradient of the trajectory's evolution; (b) Phase space for $\delta B = \delta B^+$ (grey on the background) along with the same orbit. The inset shows the region around the small periodic island.} 
\label{fig:REC_um_highest_rr}
\end{figure}

Moreover, Fig.\ \ref{fig:REC_um_rr_ps_b+} shows the same results now considering the configuration $\delta B^+$. The 2.5-$\sigma$ detection in (a) selects 228 ICs, now labelled $(x_0^S,y^S)$, that are evolved up to $10^4$ iterations in panel (b). Additionally, Fig.\ \ref{fig:REC_um_rr_ps_b+} (a) highlights a special IC $(x_0^h, y_0^h)$ that presents a computed $RR \approx 3.6\%$ about forty times the average over the ensemble $\langle RR \rangle = 0.09\%$, meaning that an orbit started from $(x_0^h, y_0^h)$ is an extremely recurrent chaotic trajectory that will be further inspected in Fig.\ \ref{fig:REC_um_highest_rr}.

The phase space with the sub-ensemble's $S$ evolution depicted in Fig.\ \ref{fig:REC_um_rr_ps_b+} (b) is again marked by trajectories that spend much time around periodic structures, all highlighted by the yellow points, indicating stickiness phenomena. Furthermore, it is interesting to note a pronounced separation between the upper region, mostly in purple-like colours, and the lower region in red/yellow. This separation is often associated with strong transport barriers found in phase spaces of general non-linear symplectic maps \cite{Caldas2012}, which once analysed from the magnetic confinement point of view, have important implications for tokamaks and plasma-wall interactions \cite{Janeschitz2001,Samm2010}.   

Finally, Fig.\ \ref{fig:REC_um_highest_rr} presents the final numerical observation from the recurrence-based detection approach for the UM. Analogous to what was presented for the BM in Fig.\ \ref{fig:REC_bm_highest_rr}, the RP related to the special IC $(x_0^S,y^S)$ is shown in panel (a), along with its evolution on the phase space for $\delta B^+$ in panel (b). 

The RP in Fig.\ \ref{fig:REC_um_highest_rr} (a) was constructed considering the same recurrence distance threshold $\varepsilon = 0.005$ and the maximum number of iterations $N = 10^4$ established in all analyses for the UM. The strong quasi-periodic is once more evident, along with the chaotic transient-like region observed in the first thousand iterations, highlighted in the initial inset. After this transient time, the stickiness region dominates the entire RP and, in the final inset, we depict the periodic patterns for the last five-hundred iterations. In Fig.\ \ref{fig:REC_um_highest_rr} (b) we observe the behaviour of the same trajectory on the phase space, where the trapping region is highlighted in yellow. The inset shows more clearly the narrow chaotic vicinity around the island.

\chapter{Conclusions}
\label{chap6:conclusions}

In this research, we investigated aspects of chaotic transport in mixed phase spaces of two symplectic maps that model the magnetic field lines of tokamaks under different magnetic configurations; The single-null divertor map, or Boozer map, models the field lines of tokamaks with poloidal divertors and; The ergodic magnetic limiter map, or Ullmann map, that models the magnetic configuration of tokamaks assembled with ergodic magnetic limiters.

Our numerical investigation initially relies on an escape analysis that determines values for the models' perturbation strengths which produces configurations that either enhance or restrict escaping magnetic field lines. Once these values are determined, we employ two original numerical methods to analyse the mixed phase spaces, comparing and illustrating differences between the general behaviour of open field lines in these systems with distinct magnetic configurations.

The first phase space analysis, based on a practical numerical method that visually illustrates distinct transient behaviour of escaping field lines, shows how the spatial organisation of relevant invariant manifolds can be linked to the average dynamical evolution considering the selected magnetic configurations on both models. For configurations that enhance the escape, the computed invariant manifolds present much more erratic and intertwined, creating notable transport channels that are absent when compared to the behaviour of the manifolds for configurations that restrict escaping field lines. These results suggest that the manifolds' complex organisation may construct suitable transport channels that can exhaust unwanted particles from the plasma edge in a controlled manner, preventing the thermal load in locations at the inner wall that are not prepared for extracting the high heat flux from the plasma.   

For the second analysis, it was proposed a recurrence-based approach proved to be suitable for identifying trajectories that widely differs from the average chaotic behaviour, specifically detecting the stickiness phenomena in mixed phase spaces. The numerical observations compared the presence of sticky trajectories between the magnetic configurations in both models. Based on these investigations, it is possible to infer that the stickiness is relatively more noticeable on configurations that restrict escaping field lines. This result has important implications while considering that, due to the configuration of the magnetic field lines, escaping particles from the plasma may access additional confinement regions in the nearest surroundings of magnetic islands in the plasma edge.    

The two selected phase space analyses explore detailed features of the behaviour of open field lines, modelled via chaotic trajectories that either experience stickiness, or are constantly influenced by underlying invariant manifolds present in the mixed phase spaces of the models. These two aspects are inherently connected to transport and diffusion properties that change depending on the magnetic configuration of interest. These analyses may, ultimately, improve our understanding of the general behaviour of magnetic field lines that confines fusion plasma in tokamaks.   

\section*{Perspectives}
As briefly mentioned in previous chapters, there are a few points, raised during specific discussions throughout the text, that would be interesting to pursue in upcoming opportunities.   

First, regarding the analysis of the escape rate, while varying the control parameter, we established a suitable methodology to identify specific values of interest for the control parameter of any system with escaping trajectories. In that sense, it would be possible to apply this methodology in other systems with escape, especially in billiard systems.

Still related to the escape analysis, it would be interesting to further investigate and understand the mechanisms behind the behaviour of the escape rate as a function of the parameter. Figure \ref{fig:MTD_ER_parameter} displays unexpected behaviours for both the BM and the UM, and deeper knowledge regarding what is happening to the dynamics while varying the parameters could be useful for further interpretations.

Concerning the transient motion investigated in Chapter \ref{chap4:transient}, an approach to explicitly quantify the relation between the mean transient behaviour, specifically the preferable paths taken by the escaping trajectories, and the spatial organisation of the invariant manifolds, associated with relevant UPOs in their respective phase spaces, would provide further support to our numerical observations.

Finally, in regards to the stickiness phenomena, the recurrence-based detection discussed in Chapter \ref{chap5:recurrence} is a general approach for near-integrable Hamiltonian systems. In that sense, now that it is possible to identify trajectories that would always experience stickiness in their finite time evolution, it would be suitable to study other open questions related to stickiness, mainly why, when and for how long chaotic trajectories might experience the stickiness phenomena.

\appendix
\setcounter{chapter}{-1}
\chapter{Scientific production}

This appendix is devoted to a compilation of our scientific production during this period of research. We present and briefly discuss the developed manuscripts and the open-source code that resulted from our numerical investigations. 

\section*{Manuscripts}
During the PhD it was developed six manuscripts, five of them were published and one is currently submitted for publication. The first three are related to parallel projects, and the next three are linked to the results presented in this thesis. 

\begin{enumerate}[leftmargin=*]
\item[1st.] {\it Diffusion entropy analysis in billiard systems}. Gabriel I. Díaz, Matheus S. Palmero, Iberê L. Caldas e Edson D. Leonel. Phys.\ Rev.\ E {\bf100}, 042207. Published in October 2019;
\end{enumerate}

\begin{figure}[h]
    \centering
    \includegraphics[scale=0.87]{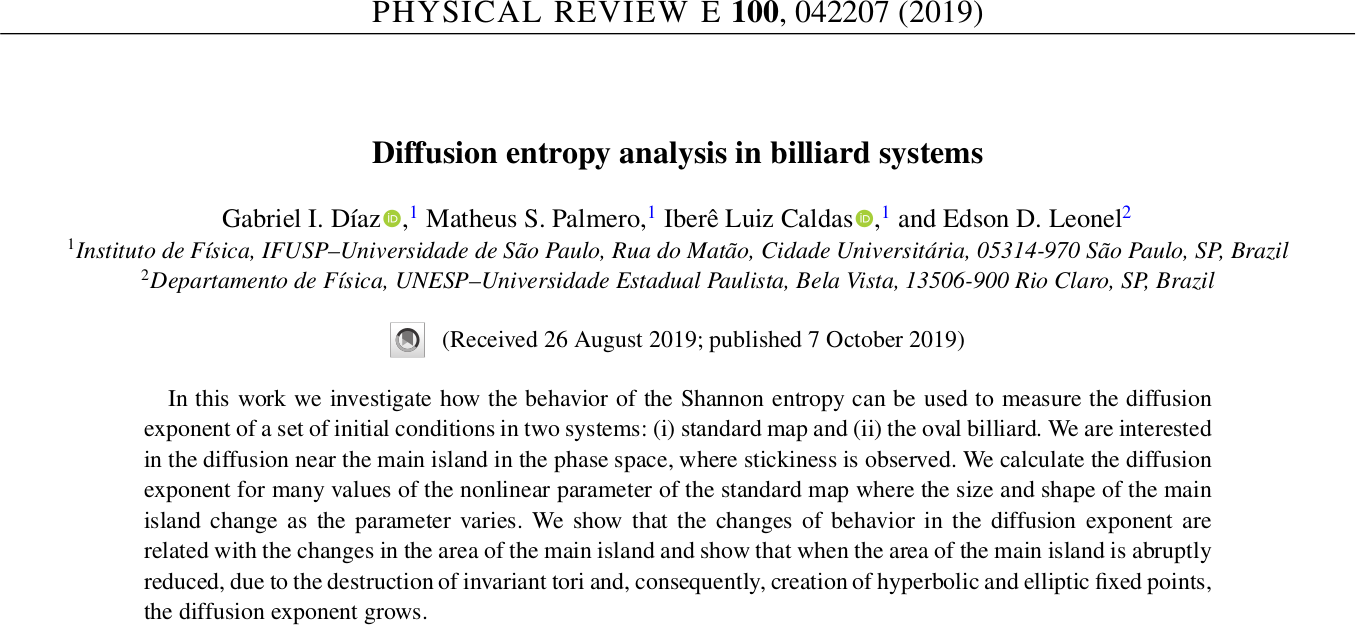}
    \caption*{\url{https://doi.org/10.1103/PhysRevE.100.042207}}
\end{figure}

\vspace{-0.75cm}

\begin{enumerate}[leftmargin=*]
\item[]In this work, we developed a numerical method for calculating the diffusion exponent of chaotic trajectories in the mixed phase space of near-integrable Hamiltonian systems. We apply this method to both the Chirikov standard map and the ovoid billiard. The calculation of the diffusion exponent for different regions of the phase space is based on Shannon entropy as a function of the number of iterations of the map.
\end{enumerate}

\begin{enumerate}[leftmargin=*]
\item[2nd.] {\it Tilted-hat mushroom billiards: web-like hierarchical mixed phase space}. Diogo R. da Costa, Matheus S Palmero, José A. Méndez-Bermúdez, Kelly C. Iarosz, José D. Szezech Jr. e Antonio M. Batista. Communications in Nonlinear Science and Numerical Simulation {\bf 91}, 105440. Published in December 2020;
\end{enumerate}

\begin{figure}[h]
    \centering
    \includegraphics[scale=0.7]{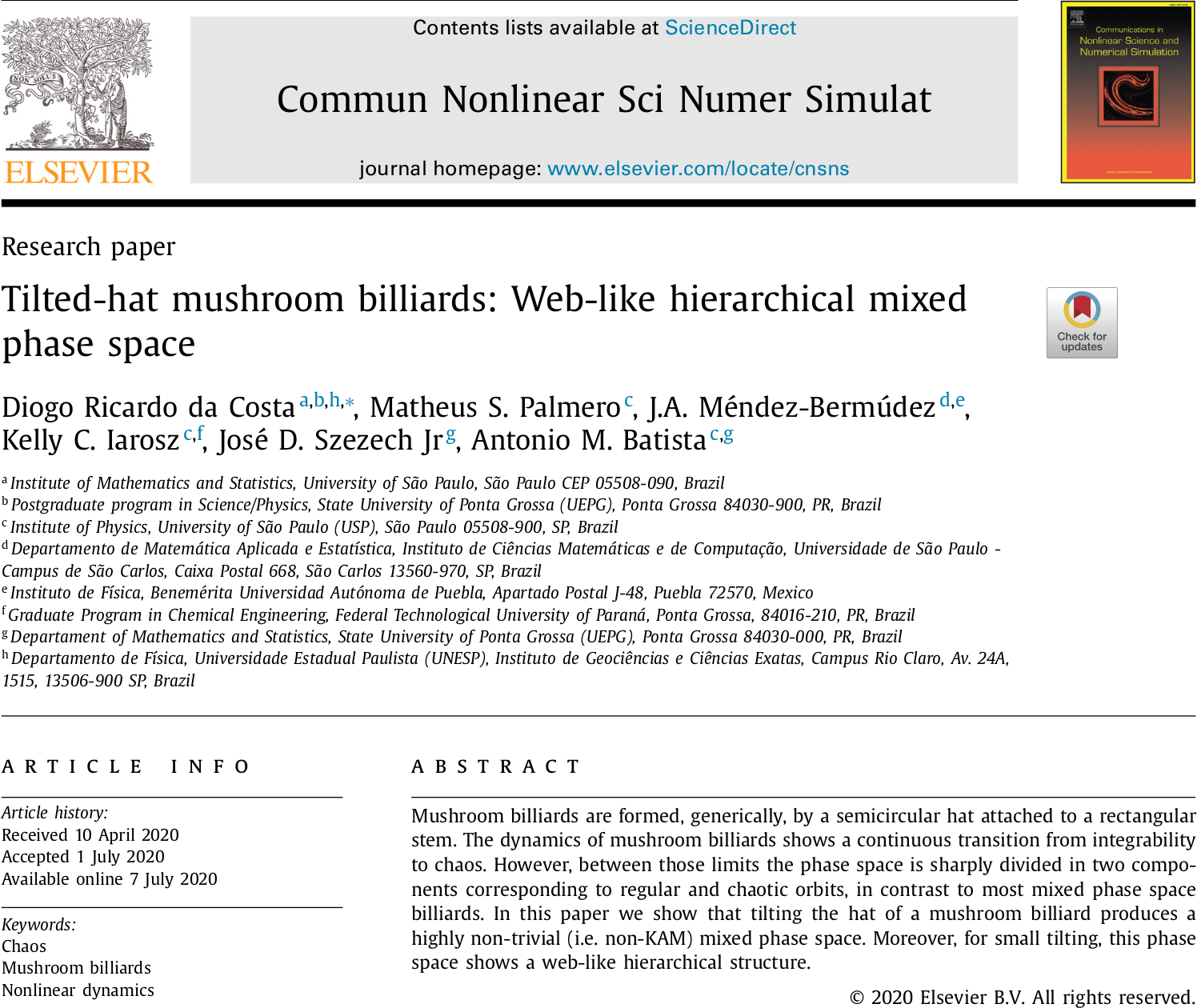}
    \caption*{\url{https://doi.org/10.1016/j.cnsns.2020.105440}}
\end{figure}

\vspace{-0.75cm}
    
\begin{enumerate}[leftmargin=*]
\item[]In this work, we introduced a novel version of the mushroom billiard, namely tilted-hat mushroom billiard. We show that tilting the hat of the billiard produces a highly non-trivial (i.e. non-KAM) mixed phase space. Moreover, for small tilting, the phase space of the model shows a web-like hierarchical structure.
\end{enumerate}

\begin{enumerate}[leftmargin=*]
\item[3rd.] {\it Sub-diffusive behavior in the Standard Map}. Matheus S. Palmero, Gabriel I. Díaz, Iberê L. Caldas e Igor M. Sokolov. The European Physical Journal Special Topics (EPJ-Special Topics) {\bf 230}, 2765-2773. Published in June 2021;
\end{enumerate}

\begin{figure}[t!]
    \centering
    \includegraphics[scale=0.8]{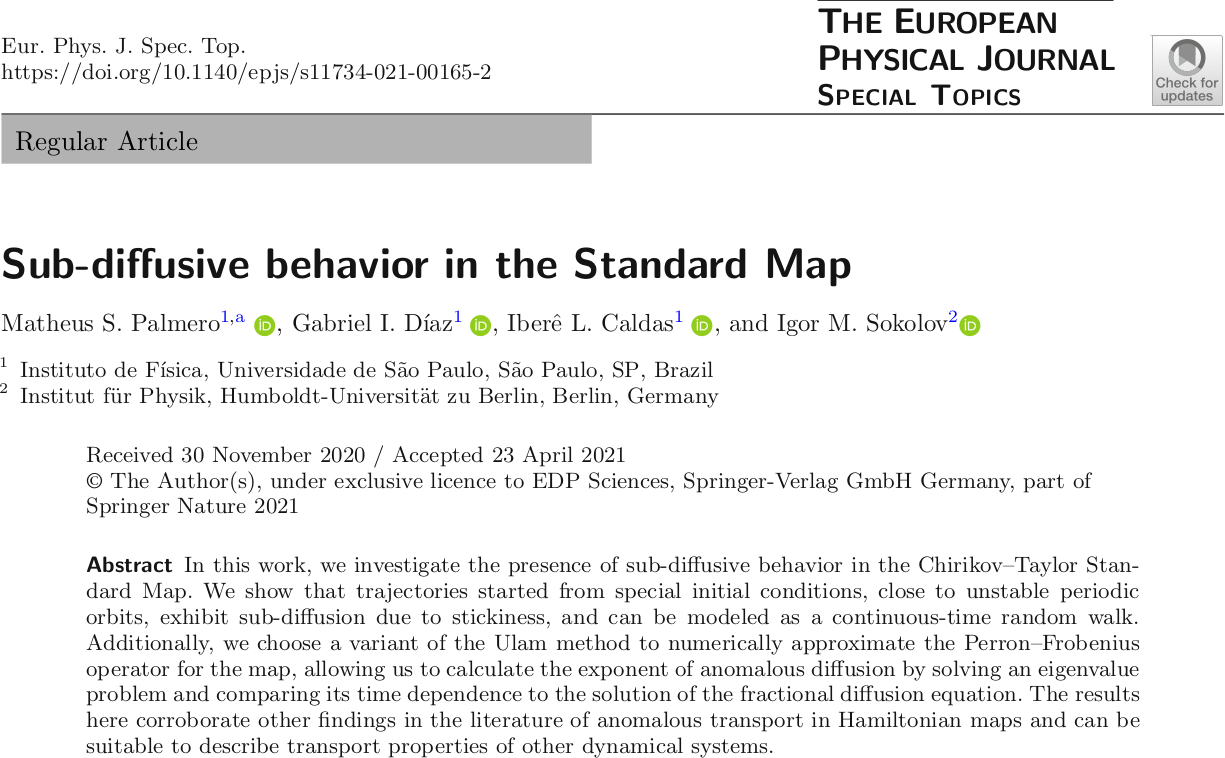}
    \caption*{\url{https://doi.org/10.1140/epjs/s11734-021-00165-2}}
\end{figure}

\vspace{-1cm}
    
\begin{enumerate}[leftmargin=*]
\item[]In this work, we investigated the presence of sub-diffusive behavior in the Chirikov standard Map. We show that trajectories started from special initial conditions, close to unstable periodic orbits, exhibit sub-diffusion due to stickiness, and can be described by a continuous-time random walk model. Additionally, we numerically approximate the Perron–Frobenius operator for the map, allowing us to calculate the anomalous diffusion exponent comparing to the solution of the fractional diffusion equation.
\end{enumerate}

\vspace{-0.5cm}
    
\begin{enumerate}[leftmargin=*]
\item[4th.] {\it Measure, dimension, and complexity of the transient motion in Hamiltonian systems}. Vitor M. de Oliveira, Matheus S. Palmero, Iberê L. Caldas. Physica D {\bf 431}, 133126. Published in March 2022;
\end{enumerate}

\vspace{-0.5cm}

\begin{figure}[h!]
    \centering
    \includegraphics[scale=0.68]{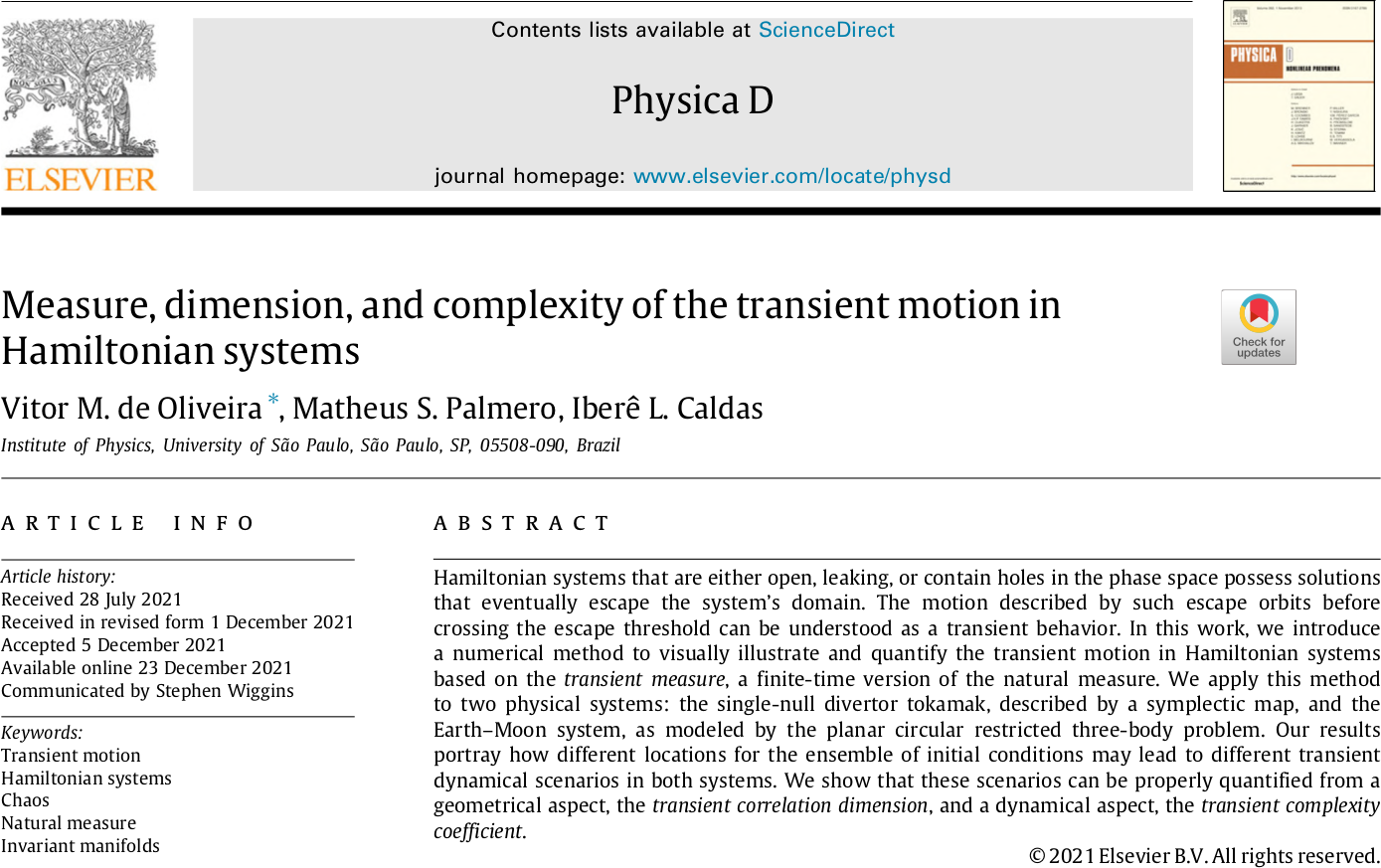}
    \caption*{\url{https://doi.org/10.1016/j.physd.2021.133126}}
\end{figure}
    
\begin{enumerate}[leftmargin=*]
\item[]In this work, we proposed a numerical method to visually illustrate and quantify the transient motion in Hamiltonian systems based on the transient measure, a finite-time version of the natural measure. We apply this method to two systems: the single-null divertor tokamak, described by a symplectic map; and the Earth-Moon system, as modelled by the planar circular restricted three-body problem. Our results portray how different locations for the ensemble of initial conditions may lead to different dynamical scenarios.
\end{enumerate}

\begin{enumerate}[leftmargin=*]
\item[5th.] {\it Finite-time recurrence analysis of chaotic trajectories in Hamiltonian systems}. Matheus S. Palmero, Iberê L. Caldas, Igor M. Sokolov. Chaos {\bf 32}, 113144. Published in November 2022;
\end{enumerate}

\begin{figure}[h]
    \centering
    \includegraphics[scale=0.75]{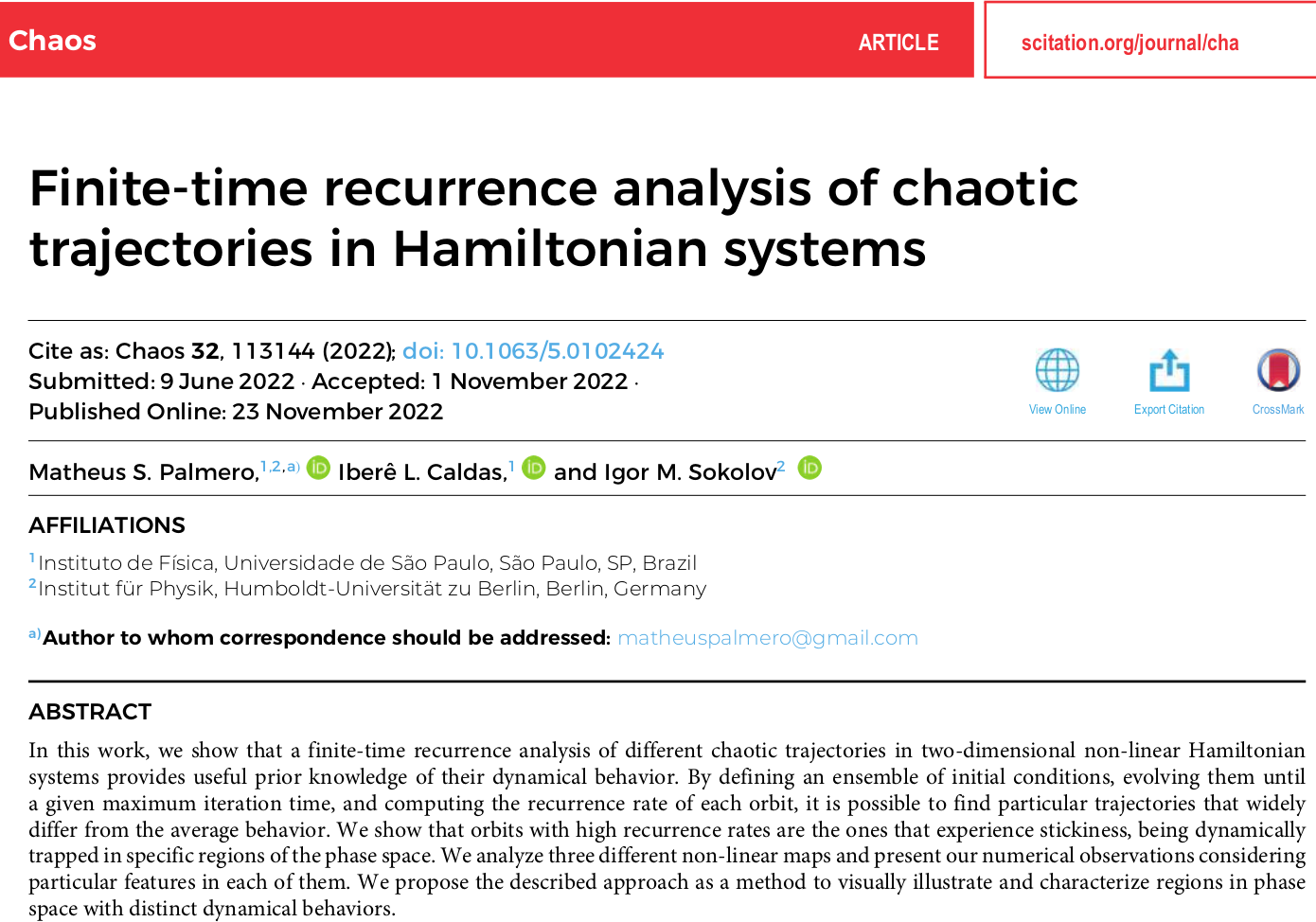}
    \caption*{\url{https://doi.org/10.1063/5.0102424}}
\end{figure}
    
\begin{enumerate}[leftmargin=*]
\item[]In this work, we showed that recurrence analysis of different chaotic trajectories in Hamiltonian systems provides useful prior knowledge of their dynamical behaviour. By defining an ensemble of initial conditions, evolving them until a given maximum iteration time and computing the recurrence rate of each orbit, it is possible to find particular trajectories that widely differ from the average behaviour. We analysed three different non-linear maps and detect the stickiness phenomena.
\end{enumerate}

\begin{enumerate}[leftmargin=*]
 \item[6th.] {\it Confining and escaping magnetic field lines in Tokamaks: Analysis via symplectic maps}. Matheus S. Palmero, Iberê L. Caldas. Submitted to Fundamental Plasma Physics;
\end{enumerate}

\begin{figure}[h]
    \centering
    \includegraphics[scale=0.8]{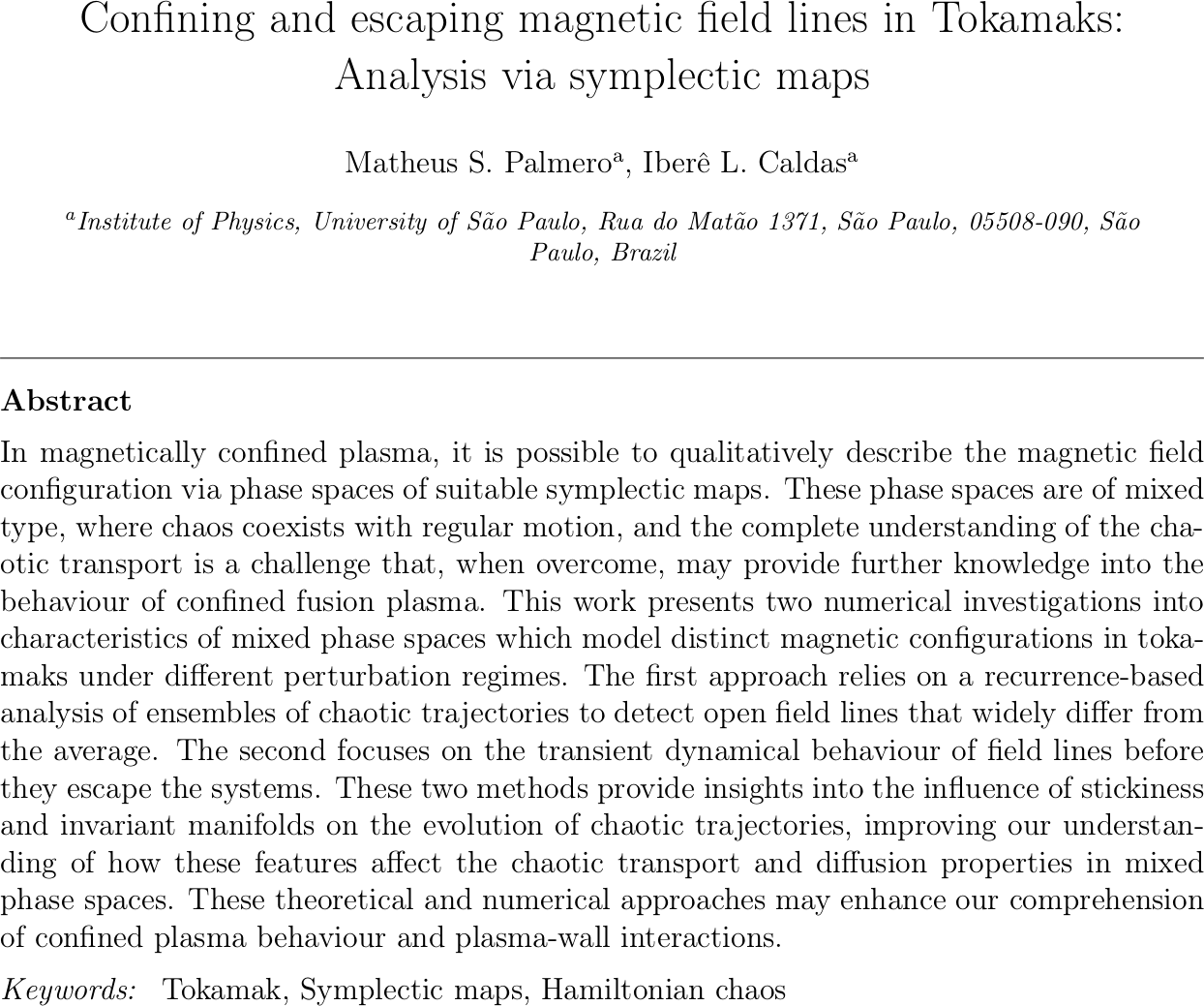}
    \caption*{\url{https://arxiv.org/abs/2304.13810}}
\end{figure}
    
\begin{enumerate}[leftmargin=*]
\item[]In this work, we compiled our numerical investigations on the key features of mixed phase spaces, which model the configuration of magnetic field lines in tokamaks under two different perturbation regimes. The first approach involves recurrence analysis of ensembles of chaotic trajectories to detect stickiness, while the second approach focuses on the transient dynamical behaviour of open field lines before they escape. These two methods provide insights into the influence of stickiness and invariant manifolds on the evolution of chaotic trajectories.
\end{enumerate}

\section*{Open source code}

The computer program that we developed for our numerical simulations was written in C programming language. Its code is fully available at the Oscillations Control Group Wiki page (\url{http://yorke.if.usp.br/OscilControlWiki/index.php/Main_Page}) and also at Github (\url{https://github.com/m-palmero/}).
\renewcommand{\thechapter}{A}

\singlespacing
\bibliographystyle{ieeetr}
\bibliography{phd_thesis}

\end{document}